\documentclass[epj]{svjour}

\usepackage{graphics}
\usepackage[square,sort,comma,numbers]{natbib}
\usepackage{epstopdf}
\usepackage{color}
\usepackage{hyperref}
\hypersetup{%
  pdfpagemode=None, %FullScreen,n
  pdfstartpage=1,
  pdfstartview=FitH,
  pdftoolbar=true,
  colorlinks = true,
  linkcolor=blue,
  citecolor=blue,
  bookmarksopen=false
}

\begin{document}

\bibliographystyle{plain}

\title{Training Schr\"odinger's cat: quantum optimal control}
\subtitle{Strategic report on current status, visions and goals for research in Europe}
\author{Steffen J. Glaser\inst{1} \and
Ugo Boscain \inst{2} \and
Tommaso Calarco \inst{3}\and
 Christiane P. Koch\inst{4} \and
 Walter K\"ockenberger\inst{5} \and
 Ronnie Kosloff\inst{6} \and
 Ilya Kuprov\inst{7} \and
 Burkhard Luy\inst{8} \and
 Sophie Schirmer\inst{9} \and
 Thomas Schulte-Herbr{\"u}ggen\inst{1} \and
 Dominique Sugny\inst{10,11} \and
 Frank K. Wilhelm\inst{12}
}

%SGlaser: Check: Additional authors and institutions to be added: Ugo Boscain ...

\institute{
  Department Chemie, TU-M{\"u}nchen (TUM), Lichtenbergstrasse 4, 85747
  Garching, Germany  \and
  CNRS, CMAP Ecole Polytechnique, Palaiseau, France and Team GECO, INRIA Saclay Ð Ile-de-France \and
  Institute for Complex Quantum Systems and Center for Integrated Quantum Science and Technology,
University of Ulm, Albert-Einstein-Allee 11, D-89069 Ulm, Germany \and
  Theoretische Physik, Universit\"at Kassel, Heinrich-Plett-Str. 40,
  24132 Kassel, Germany \and
  Sir Peter Mansfield Magnetic Resonance Centre, University of Nottingham, University Park, NG7 2RD, United Kingdom \and
  Institute of Chemistry the Fritz Haber Research Center for Molecular
  Dynamics, The Hebrew University, Jerusalem 91904, Israel \and
  School of Chemistry, University of Southampton, University Road,
  Southampton, SO17 1BJ, United Kingdom \and
  Institut f\"ur Organische Chemie and Institut f\"ur Biologische
  Grenzfl\"achen, Karlsruher Institut f\"ur Technologie (KIT),
  Fritz-Haber-Weg 6, 76131 Karlsruhe, Germany\and
College of Science, Swansea University, Singleton Park, Swansea SA2
8PP, Wales, United Kingdom  \and
  Laboratoire Interdisciplinaire Carnot de Bourgogne-Franche Comt\'e (ICB), UMR 6303 CNRS-Universit\'e de Bourgogne, \\9 Av. A.
Savary, BP 47 870, F-21078 Dijon Cedex, France \and Institute for Advanced Studies, Technische Universit\"at M\"unchen, Lichtenbergstrasse 2a, D-85748 Garching, Germany \and
 %Theoretical Physics, Saarland University, 66123 Saarbr{\"u}cken, Germany
 % we got a recent memo that this is in fact the address we should use
 Theoretische Physik, Saarland University, 66123 Saarbr{\"u}cken, Germany
}

\date{Received: \today / Revised version: date}

\abstract{
It is control that turns scientific knowledge into useful technology: in physics and engineering it
provides a systematic way for driving a system from a given initial state into a desired target
state with minimized expenditure of energy and resources -- as famously applied in the Apollo programme.
As one of the cornerstones for enabling quantum technologies, optimal quantum control keeps evolving and
expanding into areas as diverse as quantum-enhanced sensing, manipulation of single spins, photons,
or atoms, optical spectroscopy, photochemistry, magnetic resonance (spectroscopy as well as medical imaging),
quantum information processing and quantum simulation. --- Here state-of-the-art quantum control
techniques are reviewed and put into perspective by a consortium uniting expertise in optimal control theory
and applications to spectroscopy, imaging, quantum dynamics of closed and open systems.
We address key challenges and sketch a roadmap to future developments.
\PACS{
      %{please select by commenting, or suggest more/other:}{}\and
      {02.30.Yy}{Control theory in mathematical physics}   \and
     % {03.67.Ac}{Quantum algorithms, protocols, and simulations} \and
      {03.67.Lx}{Quantum computation architectures and implementations} \and
      {07.57.Pt}{Submillimeter wave, microwave and radiowave spectrometers; magnetic resonance spectrometers, auxiliary equipment, and techniques} \and
   %   {03.67.Pp}{Quantum error correction and other methods for protection against decoherence} \and
      %{07.57.-c}{Spectroscopy in atomic and molecular physics} \and
%%% chr replaced by Coherent spectroscopy
%      {61.05.Qr}{Nuclear Magnetic Resonance (NMR) in structure
  %      determination} \and
  %    {82.50.Nd}{Control of photochemical reactions}\and
      {82.53.Kp}{Coherent spectroscopy of atoms and molecules}
      %{82.56.Jn}{Pulse sequences  in NMR} \and
 %     {85.25.Dq}{Superconducting quantum interference devices (SQUIDs)}\and
%      {87.61.-c}{Magnetic Resonance Imaging (MRI) in medical physics} \and
  %    {87.64.kh}{Electron Spin Resonance (ESR) in biophysics} \and
    %  {89.20.Bb}{Technological research and development}
     }
}

\maketitle

%%%% TABLE OF CONTENTS %%%%
\setcounter{tocdepth}{2}
\tableofcontents
%%%%%%%%%%%%%%%%%%

%-----------------------
%\color{blue}
%Comments/corrections by Steffen (July14)
%This extended file is based on version 67 of the roadmap from July 1.
%
%
%\color{black}
%-----------------------

\section*{Foreword}

%\fwcomment{Frank comments in this colour}
%\fwcomment{Advocating Pacs 02.30.Yy,03.67.Lx,07.57.-c,61.05.Qr}

%\fwcomment{We should have an update to our conversation on authorship. As it stands, we are using what was decided in Nottingham
%but with some heavy lifters within the help, we may want to extend it}

% \color{blue}
The authors of the present paper represent the QUAINT consortium,
%\cite{QUAINT_Webpage},
a European Coordination Action on Optimal Control of Quantum Systems, funded by
the European Commission Framework Programme 7, Future Emerging Technologies
FET-OPEN programme and the Virtual Facility for Quantum Control (VF-QC).
%\cite{VF-QC_Webpage},
This consortium unites expertise in optimal control theory
and applications to quantum systems both in existing and widely used areas such
as spectroscopy and imaging and in emerging quantum technologies such as quantum
information processing, quantum communication, quantum simulation and quantum
sensing.
%quantum dynamics of closed and open systems.
Challenges to quantum control
have been gathered by a broad poll of leading researchers across the
communities of general and mathematical control theory, atomic, molecular, and chemical
physics, electron and nuclear magnetic resonance spectros\-copy as well as medical imaging,
quantum information, communication and simulation.
144
experts
of these fields have provided feedback and specific input on the
state of the art, as well as mid-term and long-term goals. These have been
summarized in this document, which can be viewed as a perspectives paper,
providing a roadmap for the future development of quantum control. As such an endeavour can hardly
 be complete and there are many additional areas of
quantum control applications, such as spintronics,
%\color{black}
nano-optomechanical technologies etc.,
this roadmap is designed as a living document that is available at
at the homepage of the VF-QC web page, %Virtual Facility for Quantum Control
\cite{quantumcontrol},
 where additional aspects as well as new developments and ideas will be included.

% \color{black}

\section{Introduction}
\label{sec:intro}

It is control that turns scientific knowledge into technology.
The general goal of quantum control is to actively manipulate dynamical
processes at the atomic or molecular scale, typically by means of external electromagnetic
fields or forces.
The objective of
quantum optimal control is to devise and implement shapes of pulses of
external fields or sequences of such pulses, that reach a given task in a
quantum system in the best way possible.
Quantum control builds on a variety of theoretical
and technological advances from the fields of mathematical control theory and
numerical mathematics to better electronic devices such as arbitary-waveform generators
with sub nanosecond time resolution or stronger magnetic fields.

The challenge to manipulate nature at the quantum level offers a huge potential
for current and future applications.
Quantum systems and processes cover a wide range from atomic and molecular physics, chemistry, materials (such as semiconductors, superconductors) to biosystems and medicine.
Useful applications range from magnetic resonance imaging and spectroscopy and
the precise control of chemical reactions to emerging second generation quantum
technologies.
Quantum optimal control is part of the effort to engineer quantum technologies from the bottom up,
and
many striking examples of surprising and non-intuitive - but extremely efficient and robust - quantum control
techniques
have been discovered in recent years.
Examples of important current applications
are the precise measurement of magnetic fields with nanometer-scale resolution using NV centers in diamond,
state engineering of Bose-Einstein condensates
and high-fidelity quantum gates in superconducting quantum processors.
%dramatic gains
Similar to the first generation of
quan\-tum-based technology that brought forward
 the semiconductor transistor, the laser, magnetic resonance imaging and
 spectroscopy in the last century, also the currently developing second
generation of quantum technology based on superposition, entanglement and many
body quantum systems are expected to expand on the potential for new disruptive
technologies -- from spintronic devices, quantum metrology, computing
technology, to elucidating chemical reaction dynamics and material properties to
biophysics. Quantum control is a strategic cross-sectional field of research,
enabling and leveraging current and future quantum technology application.

While the precise way to manipulate the behavior of these systems may differ --
from ultrafast laser control to radio waves, the
control, identification and system design problems encountered share
commonalities, while at the same time being quite distinct from classical
control problems. Advancing quantum control therefore requires bringing together researchers from different
application areas to forge a community, create a common language and identify
common challenges.
The further development of this field of research offers many beneficial effects
for today's and tomorrow's society, related to health through
faster, better, safer diagnostics and treatment,
secure communication in a digital world, % based on optimal-control enabled quantum repeaters,
highly accurate navigation systems,
more efficient and clean harvesting of solar power,
the search for resources,
efficient energy storage and transportation,
quantum machines and precision sensing
and monitoring of the environment
\cite{Warren1993,Rice2000a,Shapiro2003,Tannor2007,DAlessandro2008,
Brif2010}.
%Advanced numerical methods
%\cite{Bryson1975b,CNM86,Peirce1987,Shi1988,Krotov96,OZR99,Konnor1999,
%Palao2003,Ohtsaki2004,
%GRAPE
%,
%TVKKGN09,Machnes2011,Fouquieres2011,Eitan2011} can successfully search for
%controls which achieve efficient polarization transfers. In addition,
%analytical methods
% \cite{Pontryagin1964,Jurdjevic1997,Bonnard2003a,Boscain2004a}
%provide bounds on possible transfers.
%
%\color{blue} \noindent $\bullet$ Insert current and future fields of applications and discuss impact on society etc.
%
%Advanced numerical methods:
%
%\cite{Bryson1975b,CNM86,Peirce1987,Shi1988,Krotov96,OZR99,Konnor1999,
% Palao2003,Ohtsaki2004,
%GRAPE ,
%TVKKGN09,Machnes2011,Fouquieres2011,Eitan2011}

The paper is organized as follows: Section 2 is focussed on mathematical optimal
control theory, which is followed by a description of the state of the art as well as mid-and long-term
perspectives of quantum control applications in
atomic, molecular, and chemical physics (section 3),
magnetic resonance (section 4) and
quantum information and communication (section 5). Prospects for applications and commercial exploitation are outlined (section 6) and, in the end, conclusions are given (section 7).
%\color{black}

\section{General aspects and mathematics of optimal control}
\label{sec:math}

The recent advances of quantum control, by now recognized to be
essential for the  further development  of quantum technologies (see,
e.g.,~\cite{Dowling2003,Devoret2014}) and reviewed in
sections~\ref{sec:Chem} to~\ref{sec:QICT} below, are based on
powerful tools from
mathematical control theory
\cite{Pontryagin1964,Lee1967,Kalman1969,Bryson1975,%
Sontag1998,Kirk2004,Coron2007}.
As often in the interplay of mathematics and physics, mathematical
concepts not only proved fruitful for the solution of physics
problems, but in turn specific physical features required further
mathematical development. In the context of quantum control, these are
for example  entanglement and the nature of quantum
measurements. Eventually, a new research domain, \emph{mathematical
  quantum control theory}
\cite{DAlessandro2008,Wiseman2010,Cong2014,Brif2010,Dong2010,Altafini2012,Borzi2012}
has emerged.
% Cong was recommended by Buzek in Southampton %

In general, quantum control theory needs to answer two fundamental
questions, that of controllability, i.e., what control targets are
accessible, cf. section~\ref{subsec:controllability},
and that of control design, i.e. how can a target be reached.
Approaches for control design can be open-loop or
closed-loop. In the latter case, the specific nature of quantum
measurements needs to be taken into account. This is covered in
section~\ref{subsec:feedback} whereas the remaining control design
approaches are  reviewed in section~\ref{subsec:design}.
Open loop techniques include approaches based on the Pontryagin
maximum principle \cite{Pontryagin1964}, i.e., quantum optimal
control, with solutions
obtained analytically (section~\ref{subsubsec:geometric}) or
numerically (section~\ref{subsubsec:oct}). Optimal
control does not make any assumptions on the quantum system
%i.e., it is generally applicable.
%On the downside, it is not always
%straightforward to account for constraints or ensure robustness of the
%solution. The latter is a hallmark of adiabatic control and its
%variants which are covered in section~\ref{subsubsec:adiabatic}.
and also experimental constraints and robustness requirements can be fully taken into account,
i.e. it is generally applicable.
%It is also possible to take into account constraints and robustness of the solution.
%Although adiabatic control and its
%variants (cf. section~\ref{subsubsec:adiabatic})
%are less general, 
However in some cases,
adiabatic control and its variants (cf. section~\ref{subsubsec:adiabatic})
can provide a more straight-forward approach to
consider robustness issues and constraints.

\subsection{Controllability and simulability}
\label{subsec:controllability}

\subsubsection{State of the art}

Controllability analysis determines whether a
quantum system can be brought from any given initial state to any
desired target state, or, more generally, from a given set of initial
states to any set of target states.
%(where closed systems preserve the eigenvalue spectrum).
Adapting results from classical linear control systems
\cite{Kalman1969,Sontag1998} to bilinear
systems with a non-switchable drift term, a rigorous Lie-framework was
developed for closed systems
\cite{Jurdjevic1972, Sussmann1972, Brockett1972, Brockett1973, Jurdjevic1997,
Elliott2009}.
Based on this work, for quantum systems with finite dimension,
controllability by now is  well understood
\cite{Albertini2003,DAlessandro2008,Dirr2008,Kurniawan2012}~\footnote{Note
  that only the
  standard bilinear situation (as in the Schr{\"o}dinger
  equation) is discussed here.}.
Different notions of controllability have been
introduced for pure states, mixed states, and evolution operator
dynamics \cite{Schirmer2001}. The main controllability
test is based on  the rank of the dynamical Lie algebra, which
is generated by the drift and the different control Hamiltonians.
The difficulty of using the rank condition in large systems has
motivated
a geometric approach based on graph theory \cite{Altafini2002},
yielding eventually a
complete set of symmetry criteria for controllability
\cite{Zeier2011}.

In infinite-dimensional systems, the mathematics is much more
intricate and the few existing results are confined to
quantum systems with a discrete spectrum. A first result is a
general obstruction property to exact controllability
\cite{Turinici2000,Mirrahimi2004}. %%% chr: is it clear what
                                %%% obstruction property is?
%% Dom: no but too long to explain, it is a mathematical obstruction                                
This was recently amended by positive results
about exact \cite{Beauchard2005,Beauchard2010} and approximate
%%% chr: citation for approximate controllability seems to be missing
%% Dom: the main result is the paper by the group of ugo
controllability \cite{Chambrion2009},
based on Galerkin techniques.
For the specific case of a generalized Jaynes-Cummings
model, i.e., several two-level systems
coupled to  a harmonic oscillator, symmetry methods were used to assess
controllability  \cite{Keyl2014}. %%% chr: with which result?
%% Dom: with Lie algebra techniques

Simultaneous controllability concerns the control
of a continuum of finite dimensional quantum
systems by only a few control fields. This is also known as ensemble
controllability.
Approximate and exact controllability results have been obtained recently in this direction
\cite{Beauchard2010,Li2006,Li2009}.
%%% chr: the previous sentence should be made more precise (what results?), Dom:OK
This analysis is important for designing control fields which are
robust to experimental
imperfections \cite{Turinici2004,Kobzar2004,Kobzar2008,Skinner2011,Kobzar2012,Goerz2014}.

In addition to experimental imperfections and fluctuations,
decoherence may pose an obstacle to control.
For open quantum systems, the control field usually cannot
fully compensate dissipation, as rigorously shown for the case of
Markovian dynamics \cite{Altafini2003}. These
results for Markovian dynamics were recently generalized to a complete
Lie semi-group picture~\cite{Dirr2009,OMeara2012}. In contrast,
controllability of systems with non-Markovian dynamics presents by and
large uncharted territory~\cite{Reich2015}.
Reachability
is expected to be larger for non-Marko\-vian dynamics since
non-Markovianity implies information back-flow from the environment to
the system~\cite{Breuer2009}.  Indeed, an
explorative study showed non-Markovian map synthesis to be
stronger than its restriction to the Markovian case (as anticipated in
Ref. \cite{Lloyd2001}). In contrast, and  surprisingly so, for the simpler
problem of state
transfer under swichable Markovian noise this is not the case
\cite{Bergholm2012}.

Finally, even if a system (A) is not fully controllable, its
controlled dynamics
may still suffice to generate a desired effective evolution
as brought about by another quantum dynamical system (B). This is the
paradigm of finite-dimensional quantum simulation. Adapting the tools from
controllability analysis, one readily sees that system A can simulate
system B if the  system algebra of A encompasses that of system B. A
recent generalisation of the results in Ref. \cite{Zeier2011}
provides a complete set of symmetry criteria
together with an algorithm in order to decide simulability on system-algebraic
grounds \cite{Zimboras2015}. Resorting to exactly solvable problems
may help this case \cite{Burgarth2010a}.

\subsubsection{Mid-term prospects: goals and challenges}
%\sgcomment{Dominique, in sections 3, 4, and 5, "mid-term prospects" and "long-term visions" are given separately. Could a similar separation also be done in sections 2.1, 2.2, 2.3 and 2.4? However, we don't have to force this if it does not make sense. Dom: I change the organization and I add a long term vision part for the mathematical section}

The major challenge is to better understand controllability in open quantum
systems. For  infinite-dimensional open
systems with a discrete spectrum which undergo Markovian evolution, a
rigorous  understanding is non-trivial but may be pursued by
extending the standard Lie-Galerkin techniques
to non-unitary evolution.
With the final goal of identifying sufficient conditions for
approximate controllability of open systems analogous to
closed systems \cite{Chambrion2009}, a mid-term
perspective is to perform a sound regularity analysis of a
well-posed and defined mathematical control problem. %%% chr specify what is meant
                                %%% by well-posed mathematical control problem
                                %% Dom: well-defined with the right mathematical assumptions

No rigorous controllability analysis so far has tackled open quantum
systems which undergo non-Markovian dynamics.
From a control point of view, non-Markovianity may be connected  to
beneficial aspects of the system-environment interaction, whereas
the detrimental part is linked to those Markovian dynamics that cannot
be remedied -- an aspect important not only for controllability but also
for dissipative state engineering and quantum memories.
Along similar lines, projective measurements may entail Zeno-type
dynamics exploring directions that do not show up in the unobserved
system~\cite{Burgarth2014}.

Even for closed quantum systems, several highly relevant questions are
still open. These include the
exact or approximate controllability of the
Schr{\"o}dinger equation with mixed or continuous spectrum.
This question is important since it covers
dissociation and ionization processes, the control of which is a major
goal in chemical  physics, cf. section~\ref{sec:Chem}.

%A rigorous formulation in the
%general case of the simultaneous controllability of an ensemble of quantum
%systems is still lacking and will be an important prerequisite for the
%applications of quantum control.
%In addition of a mathematical information about
%the dynamics of the system, controllability studies could also provide some
%other interesting results about the way to manipulate the system. These results
Beyond controllability, a precise description of the reachable set (in
the open case) and upper bounds on minimal time to reach target states
are largely unknown. %and any information
%about the control field to use. Such aspects have not been extensively studied
%up to date.
Also a
universal estimate time to control finite-dimensional quantum systems (with
necessary drift) is still an open challenge.

\subsection{Control design}\label{subsec:design}

%\subsection{Open-loop  analytical control strategies}
%\subsection{Quantum control by analytic control fields}\label{secana}
%\sgcomment{Dominique: This section is related to control fields para\-metrized by analytical functions, whereas the title implies that it is about analytical solutions (which are only described as part of section 2.3) based on geometric optimal control. (A reader interested in analytical approaches could easily miss the contents of section 2.3). Perhaps the title should be changed or the analytic approaches discussed in section 2.3 could be moved to section 2.2 (?).
%Analytical approaches to be mentioned could also include
%Cartan decomposition \cite{Khaneja01b} and N. Khaneja, S. J. Glaser,
%"Cartan Decomposition of SU(2$^n$) and Control of Spin Systems",
%{\it Chem. Phys.} 267, 11-23 (2001),
%sub-Rieman\-nian geodesics based on the so-called quantum gate design metric \cite{Khaneja2002, KHSY07},
% singular arcs
% \cite{Khaneja2003,Lapert2010}, etc \cite{RKG:2002}.\\
%Dom: I change the title of this section and its content. I include in particular the results from geometric optimal control
%}

\subsubsection{Geometric optimal control---state of the art}
\label{subsubsec:geometric}

Optimal control theory can be viewed as a generalization of the
calculus of variations for problems with dynamical constraints.
Its modern version was born with Pontryagin's Maximum Principle
in the late 1950's \cite{Pontryagin1964}. Its development was
boosted by using Kalman filters
\cite{Kalman1960,Kalman1969} in the Apollo programme and it is now a key tool
in many applications  including quantum mechanics.

Solving an
optimal control problem means finding a control law, i.e., a pulse
sequence, such that the corresponding trajectory satisfies given
boundary conditions and equation of motion and
minimizes a cost criterion such as the energy or duration of the
control.  Pontryagin's Maximum Principle can be formulated in terms of
classical Hamiltonian equations.
Usually, first one obtains extremal trajectories, i.e., controls,  by
solving these equations. Secondly, one selects among the extremals those
which minimize the
cost. Although  looking straightforward, the
practical use of Pontryagin's Maximum Principle
is far from trivial and control problems require
geometric analysis and numerical methods. The latter are described
below in section~\ref{subsubsec:oct}.

If the system is sufficiently simple, for example low-dimensional,
the optimal  control problem may be
solved analytically. Typically, this requires a geometric
analysis of the control problem from which one can deduce the
structure of the optimal solution, a proof of global optimality, and
physical limits such as the minimal time to reach the target \cite{Lee1967}.   %%% chr: It would be good to add a citation here, Dom: OK
Mathematical tools that were developed recently
\cite{Agrachev2004,Bonnard2003,Jurdjevic1997,Boscain2004}
%%% chr: stronger than what? how are these stronger tools related to
%%% the geometric tools? Dom: Stronger than the old ones, I remove this word
allowed to solve problems of increasing difficulty including fundamental
control problems
for closed \cite{Khaneja2002,Boscain2006,DAlessandro2001,DAlessandro2008,
Boscain2002,Garon2013} and open quantum systems \cite{Bonnard2009,Lapert2010,
Khaneja2003a,Khaneja2004,Sugny2007,Yuan2012}. This method is able to treat quantum control problems ranging from 
two and three level quantum systems, two and three coupled spins to two-level dissipative quantum systems whose dynamics is governed by the Lindblad equation.
%%% chr: The previous sentence is not specific enough: what problems?
%%% what kind of difficulties? it would be good to provide more
%%% detail and to enlarge this paragraph to give a better idea of what
%%% this method is capable of
%%% Dom: I add some examples

Extending methods of geometric control to infinite
dimensions~\cite{Chambrion2009} was unexpectedly successful. Also, the
Cartan decomposition method turned out to be an efficient tool for the
control of spin systems \cite{Khaneja2001,Khaneja2001a}.
This decomposition leads to
a drastic reduction of the dimensionality of the quantum system, and
may enable its geometric description. A relatively little explored way
to derive analytic control fields is singular optimal arcs
\cite{Bonnard2003}. This method has  been applied to some exemplary
quantum systems \cite{Khaneja2003,Reiss2002,Lapert2010,Wu2012,Yuan2005,Yuan2007,Yuan2011}, but
many aspects remain to explored. In particular, this approach could be interesting to derive analytical optimal control fields.
%%% chr: which aspects? why would they be promising? Dom: Ok, I add a sentence
These methods can also be advantageously combined with numerical
optimization techniques in order to manipulate more complicated
quantum systems \cite{Lapert2012}.

\subsubsection{Numerical optimal control---state of the art}
\label{subsubsec:oct}
%\subsection{Open-loop numerical optimal control strategies}\label{secoct}
%\subsubsection{State of the art}
% Optimal control theory can be viewed as a generalization of the
% calculus of variations for problems with dynamical constraints.
% Its modern version was born with Pontryagin's Maximum Principle
% (PMP) in the late 1950's \cite{Pontryagin1964}. Its development was
% boosted by using Kalman filters
% \cite{Kalman1960,Kalman1969} in the Apollo programme and it is now a key tool
% in many applications  including quantum mechanics. Solving an
% optimal control problem means finding a control law (aka pulse sequence)
% such that the corresponding
% trajectory satisfies given boundary conditions, an equation of motion and
% minimizes a cost criterion (e.g.,
% energy or duration of the control).
% Usually, first one goes for extremal trajectories
% solving a generalized Hamiltonian system subject to the maximization
% condition of the PMP.
% Secondly, one selects among the extremals those
% which minimize
% cost. Although  looking straightforward, the
% practical use of the PMP is far from being trivial and control problems require
% geometric analysis and numerical methods. The geometric approach being described in Sec.~\ref{secana}, this paragraph
% only focuses on the numerical optimal control tools.

If the set of control equations, resulting from the maximum principle,
cannot be solved analytically, numerical optimal control theory
provides a viable alternative.
Algorithms comprise (i) gradient ascent algorithms \cite{Khaneja2005}
(which can be extended to second-order  quasi-Newton and Newton
methods \cite{Fouquieres2011,Machnes2011,Ciaramella2015}) and (ii)
Krotov-type methods \cite{Krotov1996,Maday2003,Reich2012}, also
permitting this extension \cite{Eitan2011}. The main difference
between these two approaches is that the control is
updated (or replaced) for all times simultaneously in case (i) and
sequentially in case (ii) which implies different convergence
properties. The algorithms are comparatively easy to use and several
program packages include optimal control modules, i.e.,
SIMPSON \cite{Tosner2009}, SPINACH~\cite{Hogben2011},
DYNAMO \cite{Machnes2011}, and
QuTiP \cite{Johansson2013}.
%to a new class of control problems without no restriction
%on the dynamics.  In recent years,
Modifications to account for experimental imperfections and
limitations and to ensure robustness of the solution have been
introduced
\cite{Werschnik2007,Skinner2004,Walther2009,Janich2011,Motzoi2011,Spindler2012,
Reich2014,Hincks2014,Lapert2009,Kobzar2008}, and numerical optimal control theory has been extended to
open quantum systems \cite{Bartana1997,Kallush2006,SchulteHerbruggen2011,Schmidt2011,Floether2012,Goerz2014a}.

It is also possible to restrict the control solution to a predefined
analytical form. In this case, the control only depends on
a limited set of free parameters which are optimized
\cite{Skinner2010,Meister2014,Machnes2015}.
Applications of numerical optimal control are discussed below in
sections~\ref{sec:Chem} to ~\ref{sec:QICT} but by now have grown too
numerous for a complete bibliography.

A caveat of numerical algorithms which are derived from the maximum
principle is that they require to solve the system's equations of
motion many times, at least twice per iteration step. This potentially
hampers application to high-dimensional quantum systems. So far, three
approaches have been pursued to cope with this issue.
(i) Optimization of the state space representation:
For the  high-temperature regime of
ensemble NMR spectroscopy, the SPINACH package~\cite{Hogben2011}
uses state propagation with interactions limited to pertinent
short-range ones (thus leading to efficient truncation of the
underlying spin Lie algebra) for simulating unprecedentedly large
spins systems  such as entire biomolecules \cite{Savostyanov2014} with
striking precision.
(ii) Gradient-free optimization: For many-body quantum systems,
the chopped random basis (CRAB) method uses a parametrization of the
control and
interfaces the tensor compression of the time-dependent density matrix
renormalization group with parameter
optimization~\cite{Doria2011a,Caneva2011}. The number of propagations
may be significantly reduced compared to gradient-based optimization,
provided a small basis is sufficient to represent the control.
(iii) Local control theory: It
determines the control field from instantaneous dynamical
properties of the system by requiring monotonic increase or decrease of a
performance index \cite{Engel2009a,Sugawara2003}. Under well-established
conditions \cite{Wang2010b}, the system converges
asymptotically towards the target. Local controls are
reminiscent of the closed-loop Lyapunov method in stabilization; and
correspondingly the approach has also been termed tracking.
%%% chr: There is a paper by Rabitz from the 1990s that has tracking
%%% control in the title that needs to be cited here, please add 
%% Dom: I do no find it ??? 
Due to the inherent nature of the  quantum measurement process,
this approach has been transformed into an open-loop
control law in the quantum world. In this approach, the equation of
motion needs to be solved only once.

Mathematical quantum optimal control theory  shows how to design
optimal control fields and describes under which
conditions they exist \cite{Peirce1987,Albertini2003}. It does,
however, not investigate the complexity
inherent to the search \cite{Lloyd2014}.
%%% chr: we need to refer here to recent work by Simone Montangero on
%%% the relation between search complexity and effective state space
%%% dimension (see also the discussion in Shai's GOAT paper)
The search complexity is related to the
description of the control landscape which specifies the control
objective as a function of the control variables. Different results on the structure of the control landscape 
%%% chr: explicitly spell out which results
have been established  recently
\cite{Chakrabarti2007,Fouquieres2013,Pechen2011,Rabitz2012}. One of the
main claims has been that, provided there are no constraints on the
control, the control landscape does not contain local traps in all
controllable closed
and some specific open quantum systems \cite{Brif2010}. This proof
has recently been  contested and is still the object of a vivid debate
\cite{Pechen2011,Rabitz2012,Pechen2012}.
%%% BEGIN COMMENT
% \sgcomment{Dominique, Sophie: Can these conditions be more detailed, does the
% analysis assume continuous and unlimited controls (???) Dom: OK}
%%% END COMMENT
Once constraints are present, such as finite control field amplitude,
this may induce  traps in the control landscape
\cite{Riviello2015,Moore2015}.
%%% chr: there are one or two earlier papers by the Rabitz group one
%%% this, I think Katharine Moore Tibbetts (possibly with only one
%%% last name at the time) is the first author, please add Dom: OK
Whether and how imposing robustness of the control influences the
control landscape is an open question.

\subsubsection{Control via adiabatic dynamics---state of the art}
\label{subsubsec:adiabatic}

While optimal control theory is a general approach that is
particularly well suited for identifying fast controls, the target can
also be tackled by adiabatic
techniques \cite{Bergmann1998,Vitanov2001,Guerin2003,Teufel2003}. The
most widely known among these is Stimulated Raman Adiabatic Passage
(STIRAP)\cite{Bergmann1998}.
%%% chr: STIRAP should be explicitly mentioned because it is used
%%% later
Adiabatic techniques usually employ a sequence of very intense pulses
over a comparatively long timescale
which enforces adiabatic following of the system dynamics. The pulses
can be frequency chirped according to the
structure of the energy levels. Such processes are inherently robust
to small variations of laser or system parameters and thus
well-suited to open-loop control. 
%Adiabatic arguments can also be used
%to analyse  controllability, see Ref.  \cite{Boscain2015} for
%systems with an infinite-dimensional Hilbert space.
%%% chr: I don't see how the remark on controllability is connected to
%%% the rest of the paragraph Dom: I remove this sentence
A main drawback of adiabatic control is the total time and energy,
requirements which cannot always be met in an experiment.

To address this issue, shortcuts to adiabaticity were developed
\cite{Torrontegui2013}. %, where
%the adiabatic approximation is modified by changing the correction term in order to derive
%exact analytical solutions in finite time. More generally, the shortcut to  adiabaticity
This approach  can be viewed as an inverse engineering technique based on
Lewis-Riesenfeld invariants. It has been applied to a variety
of quantum systems \cite{Ruschhaupt2012,Chen2010,Daems2013,Ban2012}
for which such an invariant exists.
Even for an ensemble of quantum systems, analytical solutions based on
this technique are not
out of reach \cite{Daems2013}. Similarly to enforcing robustness in
ensemble optimization using optimal control
\cite{Kobzar2008,Goerz2014}, the basic idea  consists in selecting
among a family of (exact) solutions the ones
that are effectively robust to some extent to variations of the system
parameters.

%  Secondly, adiabatic
% techniques \cite{Bergmann1998,Vitanov2001,Guerin2003,Teufel2003}
% are usually achieved by a series of intense pulses which can be
% frequency chirped according to the
% structure of the energy levels and sufficiently slow so as to fulfill
% adiabatic conditions. Such processes are robust
% to small variations of laser or system parameters and thus
% well-suited to open-loop control. Note the recent controllability
% results obtained for systems with an infinite-dimensional Hilbert space %\cite{Mirrahimi2009}
% by using adiabatic
% arguments \cite{Boscain2015}. A main drawback of adiabatic control
% is the total time and energy.
% %These conditions are only satisfied for some experimental setups.
% To improve this aspect, efforts were made
% to develop a shortcut to adiabaticity
% \cite{Torrontegui2013}, %, where
% %the adiabatic approximation is modified by changing the correction term in order to derive
% %exact analytical solutions in finite time. More generally, the shortcut to  adiabaticity
% an approach that can be viewed as an inverse engineering technique based on
% Lewis-Riesenfeld invariants as recently applied to a variety
% of quantum systems \cite{Ruschhaupt2012,Chen2010,Daems2013,Ban2012}.

\subsubsection{Mid-term prospects: goals and challenges for control design}

As with controllability analysis, the major challenge for control
design is posed by open quantum systems since most systems of interest
interact with their environment to a non-negligible  extent.
An important challenge is the extension of the standard adiabatic and
shortcut  techniques to open quantum systems. Beyond some
preliminary results \cite{Joye2007,Sarandy2005}, there are many open
questions, for example, how to generalize
Lewis-Riesenfeld invariants to open system dynamics.

All open-loop control approaches discussed above assume sufficient
knowledge of the quantum system. Correspondingly, for open quantum
systems,  system identification needs to include dissipative parameters.
A first step to combine quantum control with implicit learning about the
system parameters is the ADHOC
technique~\cite{Egger2014a}.

Decoherence in open quantum systems can also be used as a resource in
what has become known as dissipative state
engineering~\cite{Buchler2008,Verstraete2009,Krauter2011,Schirmer2010}. Optimal
control theory allows for tackling dissipative state engineering for
quantum systems that are too complex for manual design of the
driven-dissipative dynamics. This naturally includes incorporation of
the noise as additional optimization
parameter~\cite{Bergholm2012,Pechen2014}. When added to coherent
controls, time-dependent Markovian noise (amplitude-damping) enables the
control system to transform any initial into any desired target
state~\cite{Bergholm2012}.
It can be easily integrated in toolboxes like
DYNAMO~\cite{Machnes2011}, yet the implementation in realistic
settings remains unexplored.

On an algorithmic level, the basics of numerical quantum optimal
control are
well established for both Markovian and non-Markovian
open system evolutions, as described above. The main challenge there
is the efficient
numerical modeling of increasingly complex dynamics.
A promising route is provided by stochastic methods which only require
propagation of (several) pure states to unravel the true
dynamics \cite{Bouten2007,parthasarathy1992,Schmidt2011}.
These methods are applicable
to both systems interacting with an environment or subjected to
measurements that can be modelled by a stochastic process.
An open question in this context is to find the most efficient way to
control these systems such that the control is robust with respect to the
stochastic parameter \cite{Annunziato2010}.

In view of numerical optimal control theory as an open-loop technique,
integration with application is a crucial issue that has only very
partially been addressed. In order to make the theory
more useful for specific experiments, the gap
between theory and experiments needs to be bridged. In spite of recent
progress, a systematic  and efficient algorithm addressing this issue
is still lacking. One option consists in combining open-loop optimal
algorithms to closed-loop control
techniques in order to reduce their respective drawbacks \cite{Egger2014a}.
%\sgcomment{add also older references, e.g. Corey, Laflamme, Emsley ??? }.
Using these two tools cooperatively
is expected to provide flexible quantum control. A current challenge is
still to improve the
computational speed and accuracy of the algorithms. A promising
approach to this end is the combined use of
numerical and geometric optimal control methods \cite{Lapert2012}.
Analogously to open quantum systems, improving the computational speed
will be an important
prerequisite to attack control problems of increasing complexity,
representing experimental settings in a realistic way. Eventually, the
goal is to enable the problem-adapted design of controls in
time-critical applications.

In view of a better understanding of control complexity, a rigorous
understanding of two crucial issues needs to be developed --
interrelation between minimum control dimension and (effective)
system dimension \cite{Machnes2015,Lloyd2014} %% chr: please add here ref to
                                %% S. Montangero
and the questions of traps in the quantum control
landscape. Even if many quantum control problems have no traps, i.e.,
their
control landscapes  are well-suited to the local numerical
optimization approa\-ches discussed above, this is
not always the case. %%% chr: it would be good to add an example
One then needs to resort to global optimization
techniques which are numerically expensive and generally do not
perform very well. Whether and how global optimization methods such as
dynamic programming may be (better) adapted to quantum control is also
an open question.
The efforts to solve more complicated control problems via geometric optimal control techniques must continue with a special emphasis on open quantum systems and on non-linear dynamics \cite{Zhang2011}.
%%% chr: There is not a single mid-term goal on geometric control, can
%%% you add one or two? Dom: Ok

%The efforts on motion planning via geometric optimal control techniques must
%continue via a special emphasis on open systems governed by the Lindblad
%equation. A better understanding of the structure of the reachable set of open
%systems will be provided by such studies.

%A final objective is to develop an explicit connection between the mathematical
%control techniques and numerics in view of a
%general formalism permitting to describe different control applications and to
%support them by rigorous mathematical results on time optimality and reachability
%as listed above.
% The used numerical algorithms are still too dependent of the considered example.
%Some of them have a limited efficiency because they are not supported by sound
%mathematical results.

%%% adiabatic

Another fundamental open question is whether it is possible
to connect the Lewis-Riesenfeld invariants, used for the shortcuts to
adiabaticity,  with the formal framework of the Pontryagin Maximum
Principle of optimal control theory. This would hold the promise of
combining the best of two worlds.

\subsection{Quantum feedback control theory}
\label{subsec:feedback}
\subsubsection{State of the art}

In contrast to the open-loop techniques discussed in
section~\ref{subsec:design},
closed-loop strategies \cite{Dong2010} are capable of coping with
unpredictable disturbances. This
approach is widely used in classical control theory, where
information from the state of the system is fed back to the controller to
correct the field in action. However the perturbing effect of
quantum measurements precludes direct application of
classical concepts. It required development of new techniques tailored
to quantum dynamics \cite{Wiseman2010,Doherty2000}. An indirect
application is found in experimental closed-loop quantum control,
where each cycle of the loop is realized experimentally, including
preparation of a new sample in each cycle \cite{Judson1992}. Different
learning algorithms, such as genetic
algorithms, have been developed to design the new control field after each
iteration \cite{Brif2010}. %This second aspect will be not discussed below.

Controlling a quantum system by feedback in a closed loop
can be realized in two different ways
\cite{Yamamoto2014} -- using the measurement-based feedback approach
\cite{Wiseman1993} or the coherent feedback method \cite{Lloyd2000}. As in
classical control theory, the first option is based on a measurement
process and real-time manipulation of the system depending on the
measurement result. The
measurement feedback approach built on quantum filtering techniques
\cite{Bouten2007} is nowadays mathematically well described
\cite{Wiseman2010,Bouten2009}. Quantum filtering consists in generating an
estimate of the state of the system from the measurements performed
on it. Measurement-based feedback has been shown to
be efficient  \cite{Combes2006,Combes2008}, making
real-time control of quantum systems experimentally feasible
\cite{Sayrin2011,Kirchmair2013}.

In the second option, no measurement
is used  and the quantum system is directly connected to a quantum
controller. The  experimental feasibility of coherent feedback control
has recently led to impressive results
\cite{James2008,Gough2009,Mabuchi2008,Iida2012}.
%%% chr: can you be more specific what these results entail?

%\sgcomment{Thomas/Dominique: Somewhere here we should include a short paragraph
%(and references) on the point raised by %Vladimir Busek during the QUAINT workshop
%concerning problems related to quantum systems that are controlled by quantum systems.
%DONE BY THOMAS IN PAR. BELOW}

\subsubsection{Mid-term prospects: goals and challenges}
It is the specific quantum nature of systems that in general makes
them susceptible to
measurement backaction. Thus in the quantum domain, finding an
experiment-class adapted  balance between open-loop control techniques
and closed-loop feedback counterparts is one of the major challenges.
Moreover, since measurement-based feedback is
restricted to the processing time of the classical components, which have the
slowest time scale, exploiting the limits of coherent feedback control is
useful in view of future technologies.
An advantage of coherent feedback over its
measurement-based counterpart is the reduced noise produced by the control
process since there is no disturbance from a system measurement. Feedback
control methods promise particular robustness and flexibility, but
some important
questions related to the quantum nature
of the dynamics remains to be solved.
In particular,
there is not a general theory showing that quantum controllers perform better
than their classical counterparts, and if so, in which cases and under which
experimental conditions.

Many other questions are still open in quantum
feedback control. They extend from a general theory about the role of weak
measurements in the control of quantum systems to feedback control of
non-Markovian dynamics and the influence of model uncertainties on the feedback
control.
Similarly, another interesting perspective is to explore whether programmable
quantum processors~\cite{Buzek2006, Hillery2009} themselves could be used or
programmed such as to steer quantum systems.

Most of the developments of quantum feedback control have been made in the
context of quantum optics \cite{Reiner2004,Sayrin2011,Wiseman1993,Wiseman2010}.
% \sgcomment{Please add references, Dom: OK}
Recent technological progresses have enabled advances in superconducting
circuits both theoretically \cite{Martin2015} and experimentally
\cite{Riste2012,Vijay2012}.
A problem which is starting to attract attention is
feedback control in quantum transport \cite{Emary2014}. In other words, the
question is to understand how the current quantum control techniques developed
for quantum optical systems can be applied to hybrid systems involving quantum
dots, superconducting qubits and opto-mechanical resonators. The existing quantum
feedback theory has to be adapted to these new dynamics.

\subsection{Long-term vision}
% Quantum control theory is nowadays a mature mathematical discipline with a wide
% range of applications in science
% and engineering. Many different theoretical techniques have been developed and
% used with success since the beginning of
% the 1990s to control quantum systems.
Quantum optimal control theory has reached a level of maturity that
enables tackling the question of quantum control in a world which is
never fully quantum, be it due to residual couplings to the
environment or due to
imperfections in the controls or system characterization. The ultimate
goal is to develop a rigorous understanding of the fundamental limits
as well as the opportunities for quantum control under real-life conditions,
in terms of both controllability and control design. This includes in
particular control of quantum systems undergoing non-Markovian
evolution, as found in condensed phase physics and
chemistry. It will most likely require a combination of the techniques
described above and a synergy of efforts, ranging from pure
mathematics via algorithm design all the way to developing techniques
for integrating theory and experiment.
Meeting these challenges would provide the mathematical and
algorithmic underpinnings for the application of quantum control in
all of the three fields described in the following.

% In addition of the specific challenging questions described above, an
% ultimate goal
% is to develop an explicit connection between the mathematical
% control techniques and numerics in view of a
% general formalism permitting to describe different control applications and to
% support them by rigorous mathematical results on time optimality and reachability
% as discussed in this section. The used numerical algorithms are still too dependent of the considered example.
% Some of them are not supported by sound
% mathematical results, resulting in limited efficiency.

\section{Atomic, molecular, and chemical physics}
\label{sec:Chem}

\subsection{State of the art}
\label{subsec:Chem:stateofart}
Within the realm of atomic, molecular and chemical physics, quantum
control was first discussed in the context of chemical reactions.
It was termed 'coherent control' at the time
and conceived as a method to determine the fate of a reaction using
laser fields~\cite{Tannor1985,Tannor1986,Kosloff1989}. The basic
idea was to employ interference of matter waves to enhance
the desired outcome and suppress all
others~\cite{Rice2000,Brumer2003}. A way to create the desired
matter wave interference is by tailoring
laser fields~\cite{Tannor1985,Tannor1986,Kosloff1989}.
A reaction is viewed
as the following sequence of events---approach of the reactants;
formation of a new chemical bond;
intermediate dynamics of the generated molecular complex;
stabilization into the target products, typically involving the
breaking of another chemical bond. Each of these steps can in
principle be controlled.

The advent of femtosecond lasers and pulse shaping
technology~\cite{Weiner2000,Monmayrant2010}
in the 1990s allowed for  experimentally testing the idea
of controlling chemical reactions. In the lab, the laser pulse
shape can most easily be determined in feedback loops combined with
e.g. genetic algorithms~\cite{Judson1992}.
Very soon after the theoretical
proposals several pioneering experimental papers were published
describing control of unimolecular dissociation or fragmentation, see
Refs.~\cite{Gordon1997,Rabitz2000,Levis2001,Daniel2003,Brixner2003,Dantus2004,
Wollenhaupt2005,Nuernberger2007,Kuhn2007}
and references therein. Yet much still remains to be understood about
the mechanisms and even whether quantum coherence is
involved~\cite{Brumer2007}.

Control strategies include both
weak- and strong-field scenarios.
In the first case, a wavepacket is launched, and its ensuing
(ro-)vibrational dynamics is exploited.
In the strong-field regime, the laser pulse
coherently controls the dynamics during the pulse while
utilizing the effective modification of the energy levels of
atoms~\cite{Bayer2009,Bayer2014,TralleroHerrero2005,TralleroHerrero2006} or,
respectively, the potential energy landscape experienced by the
molecules, via the dynamic Stark
effect~\cite{Underwood2003,Sussman2006,Bayer2013}.

Weak field control of non-resonant excitation may employ optical
interferences to control e.g. the population in a final
state~\cite{Meshulach1998}. Weak field here implies low order perturbation theory to be applicable.
It was proven early in the development of
quantum control, that for weak fields in an isolated quantum
system, phase-only control is impossible for an objective which
commutes with the free, or drift, Hamiltonian~\cite{Brumer1989}.
A qualitative explanation is that under such conditions there are no
interfering pathways leading from the initial to the final stationary
states. Experimental evidence has challenged this assertion, claiming
demonstration of weak-field phase-only
control for an excited state branching
ratio~\cite{Prokhorenko2006}.
The phenomena were attributed to the influence of the environment.
A subsequent study showed that such
controllability is solvent dependent~\cite{Walle2009}.
A theoretical demonstration that phase-only control is possible in weak
field for an open quantum system soon followed~\cite{Katz2010}.
It is still an open question if weak-field phase-only control is
possible for targets
that commute with the Hamiltonian in open quantum systems. For example,
can population transfer from a ground to the excited surface be
phase-controlled  for a dye molecule in solution? While
this has theoretically been shown to be
impossible if the time evolution is Markovian~\cite{AmShallem2014},
most solvents lead to non-Markovian system dynamics.

In the context of coherent control of a chemical reaction,
the bimolecular process of bond
formation using femtosecond lasers remained much more
elusive~\cite{Marvet1995,Gross1997,Salzmann2008,Nuernberger2010,Merli2009,
Rybak2011,Koch2012,Amaran2013} than bond breaking. Its coherent
control was demonstrated only very recently~\cite{Levin2015}.
Full control of a binary reaction---from its entrance channel of
scattering reactants to the targeted products in a selected internal
state---is still an open goal.
Realizing this dream would create a new type of photochemistry with
selective control of yields and branching ratios.

Coherent control of bond formation in the gas phase turned out to be
so difficult because it starts from an incoherent thermal
ensemble. The laser pulse then needs to pick those scattering pairs
which show some correlation in their translational or rotational
motion. Averaging over rotations can be avoided by orienting or
aligning the molecules using strong laser
fields~\cite{Stapelfeldt2003}. Adiabatic alignment occurs in-field and
is achieved with nanosecond pulses.
In contrast, femtosecond laser pulses create non-adiabatic alignment
that persists after the pulse is over.
This second option where the alignment is produced in field-free
conditions is more interesting in view of the applications since the
laser pulse does not disturb the molecular
system~\cite{Seideman2006,Sugny2004}.

Spatial averaging in the gas phase implies an integral over the beam
profile and blurs coherent effects as the atoms or molecules are
exposed to different intensities. Spatial resolution is thus a
prerequisite for control~\cite{Katsuki2006}.
Subwavelength dynamic localization of the laser
intensity can be achieved on the nanometre
scale~\cite{Aeschlimann2007}, lifting the restrictions of
conventional optics.  Optimal control theory can be used for spatial
shaping of the laser fields used in
nanoexcitation~\cite{Pfeiffer2013}.

Another way to overcome thermal and spatial averaging is by cooling and
trapping the atoms or molecules. Coherent control was suggested as a method
to cool internal molecular degrees of freedom~\cite{Bartana1997,Bartana2001}.
This has been realized experimentally for cooling vibrations of ultracold
cesium dimers~\cite{Sofikitis2009}  and rotations of trapped
aluminum hydride ions~\cite{Lien2014}. While restrictions imposed on
the cooling efficiency by the molecular structure can be circumvented
using optimal control~\cite{Reich2013}, a persistent challenge to cooling
internal molecular degrees of freedom are the timescale separation
between vibrations and rotations as well as the enormous bandwidths
that are required for strong bonds. In order to prepare molecules that
are cold in their translational degrees of freedom, molecules are
assembled from atoms which are much easier to cool. However,
a major problem in creating molecules from
atoms is the extreme change in time and length scale. Optimal control
was studied to overcome this issue~\cite{Koch2004}.
Initial experiments employed the simple scheme of a chirped pulse to
compress two atoms to closer proximity~\cite{Wright2005}. This
can be viewed as a first step in ultracold laser-induced, i.e.,
photo-association.
Association yielding ultracold molecules in a single internal quantum state has
been demonstrated employing magnetic field ramps (magneto-association)
followed by STIRAP-type protocols~\cite{Danzl2008,Ni2008}.
An experimental challenge still unfulfilled is complete coherent control
of ultracold photoassociation.

One of the important issues here, particularly in polyatomics, is the
need for accurate ground and excited potential energy surfaces to
design optimal pulses a priori, as well as to interpret the mechanism
of pulses that are found by experimental optimization. For this reason
there has been significant interest in recent years in inverting the
information in optical spectroscopies in order to reconstruct excited
state potential energy surfaces~\cite{Li2010,Avisar2011,Avisar2012}.

In addition to making or breaking chemical bonds, coherent control
has demonstrated its versatility in studying energy
transfer~\cite{Herek2002}  and for spectroscopy. The latter includes
in particular non-linear and multi-dimensional
spectroscopies~\cite{Dudovich2002,Hamm2011,Serrano2015,Frostig2015}
and microscopy~\cite{Dudovich2002,Pastirk2003,Cruz2004}. In this
context, coherent control allows to increase both the resolution and
the specificity to a single molecule. Originally developed with
phase-locked time-delayed
laser pulses, non-linear and multi-dimensional spectroscopies utilize
the interference created by these pulses. Laser pulse shaping not only
lifts the constraint of having separate pulse beams in non-linear
spectroscopy and
microscopy~\cite{Dudovich2002,Hamm2011,Serrano2015}, it also
addresses naturally the requirement
to decipher the interference which is at the basis of the
non-linear spectroscopies. Eventually, this understanding has lead to
the use of laser pulse shaping in optical imaging with compensation
of the effect of scattering media such as biological
tissue~\cite{Katz2011,Katz2012}; optical microscopy with enhanced
chemical sensitivity and
contrast~\cite{Oron2005,Ogilvie2006,Pillai2009,Rehbinder2014};
chemical analysis and detection via
optical
discrimination~\cite{Li2008,Courvoisier2008,Kiselev2011,Steinbacher2015};
cancer diagnosis~\cite{Matthews2011}; and
material processing~\cite{Englert2008}.
Similarly, laser pulse shaping is expected
to enhance chemical sensitivity in other detection methods such as mass
spectroscopy~\cite{Bohinski2014,Tibbetts2015}.

Initially, the coherent control of molecules, be it in the context of
chemical reactions or non-linear spectroscopies or energy transfer,
considered the dynamics of the nuclear degrees of freedom, using
femtosecond laser pulses as the main workhorse. More
recently, the focus has shifted to controlling electron dynamics. This
is being made possible by the development of advanced x-ray sources which
probe the dynamics of electrons within atoms and molecules on
attosecond time scales. Their potential for exploring the quantum
nature of the nanoworld is unprecedented. For example, using xuv and
x-ray light for multidimensional spectroscopy could probe valence
excitations locally on different atomic sites in a molecule. This
would be invaluable for
understanding energy transfer in biological systems and quantum
devices. The use of x-ray light sources is currently facing a number of
challenges that can be tackled by quantum optimal control. First of all,
the large energy of xuv and x-ray pulses results in a high
probability of ionization, reflecting the problem of controllability
when a continuum of states is involved. This has  been addressed in a
recent study, where optimal control theory was
used to predict experimentally feasible pulses to
drive xuv-Raman excitations through the ionization
continuum~\cite{Greenman2015}.

Another control problem is the creation of the xuv light pulses
themselves. In particular, high harmonic generation, where a very strong
near-infrared femtosecond laser pulse accelerates an electron in such
a way that it emits xuv light, is an ideal candidate for coherent
control: Theoretical predictions for optimum driving should be easy to
adapt in experiment, given the existing pulse shaping
capabilities. The challenge that high harmonic generation poses to
optimal control theory is a frequency-domain
target~\cite{Serban2005,Werschnik2007,Schaefer2012}. A conclusive
answer whether shaping the femtosecond laser pulses can improve high
harmonic generation has not yet been provided.

Optimal control theory was first applied to chemical reactions using
Krotov's method~\cite{Tannor1992,Somloi1993} and gradient
ascent~\cite{Gross1992,Zhu1998}. The theory was quickly extended to
Liouville space~\cite{Yan1993,Bartana1997} to treat condensed phase
situations and cooling. The major experimental constraint in
expimerents with shaped femtosecond laser pulses is the fixed
bandwidth. This can be accounted for by including frequency filtering
in the optimization~\cite{Gollub2008,Schroder2009,Lapert2009,Walther2009}
 or by
imposing spectral constraints~\cite{Palao2013,Reich2014}.
Optimal control techniques have also been applied with success to
molecular alignment and orientation in gas
phase~\cite{Seideman2003}. The
design of optimal solutions has allowed to reach the best possible
degree of alignment and orientation~\cite{Salomon2005} within the
experimental constraints such as temperature~\cite{Lapert2012a} or
collisions~\cite{Seideman2005}. In addition to improving
existing control strategies, optimal control has also been used to
explore new regimes of alignment dynamics such as planar
alignment~\cite{Hoque2011}.

\subsection{Mid-term prospects: goals and challenges}
\label{subsec:Chem:midterm}

Thirty years after the conception of reaction control, it is fair to ask
whether the idea of coherently controlling a chemical reaction
can work at all. In this respect it is important to realize that
the basic ingredients for achieving this goal have all been developed.
The challenge that remains to be overcome,
is their assembly and application to a specific reaction.
The dream
of coherently controlling a chemical reaction all the way from its
entrance channel to the reaction products thus seems to be within
reach. It includes the controlled formation (or photoassociation) of
a new chemical bond, the controlled dynamics of the
intermediate complex, most likely involving a conical intersection, the
controlled cleavage of another chemical bond as well as the
stabilization of the reaction product.
Realistically, demonstration of control over a complete reaction can
be expected within the next few years for a sufficiently simple
reaction complex, involving only a few atoms.

Experimental techniques are currently developed to trap and cool a
single molecule. These are particularly advanced for single molecular
ions~\cite{Hansen2012,Rugango2015}.
The molecules can be simple diatomics or large biological chromophors
such as rhodopsin. Such trapped species are ideal candidates for
control, inducing for example isomerization in rhodopsin.
Instead of trapping and cooling molecules directly, they can also be
assembled from cold atoms one by one.
The number of different atoms which have been cooled and trapped has
been increasing steadily. Atoms other than the first row alkali metals
have the potential for complex chemistry with multipole bonds.
This raises the challenge of assembling molecules from these
constituents. The experimental obstacle is the development of pulse
shaping techniques in the picosecond and nanosecond range.

Another mid-term goal for quantum control is the control of electron
dynamics.
The capability to control electrons implies Angstrom-scale ultrafast
imaging methods which can be realized in the form of laser-induced
electron diffraction and high-harmonic spectroscopy. Specific mid-term
goals that seem within reach using these tools
include the control of subfemtosecond charge
migration; the controlled generation of spin-polarized electrons from laser
ionization; recognition of the absolute configuration of chiral
molecules with shaped laser pulses; and, using high-harmonic
spectroscopy of molecules,
ultrafast imaging of structure and dynamics on sub-atomic length
scales.

\subsection{Long-term vision}
\label{subsec:Chem:longterm}

\subsubsection{Synthesis}
The ultimate chemical synthetic challenge is to assemble large
chiral molecules from elementary building blocks.
Currently such syntheses are carried out in solution where the chemical
products are stabilized by entropy generation
caused by heat transfer to the environment. The vision would be
composed of synthesis by photoassociation
via polarization shaped light where the final product is stablized by
laser cooling and trapped by light.

\subsubsection{Analysis}

The vision is  a light field tailored to a specific molecule or
functional group generating a specific physical outcome such as
light emission or ionization. Such a capability will enhance the
threshold of detection of a specific hazard or in medical applications.
Combined with spatial super-resolution, the analytic methods are expected to
find applications in molecular-scale microscopy.

\section{Magnetic resonance} \label{sec:MRS}

The optimal control of spin dynamics is at the heart of well established
magnetic resonance technologies and of emerging new fields of quantum
technologies.
%Magnetic resonance is one of the most impressive
%success stories of quantum control and technology.
Nuclear magnetic
resonance spectroscopy (NMR)
\cite{Ernst1987,Abragam1961, Levitt2001},
electron spin resonance spectroscopy (ESR)
\cite{Schweiger2001}
and magnetic resonance imaging
(MRI)
\cite{Bernstein2004}
are based on the control of
%exploit the quantum mechanical nature of
nuclear or electron spins
%  - NMR and MRI control nuclear spins and ESR controls electron spins
with the help of time-dependent electromagnetic fields.
%in the form of radiowave or microwave pulses.
%Spin is a property of elementary particles related to the quantization of magnetic
%moments.
%- a single spin is the smallest permanent magnet possible, essentially a
%quantum of magnetic material
In fact, magnetic resonance is one of the most impressive
success stories of quantum control and technology.

The mathematical description of the state and dynamics of nuclear spins or electron spins is
essentially identical to the canonical description of abstract quantum bits
(qubits). In terms of the theoretical and experimental control of spins or qubits, NMR had a long head start
compared to other quantum technologies as the community has actively explored
and developed quantum-control methods for more than 60 years. This was driven by
many very concrete and important applications in physics, chemistry, biochemistry,
biology and medicine. The interdisciplinary impact of quantum-control enabled
magnetic resonance is impressively reflected by Nobel prizes in physics
(Felix Bloch, Edward Purcell, 1952), chemistry  (Richard Ernst, 1991; Kurt
W\"uthrich; 2002), and medicine (Paul Lauterbur, Peter Mansfield, 2003)
\cite{Ernst1992, Wuthrich2003, Mansfield2004}.
%The tools of quantum control developed in magnetic resonance have also given NMR a head start in the field of experimental quantum information processing \cite{Vandersypen2005,Ryan2008,Jones2011} and have been adopted by many other quantum technology plattforms.

Today, NMR is arguably the most important tool
in chemistry to determine the molecular structure and dynamics of molecules.
ESR is an essential technique in radical reaction
chemistry, catalysis, electrochemistry and photosynthesis research.
MRI is one of the most informative and frequently used modalities in
medical diagnostics.
The huge range of practical applications has generated multi-billion
dollar branch of the instrument manufacturing industry (Bruker,  Sie\-mens, Phillips, General
Electric, JEOL, etc.). This in turn has resulted in the continuous development
of more and more sophisticated instruments with superb flexibility in terms of
the available control schemes: For example, arbitrary waveform generators and
linear amplifies are standard NMR and MRI equipment for more than three
decades (and have more recently also become available
with sub-nanosecond time resolution for ESR applications \cite{Spindler2012,Kaufmann2013}). With their help,
even very complex pulse shapes can routinely be implemented with high fidelity
\cite{Kobzar2005}.
The excellent agreement between theory and experiments (as a result of
the highly accurate theoretical description of the physics of coupled spins and
the availability of very reliable hardware to implement virtually arbitrary
pulse sequences and shapes) also has made NMR an attractive testing ground for
the experimental demonstration of new control approaches for finite-dimensional
quantum systems in general. Concepts of quantum control and sophisticated
quan\-tum-control design tools developed in the field of NMR have found many
applications in other fields, such as in quantum information processing \cite{Vandersypen2005,Ryan2008,Jones2011}, optics
(photon echos)
\cite{Kurnit1964},
in neutron scattering or in the control of nano devices based on quantum dots,
 artificial atoms etc.

Two important emerging fields of magnetic resonance are hyperpolarization methods for bulk NMR
and the control and measurement of individual spins or spin systems, e.g. of NV
centers in diamond with many potential applications in sensing and quantum
information processing
\cite{Jelezko2006,Haberle2013,Dolde2014}.
Hyperpolarization techniques (also known as spin cooling),
\cite{Viale2010,Duckett2012,Kuhn2013}
can generate highly polarized non-thermal spin states. Hence, the relatively low
sensitivity of NMR (due to the small magnetic moments of nuclear spins and the
resulting weak thermal polarization) can be overcome by using a variety of
approaches. In particular, there are two different methods that have recently
become increasingly popular in practical applications. The first one is based on
transfer of the much higher thermal polarization from unpaired electrons onto the
nuclear spin ensemble in a process  called dynamic nuclear polarization
(DNP) \cite{Overhauser1953,Ni2013,Griesinger2012}. The second method
involves the
use of parahydrogen and a transfer of its highly populated singlet spin state
onto nuclear spins in receptor molecules \cite{Bowers1986}.
Both methods have been known already for many years but only recently
significant progress has been made in terms of a full quantum description of the
underpinning spin physics and the optimization of the required experimental
hardware.

The detection and control of individual nuclear spins close to Nitrogen-vacancy
(NV) centers in diamond
\cite{Jelezko2004,Muller2014}
is a premier example of a new area of optimal control of individual spin systems.
In these experiments, single nuclear-spin detection is achieved by an
efficient readout based on couplings of the nuclear spins to electron spin
states and their detection using optical techniques.
In the field of spintronics, spins
of individual atoms (such as $^{31}P$ donors) or quantum dots
can be electrically detected and controlled by radio- or microwave pulses with high fidelity\cite{Awschalom2013, Pla2013}.

%Other Hyperpolarozation Techniques? Parahydrogen etc. optical->(electronic)->nuclear polarization: NV centers in diamond
%transfer to singlet states: long-lived states ...

%More on individual spin systems, NV centers etc. !!!!!

\subsection{State of the art}
\label{subsec:MRS:stateofart}
The design of pulse sequences that provide maximum sensitivity, selectivity,
as well as maximum resilience to instrumental imperfections, is
central in the ongoing effort to improve magnetic resonance technologies.
Both analytical and numerical methods have been used to design multiple-pulse
sequences (including composite or shaped pulses).

Since the 1950s, the field of NMR has been a highly active
"breeder" for methods to control finite-di\-men\-sio\-nal quantum systems.
The wide spectrum of applications has motivated the development of more and
more sophisticated tools for quantum control in a quest to push the
experimental performance
%, selectivity and resilience to experimental limitations and imperfections
to their limits.
Pulsed magnetic resonance techniques such as the two-pulse Hahn echo
and the three-pulse stimulated echo
\cite{Hahn1950}
were important milestones on the path to
%ever more complex and more powerful pulse sequences which today
multiple-pulse sequences, which
often consist of tens or hundred thousands of individual pulses with defined
amplitudes and phases. This was enabled by the development of theoretical tools
such as Average Hamiltonian Theory (AHT)
\cite{Haeberlen1976}
and the symmetry analysis of composite pulses and multiple-pulse cycles and supercycles
 \cite{Levitt1986, Levitt1996, Freeman1998, Levitt2001}.

More
recently, sophisticated theoretical approaches of geometric optimal control theory
\cite{Pontryagin1962a,Jurdjevic1997,Khaneja2001,Khaneja2002,Khaneja2003,Bonnard2003,Boscain2004}
have been applied to
quantum control problems motivated by magnetic resonance applications.
Although these analytical approaches are typically limited to low-dimensional
quantum systems,
they are able to provide the best possible solutions to a given problem by
proving global optimality. The resulting physical performance limits are
important theoretical results in their own right and also
provide benchmarks for numerical and experimental optimization techniques.
Geometric optimal control has been very successfully applied to highly
non-trivial examples of uncoupled and coupled spins both in the absence and
presence of relaxation (dissipation and decoherence). For example, based on
geometrical optimal control analysis,
the minimal time for quantum gates in systems of two and three coupled spins
(qubits) have been determined and experimentally implemented
\cite{Khaneja2001, Khaneja2002, Reiss2003}.
And in this context, the so-called quantum-gate design metric was discovered,
which plays an important role for the design of so-called geodesic unitary
gates
\cite{Khaneja2002,Khaneja2007a}.
For example, time-optimal control schemes for the simulation of effective
trilinear couplings in systems with only next-neighbor couplings were derived
based on geometric optimal control and experimentally demonstrated using NMR.
This is a particularly impressive example, because in the limit of small time
increments $\Delta t_{\rm sim}$ under the simulated trilinear Hamiltonian, the
actual duration of the controls also approaches zero, whereas previous
approaches based on AHT have a minimal duration
of $1/(2J)$, i.e. for $\Delta t_{\rm sim} \rightarrow 0$, the gain of the
optimal-control schemes approaches infinity
\cite{Khaneja2002,Khaneja2007a}.
Examples of geometric optimal control applications involving singular arcs
\cite{Bonnard2003} in NMR are
relaxation-optimized polarization transfer experiments
\cite{Khaneja2003,Khaneja2003a,Stefanatos2004, Khaneja2004},
minimal-time controls
for the saturation of spins,
 \cite{Lapert2010},
for maximizing contrast in MRI
\cite{Lapert2012}
and for maximizing the achievable signal-to-noise ratio per unit time
\cite{Lapert2014,Lapert2015}.

Since the 1980s, numerical algorithms such as conjugate gradients, downhill
simplex (Nelder Mead), genetic algorithms and simulated annealing
have been applied to find efficient and robust quantum-control schemes
for magnetic resonance applications
\cite{Murdoch1987, Glaser1989,Liu1990,Freeman1987, Freeman1991,Untidt1998}.
Powerful optimal-control algorithms based on the Pontryagin Maximum Principle
(PMP) were applied already since the mid 1980s to the design of shaped pulses
for the manipulation of nuclear spin ensembles
\cite{Conolly1986,Mao1986,Rosenfeld1996}.
Potent variants of these PMP-based algorithms were also developed in the context
of NMR applications for coupled spins in NMR spectroscopy and quantum
information processing
\cite{Khaneja2005,Fouquieres2011,Li2011}.
They are able to optimize pulse sequences
for an ensemble of spin systems with realistic parameter ranges for detunings,
scalings of the control field etc., i.e. they can
take into account variations or uncertainties of experimental parameters.
Numerical ensemble-control methods have resulted in extremely powerful control
schemes with unprecedented resilience to pulse imperfections, while at the same
time taking experimentally bounded control amplitudes and pulse energy limits
into account.
%\cite{...} \sgcomment{REF}.

%Because most of modern optimal control research is essentially numerical,
%simulation software was universally viewed by the poll respondents as essential
%scientific infrastructure: its development should receive the same level of
%support (and respect) as scientific instrument design.
%The complexity is
%comparable -- efficient implementation of advanced algorithms on modern HPC
%hardware can be very difficult, but the benefits of someone releasing a powerful
%simulation package are felt across the community.
In magnetic resonance, an efficient and user-friendly software landscape is starting
to emerge -- there are a few software packages that support sophisticated
quantum mechanical magnetic resonance simulations as well as optimal control.
Spinach \cite{Hogben2011} %is unique in that it
supports all forms of magnetic resonance spectroscopy under the same roof,
implements sophisticated spin relaxation theories as well as most mainstream
optimal control algorithms that are pre\-sent\-ly used in magnetic resonance
spectroscopy: GRAPE \cite{Khaneja2005}, GRAPE-BFGS
\cite{Fouquieres2011,Machnes2011}, Krotov
\cite{Krotov1996}
and Krotov-BFGS
\cite{Eitan2011}.
Another
advantage of Spinach is the availability of polynomially scaling spin
dynamics simulation algorithms that make previously intractable NMR and EPR
simulation (and therefore control) problems accessible
\cite{Savostyanov2014,Edwards2014}.
 In the solid-state NMR community, the
SIMPSON
software package (SIMulation Program for SOlid state Nuclear magnetic resonance)
has been the most extensively used
general-purpose software and also
includes an optimal control toolbox to facilitate robust experiment design
\cite{Tosner2009}.
MATPULSE \cite{Matson1994,Liu2011},
DYNAMO %(acronym ...)
\cite{Machnes2011}
and
QuTiP
%(Quantum Toolbox in Py\-thon)
\cite{Johansson2013}
are versatile Matlab and Python based
simulation and optimization programs. Other  software packages for the
simulation and/or optimization of spin dynamics include
SIMONE \cite{Glaser1990}, OCTOPUS(SI) \cite{Kobzar2007,Ehni2013a}, Gamma
\cite{Smith1994} and
SPIN\-EVOLUTION \cite{Veshtort2006}.
%and OCTOPUS \sgcomment{REF}.

 Both analytical and numerical methods
have been pro\-ven very effective for two- and three-level systems
\cite{Tannor1999,Boscain2002,Boscain2005,Boscain2006,Sugny2007,Bonnard2009,
Bonnard2009,Lapert2010,Lapert2011,Zhang2011,Boozer2012,Mukherjee2013,Garon2013,
Lapert2013a,Albertini2015,Skinner2012,Gershenzon2008,Gershenzon2007,Skinner2006,
Skinner2005,Skinner2003}, two uncoupled spins
\cite{Assemat2010,Assemat2012},
and  two coupled spins
\cite{Khaneja2001,Bennett2002,Vidal2002,Hammerer2002,Yuan2005,Reiss2002,
Khaneja2005a,Ehni2013}.
Moreover, significant progress has been made in understanding how to optimally
control
coupled spin systems with more than two spins
\cite{Khaneja2002,Khaneja2002a,Reiss2003,Bose2003,Zeier2004,Stefanatos2004,
Stefanatos2005,SchulteHerbruggen2005,Carlini2006,
Khaneja2007a,Yuan2007,Yuan2008,
Schirmer2009,Burgarth2010a,Wang2010,Wang2010a,Carlini2011,Yuan2011,
Lapert2012b,
Carlini2012,Nimbalkar2012,Carlini2013,Bonnard2013,Bonnard2014,Yuan2014,
Damme2014}.
Recent advances include robust broadband and band-selective pulses
in NMR and ESR.
Depending on their role in a given experiment,
universal rotation (UR) pulses (e.g. for refocussing) or point-to-point (PP)
pulses (e.g. for excitation or inversion of spins) are required. Systematic
studies of the offset bandwidth (range of detunings) and robustness with respect
to scaling of the control amplitude (width of $B_1$ inhomogeneity distribution)
of optimized PP  \cite{Kobzar2004,Kobzar2008} and UR pulses have been performed \cite{Kobzar2012}Êfor experiments where the
maximum control amplitudes (e.g. in many applications of NMR or ESR
spectroscopy) or the pulse energy (e.g. in MRI) are experimentally limiting
factors. For a desired fidelity, the bandwidth covered by a
(composite or shaped) pulse
can be made much larger than the maximum control amplitude (maximum Rabi frequency) at the price
of increased pulse duration.
Surprisingly, it was found empirically that the
pulse duration typically scales only linearly with the desired
 offset range of operation.
% applications in NMR \cite{FPBT02,Nielsen2010} such as
%broad-band pulses \cite{SRLKG03,Skinner2004,KSKGL04},
 Hence, unprecedented fidelities can be achieved that are e.g. required for
quantitative NMR applications or for
quantum error correction schemes.
%\cite{XXX} \sgcomment{REF}.
In addition, it was demonstrated that
coherence transfer elements can be designed that are not only resilient to
offset and control amplitude variations but also to variations in coupling
constants
\cite{Ehni2013,Ehni2014}.

In general, UR pulses are significantly longer than PP pulses for the same error
resilience, which was exploited in
a new strategy to build complete sequences based on standardized UR pulses when
necessary and
standardized PP pulses whenever they are sufficient
\cite{Nimbalkar2013}.
The recent application of optimal control methods to the problem of
heteronuclear decoupling yields not
only significantly improved performance,
\cite{Neves2009,Schilling2014}
but also unprecedented flexibility in the design of
tailored decoupling sequences
\cite{Schilling2012,Zhang2013,Koroleva2013}.
Individual pulses were also optimized for Carr-Purcell-Meiboom-Gill  echo train sequences
\cite{Borneman2010,Mandal2014}.
Beyond individually optimized pulses, the simultaneous
optimization of pulses provides significant performance gains by exploiting
cooperative effects
either in a single scan
\cite{Braun2010}
or in multiple scans
\cite{Braun2014}
with first applications in Ramsey-type experiments and in Hahn echo sequences.

In liquid-state NMR, optimized broadband pulses have found applications in a number of experiments.
Early examples are the CLIP-HSQC \cite{Enthart2008} and P.E.HSQC
\cite{Tzvetkova2007} for the measurement of
one-bond and geminal couplings as well as more recent experiments involving
homonuclear decoupling \cite{Reinsperger2014} and the rapid acquisition of
heteronuclear correlation experiments \cite{SchulzeSuenninghausen2014}. All
these experiments are of special interest as they are recorded in large numbers
every day in chemical laboratories and improvements in robustness, accuracy, or
acquisition time are of high interest.

For the specific conditions of solid-state NMR and experiments in oriented systems, specific pulse sequences have been optimized \cite{Kehlet2004,Vosegaard2005, Tosner2006,Lee2008,MacGregor2011,Yi2012,Loening2012}.
In ESR spectroscopy, the first broadband optimal-con\-trol pulses have recently
been developed and
experimentally implemented
\cite{Spindler2012,Kaufmann2013}.
In this context, the efficiency of novel approaches to take into account
transient effects
due to transfer line effects and limited resonator bandwidth was both
numerically
and experimentally demonstrated
\cite{Spindler2012}.

Motivated by magnetic resonance imaging applications, optimal control pulses
have been
developed
\cite{Conolly1986,Mao1986,Rosenfeld1996,King2008,Grissom2009,Liu2011,Lapert2012,Massire2013, Vinding2013}.
Applications include improved spatially selective excitation schemes
\cite{Janich2011,Skinner2012a,Janich2012},
pulses with minimal radio-frequency (rf) power and
pulses to counteract rf inhomogeneity in parallel transmission at ultra-high
field
\cite{Grissom2009,%Grissom
Liu2011%Matson}
}.
%\sgcomment{REF Check: Stollberger: ISMRM Abstract 1448, 2014 (2015)}.
For  chemical exchange saturation transfer
(CEST) imaging, chemical exchange effects were taken into account in pulse sequence optimizaion
\cite{Rancan2015}.

In DNP, optimal polarization transfer schemes from electron to nuclear spins
have been investigated for a small set of relatively simple model systems
\cite{Maximov2008,Pomplun2008,Pomplun2010,Griesinger2012,Hovav2014}.
Optimal control was also used in conjunction with the use of parahydrogen to
generate high nuclear spin polarization. In particular, it was shown that the
initial longitudinal two-spin order arising from the hydrogenation reaction can
be equally distributed between several spins and converted into detectable
magnetization \cite{bretschneider2012}.

Geometric and numerical tools from optimal control theory
have not only provided pulse sequences of unprecedented quality and
capabilities, but also new analytical insight and a deeper understanding both of
the mode of action of optimal pulses. Numerically optimized pulses can often be
interpreted as robust implementations of analytically derived optimal
trajectories
\cite{Lapert2010}.
These can be understood based on geometrical concepts
\cite{Lapert2013a}.
In addition, useful tools have been developed to
analyze complex pulses \cite{Kocher2014} and the resulting dynamics in coupled
spin systems
\cite{Kuprov2013,Pomplun2008,Garon2015}.

%Yet another fundamental problem is that of interpretation
%\cite{Kocher2014,Kuprov2013} -- more often than not, optimal control pulse
%sequences are visually uninterpretable to humans. Current research on the
%subject has branched into two primary streams: the analysis of the pulse
%waveform itself \cite{Kocher2014} and the analysis of the spin dynamics induced
%by the waveform \cite{Kuprov2013}. Both provide critical insight into the
%mechanics of the optimal solutions \cite{Kehlet2011} that, on the one hand,
%satisfy the experimentalists, and on the other, occasionally enable researchers
%to glean new classes of magnetic resonance pulse sequences [ref].

%%% Deleted
%On the industrial development side, optimal control theory research in magnetic
%resonance encouraged the development of new hardware -- the recent release by
%Bruker of a sub-nanosecond arbitrary waveform generator (AWG)
%has made it possible to start exploring the physical limits of the ESR experiments \cite{Spindler2012}.
%%% End

\subsection{Mid-term prospects: goals and challenges}

An important goal is to make optimal control algorithms easier to use, generally
applicable and to further increase their speed. Depending on the applications,
very different convergence rates are encountered and a systematic
characterization of optimal control landscapes is still missing.
%\cite Sophie ...
With more efficient numerical optimizations methods,  optimal control theory
will make it possible to design  problem-, sample- and patient-specific pulse
sequences on the fly.
% -- conventionally, only the magnetic field and the tuning parameters of
%the resonant circuit had been re-optimized every time a sample was changed.
The fast reoptimization of pulse sequences or of sequence elements,
e.g. in response to the presence of magnetic susceptibility
jumps, will significantly improve their performance. In addition to MRI
applications, this could be important in production or process monitoring by
NMR, where the pulse sequence should be able to adapt to the sample in the same
way as shim currents currently do -- examples are magnetic susceptibility and tuning variations in
imaging,
metabolomics
and oil well logging \cite{Borneman2010}.

Optimal control methods are expected to reduce the time which is required to
determine structural and dynamical information of biomolecules (e.g. proteins).
In this field, coping with large coupled spin networks, especially in the
presence of relaxation, is computationally hard and further improved
numerical/analytical approaches are high\-ly desirable. Characterizing the
experimental imperfections and fine-tuning the spectrometer and setting up the
experiment is time-consuming and efficient closed loop feedback-based automatic
procedures have to be developed, implemented and integrated with the adaptive
design of
pulse sequences to automate this process.
%Numerical
%methods could be used as a first
%step, making it possible to give an accurate approximation of
%the initial adjoint state and a geometric
%analysis may then help understand the role and the
%optimality of singular extremals.

In the field of medical imaging it is expected that optimal control methods will
lead to more sensitive and more
efficient pulse sequences, such that a patient has to spend less time in a
scanner for
an examination. This may be achievable by using multi-band excitation techniques
and
optimized image acquisition based on multiple transmit and receive channels.
Optimal control methods are also
expected to help in the extension of the clinical applicability of
ultra-high-field scanners and to provide  in-vivo spectroscopy with improved
diagnostic value.
Apart from better localized excitation, progress in the field of magnetization
preparation (e.g.
reducing the B$_0$ sensitivity of fat saturation) might be achievable.
Improved saturation pulses are expected to be useful for many different task,
from chemical shift imaging (CSI) and chemical exchange saturation transfer
(CEST) imaging \cite{Rancan2015}
to single voxel spectroscopy.
Also methods for improved quantification accuracy and biomarker imaging are
highly desirable.

Small molecule NMR spectroscopy forms the central basis of chemical analysis on
a molecular level in solution.
Applications are already manifold, but are expected to increase significantly
during the next decade.
Mostly identification and quantification of compounds in complex mixtures is
needed. Especially for quantification tasks,
methods to be derived from optimal control will deliver more reliable and
robust results. Optimal control derived
sequences will also improve the detection level of side products, making
products safer and better defined.
Especially quality control of food, pharmaceuticals and other products as well
as  metabolomics-type applications
and personalized medicine
will benefit from improved experiments and thereby directly affect everyday life.

In the field of hyperpolarization,
further hardware advances are expected, in particular in generating and
modulating high-frequency microwave fields and it is envisaged that optimal
control methods will play an important role in the development of more
sophisticated experimental schemes to transfer the electron polarization to
surrounding nuclear spins. Ideas how to control electron-nuclear spin systems in a optimal way
have already been published and
discussed. This includes the manipulation of the nuclear spins using the anisotropy
of the hyperfine interaction \cite{Khaneja2007,Hodges2008} and the
exploitation of
repeated generation of dipolar spin order to enhance polarization transport by
spin diffusion \cite{Dementyev2008}.
 Since more sophisticated catalysts are being developed for parahydrogenation
 reactions \cite{Duckett2012}, it is also anticipated that pulse shapes derived from optimal
control principles will be more frequently used to maximize the achievable
polarization and to mediate polarization transfer to specific molecular site.
%_____Hyperpol_End

\subsection{Long-term vision}

Important theoretical and practical aspects of optimal control  in
magnetic resonance
are the physical limits of quantum dynamics.
On the one hand, the questions of quantum state reachability in dissipative
systems remain largely unresolved.
%\cite{...}
On the other, many practical usage
scenarios are time-constrained, and a more general understanding of the best
possible performance in a give amount of time is highly desirable.

A very desirable outcome of the continued progress in optimal control technology
could be that,
 for a given level of performance, the use of optimal control sequences could
significantly reduce the instrument costs as well as costs of sample preparation
and purification. Sophisticated shim coils, frequency locks, complicated
combinations of susceptibility matched materials in NMR probes and other
expensive arrangements had originally been introduced to maximize spectral
resolution and selectivity. If both could be achieved by tailored pulse sequences
under less than ideal conditions, the complexity could be transferred from the
instrument design to the mathematical optimization procedure. The concomitant
reduction of hardware cost could result in more affordable instruments, e.g. for
MRI
examinations.
%The same applies to sample purification -- if highly selective
%excitation techniques become available, practical thresholds for sample purity
%are likely to become less onerous.
%xxxxxxxxxxx
Also, the integration of control design with image reconstruction and spectral
calculation in NMR and with instrument design
could result in better performance.
The combination of open loop and feedback strategies may result in fast and
fully automated tune-up procedures, which would further reduce the required
experimental time.

A long-term vision of magnetic resonance techniques is the detection of
the nuclear spins of individual molecules, such as a protein. The ability to
image the shape of a single molecule similar to the way we can image humans
today would revolutionize
structural biology and the rational and efficient development of medication. The recent developments in sensing based ib NV centers in
diamonds (see section \ref{sec:QICT}) may provide a potential road to this goal, as well as completely new
application areas e.g. in medical diagnostics.

\color{black}

\section{Quantum information and communication}
\label{sec:QICT}

\subsection{State of the art}
\label{subsec:QICT:stateofart}

Quantum technologies (see, e.g., \cite{Dowling2003}) exploit quantum coherence
and entanglement as essential elements of quantum
physics. Applications include high-precision
measurements and sensing, which would reach unprecedented sensitivity,
the simulation of physical and biological systems, which would be
impossible to study
otherwise, and quantum information processing, which would allow to solve
computationally hard problems. Successful implementation of quantum
technologies faces the challenge to preserve the relevant nonclassical features
at the level of device operation. More specifically, each task of the device
operation needs to be carried out with sufficient accuracy, despite
imperfections and potentially detrimental effects of the surroundings. Quantum
optimal control not only provides toolboxes that allow for identifying the
performance limits for a given device implementation, it also provides
the protocols for realizing device operation within those
limits.

Prominent tasks include the preparation of useful quantum states as well as
implementation of quantum operations.  The power of the quantum optimal
control approach for implementing these tasks has very recently been
demonstrated in a number of impressive experiments. For example, nonclassical
motional states of a Bose-Ein\-stein condensate were prepared with optimized
control sequences for wavepacket interferometry~\cite{Frank2014}, and the
loading of an ultracold atomic gas into an optical lattice was
improved~\cite{Rosi2013}. %(improvement by factor of 3)
With respect to quantum
operations,
%minimal times for quantum gates in systems of two and three coupled spins (qubits) have been determined
%analytically using geometric optimal control methods
%\cite{...}.
quantum optimal control allowed for error resistant single-qubit
gates with trapped ions~\cite{Timoney2008} and for single qubit gates
without the need for invoking the rotating wave approximation in nitrogen
vacancy (NV) centers in diamond~\cite{Scheuer2014}. For the latter
platform, optimal
control is also at the basis of a spectroscopy protocol allowing to image
nanoscale magnetic fields~\cite{Haberle2013}.
In quantum processor candidates based on superconducting circuits,
leakage to non-compu\-tational states in the most common type of qubit, the
transmon~\cite{Motzoi2009,Chow2010a,Lucero2010a}, was avoided and
frequency crowding was accomodated~\cite{Schutjens2013,Vesterinen2014}
thanks to optimal control results. Closed-loop optimal
control \cite{Kelly2014,Egger2014a} enabled fine-tuning  of
gates that were determined manually, allowing them to reach consistent record
fidelities within this platform.
 Control of donor qubits in
Si has been achieved and characterized \cite{Muhonen2015}. Control strategies
for spins in semiconductors currently trade off conceptually simple single-spin
schemes \cite{Pla2012,Kawakami2014} with more robust and accessible two- and
three spin techniques \cite{Petta2005,Maune2012,Kim2014}. The short natural time
scales suggest adiabatic schemes to be attractive
\cite{Greentree2004,Ferraro2015,Braakman2013}
% Note that these were not straightforward
% implementations of quantum optimal control as they are performed in other
% fields, i.e., the programming of a numerically obtained pulse into a pulse
% shaper. %%% chr: there is no such thing except maybe in NMR, better
% to leave out
In view of scaling up control, the design and implementation of
unitary maps have recently been demonstrated in a 16-dimensional
Hilbert space, spanned by the electron and nuclear
spins of individual cesium atoms~\cite{Anderson2015}.

The use of control methods in a broader sense has allowed for further
significant experimental achie\-vements,
such as to improve of the coherence of a qubit, realized by the
electron spin in an NV center, using dynamical
decoupling~\cite{Cai2012}.
A famous further example of high-end control techniques is the Paris
experiment of  stabilizing a quantum state with predefined photon
number via real-time closed-loop feedback~\cite{Sayrin2011}, which
required to include the noise back-action of controls
onto the system by way of stochastic differential calculus.

These experimental achievements were preceded by a large number of theoretical
predictions on how optimal control may improve or enable quantum state
preparation, operation and readout. State preparation protocols include
transport of atoms~\cite{DeChiara2008} and
ions~\cite{Singer2010,Furst2014} as well as transport in a spin
chain~\cite{Caneva2009,Schirmer2009,Murphy2010,Khaneja2002a,Stefanatos2005,
Yuan2007,Nimbalkar2012},
photon storage~\cite{Gorshkov2008},
preparation of squeezed
states~\cite{Grond2009}, cluster-states~\cite{Fisher2009}, non-classical
states in a cavity~\cite{Rojan2014} or in spin
chains~\cite{Wang2010,Wang2010a}, as well as preparation of a
quantum register~\cite{Doria2011a} and many-body entangled
states~\cite{Caneva2012}---to name just a few.

Likewise, optimal control helped to implement high-fidelity quantum gates
such as two-qubit gates with neutral atoms
in dipole traps \cite{Calarco2004a,Dorner2005},
on atom chips \cite{Treutlein2006},
or with Rydberg atoms \cite{Cozzini2006,Goerz2011},
two-qubit gates between ions \cite{Poulsen2010},
between an ion and an atom \cite{Doerk2010},
error-cor\-rect\-ing qubit gates of electron and nuclear spins within
single NV centers~\cite{Waldherr2014},
or entangling gates between distant  NV centers~\cite{Dolde2014}.
For superconducting qubits~\cite{Clarke2008},
two-qubit gates were optimized, starting
from Cooper pair boxes \cite{Montangero2007,Sporl2007,Jirari2009,Cross2015} to
modern transmon-based schemes \cite{Egger2014b,Goerz2014a}. In these
optimizations, special attention was paid to robustness against noise
\cite{Montangero2007,Rebentrost2009} which can even be used as a tool
for control \cite{Reich2015}. Also, readout has been addressed
\cite{Egger2014c}. In order to adapt to the strong filtering of
control lines in  superconducting qubits, transfer functions had to
be taken into account \cite{Jager2013,Motzoi2011,Spindler2012,Hincks2014}
and  experimental fluctuations and noise were
accomodated~\cite{Wang2010,Goerz2014a,Mottonen2006}.
Fidelity limits on  two-qubit gates due to decoherence were studied
for Markovian~\cite{SchulteHerbruggen2011,Goerz2014,Goerz2014a}
as well as non-Markovian~\cite{Floether2012} time evolutions.
It is noteworthy in this context that non-Markovian time evolutions also
play a role in quantum simulation, e.g., of collision models~\cite{Rybar2012}.
%this is another Buzek paper%.

In view of quantum computation, it has been suggested how to retain
universality in spite of limited local
control~\cite{Wang2010,Burgarth2010a} by using
environmental degrees of freedom~\cite{Reich2015}.
The Jones polynomial, i.e., a central invariant in knot theory, can be
evaluated, using an NMR spin ensemble at ambient temperature, in an
algorithm equivalent to deterministic quantum computing with a single
pure qubit~\cite{Passante2009,Marx2010,SchulteHerbruggen2012}.

In order to obtain these results, the quantum optimal control
methodology had to be adapted
to the requirements of Quantum Technologies.
Optimization algorithms had to be derived
for specific quantum gates~\cite{Palao2002,Palao2003,Tesch2002},
dissipative evolution as seen in the reduced system
dynamics~\cite{Kallush2006,Rebentrost2009,SchulteHerbruggen2011,Goerz2014},
or exploiting invariants in system-bath models~\cite{Grace2010},
optimization up to local equivalence classes~\cite{Muller2011},
which can also be used for arbitrary perfect
entanglers~\cite{Watts2015,Goerz2015}
or optimizing for many-body entanglement~\cite{Platzer2010}.
Moreover, control techniques were adapted to non-linear dynamics as
found in a BEC~\cite{Sklarz2002,Hohenester2007,Jager2014} and to general
dynamics, functionals and couplings to be controlled~\cite{Reich2012}.
Many-body systems
can be optimized numerically  with the chopped random basis (CRAB)
method \cite{Doria2011a,Caneva2011} by  interfacing  the
time-dependent density matrix renormalization group (t-DMRG)
with parameter optimization via e.g. the simplex algorithm.
Other techniques specifically cover robustness with respect to
experimental fluctuations or noise~\cite{Wang2010,Goerz2014,Goerz2014a,Rojan2014}  or
filters in experimental implementation of
controls \cite{Motzoi2011,Spindler2012,Jager2013,Hincks2014}.

% Sophie wrote: finally, there are a ton of papers on quantum
% spintronics, control of spin networks and spin-chain quantum wires
% ... we should mention some of them,
%scholar.google.com/citations?view_op=view_citation&hl=en&user=ViCpkpYAAAAJ&cstart=20&citation_for_view=ViCpkpYAAAAJ:HDshCWvjkbEC
%%% chr: This is reference Schirmer2009, already cited above and in
%%% NMR section

%scholar.google.com/citations?view_op=view_citation&hl=en&user=ViCpkpYAAAAJ&cstart=20&citation_for_view=ViCpkpYAAAAJ:UeHWp8X0CEIC
%I'm including the links to the gscholar pages as that makes it easy
%to get a student to look at paper that cite those and find a few more
%recent relevant papers we could stick in.
%%% chr: This is reference Wang2010 already cited above

%Definitively, I'd cite this one and freely plagarize the reference
%section and maybe papers that cite this work :-)
%www.sciencemag.org/content/339/6124/1174.abstract
%%% chr: This is the Awschalom review mentioned in green in the Foreword

%There's another older science review on conventional spintronics too
%www.sciencemag.org/content/294/5546/1488
%with nice references and citations to plagiarize ;)

\subsection{Mid-term prospects: goals and challenges}
\label{subsec:QICT:midterm}

The field of quantum technologies has matured to the point that
quantum enhancement is explored beyond quantum computation
only. Devices such as quantum simulators or quantum sensors are
currently under active development. Control methods will be crucial to
operate these devices reliably and accurately. This involves the
device preparation, or reset, the execution of the desired time
evolution, and the readout of the result. These tasks set the agenda
for the next few years.

More specifically,
central mid-term milestones include the robust implementation of gates in a
multi-qubit architecture, finding solutions to readout and fast reset
limitations, automatization of key tasks of surface code error
correction and optimal as well as robust generation of multi-particle
entangled states  for a variety of quantum technology
 platforms. All of these will
require decoherence control.

A main challenge for optimal quantum control  is to
reach convergence of numerical optimal control and experimentation. To
date, either optimal control is used for computer-aided discovery of
analytical schemes that can be remodeled in an experiment or the
quantum processor itself is employed to calibrate
gates. In order to better combine numerical optimal control and
experimentation, the modeling of  qubits as well
as errors and other non-idealities of the system,
in particular for open systems, needs to be improved and the
robustness of pulses enhanced.  Also, optimized pulses should initially be
reduced to few parameters before more complicated and effective
solutions can be pursued. On the other hand, pulse shaping platforms
need improvement.

Coming to specific quantum technology platforms,
applications of quantum control in superconducting qubits should follow the
current European thrust towards analog and digital quantum simulation and lead
to the preparation of entangled ground states, fast and accurate quantum gates
and tools for quantum machines.  For some instances, compatibility with quantum
error correction is desired.
For trapped ions, it seems realistic to combine quantum gates with ion
transport in segmented traps, using optimal control
techniques, which can also be applied to the systematic optimization
of pulse sequences for efficient generation of complex operations.
In the field of cold atoms, challenges for optimal control are
twofold. On the side of quantum simulations, the goal is to enable
high-fidelity preparation and manipulation of many-body quantum states
of increasing complexity, with and without local control. On the side
of quantum communication, the goal is to enable efficient
interconversion between flying qubits and quantum memories via
coherent atom-photon coupling, with and without cavities. In the case
of color centers in diamond, with major applications in quantum
sensing, the goal is to enhance the sensitivity of the defect spins
employed as quantum probes via improved protection from environmental
noise e.g. through dynamical decoupling techniques.
%%% BEGIN COMMENT
% \fwcomment{Prati: In the case of spins in Si, more fabrication consistency is
% required \cite{Prati14} as well as advanced control electronics
% \cite{Guagliardo13}.}
%%% END COMMENT

In general, control techniques are expected to contribute
to decoupling and dissipative state-engineering
\cite{Buchler2008,Verstraete2009,Krauter2011}, for instance in view of
enhancing the lifetime of quantum memories. In order to improve the lifetime
of a quantum register, control can also be used to implement
error-correcting gates and circuits~\cite{Waldherr2014}.
Moreover, while quantum compilation, i.e., the translation of a
unitary gate into the machine
language of pulses and evolutions, can readily be done via optimal
control up to some 10 qubits~\cite{SchulteHerbruggen2008},
a scalable assembler of elementary gates (up to 10 qubits) into many
qubits is an open problem
that may benefit from tensor-contraction techniques.
% like DMRG and PEPS.

Both numerical optimal control and closed-loop control are expected to
be useful for tackling these goals. Numerical optimal control has the
advantage of versatility, whereas closed-loop control can easily be
tuned to specific tasks such as determining parameter uncertainties. A
hybridization of both approaches is conceivable as well. The main
difficulties that need to be overcome to reach the above mentioned
mid-term milestones are a sufficiently accurate modelling of
complex quantum dynamics to build control on top, integration of
tomography and system identification with optimal control,
efficient ways to take into account experimental constraints and
uncertainties,  and bridging the gap
between the quantum control community and the communities of the
respective quantum technology platforms.

\subsection{Long-term vision}
\label{subsec:QICT:longterm}

Several current quantum technology platforms show a strong
scaling potential. Thus in the long term, control schemes need to be
made scalable. This represents a severe challenge, but
meeting this challenge will make quantum control a
basic building block of every quantum technology and ensure,  at the
same time, their proper functioning in a world that is only partially
quantum.

Take the examples of superconducting qubits, NC centers or spins in Si, where fabrication is a key
task that could and should be improved by control techniques.
The controlled adjustment of fabrication parameters should be simple,
and the qubits should to a certain extent be robust to the influences
of the rest of the architecture they are placed  in.
Independent of a specific platform,
error correction at large, for instance by toric
codes~\cite{Fowler2014,Barends2014}, is one of the strategic long-term
goals that is expected to benefit from control techniques given recent
advances by randomized benchmarking~\cite{Kelly2014}. To this end,
system-identification protocols matched with optimal control modules
will be of importance.
A pioneering step in this direction was made by the ADHOC
technique~\cite{Egger2014a} that
combines open-loop control as a first step with closed-loop feedback
learning (with implicit parameter identification). Moreover, taking
quantum control algorithms to match with
tensor-contraction techniques in order to address quantum-many body systems
(where first steps have been made by CRAB~\cite{Doria2011a,Caneva2011})
%%% or in spectroscopy by SPINACH~\cite{Hogben2011})
%%% chr: I think this is not correct, the contraction there has
%%% nothing to do with tensor networs:
is expected to pave the way to more accurate handling of
experimental quantum simulation setups.
%%% chr: the following is too detailed given the previous text
% In this context, 1D topologies
% (chains) and tree-like branches will be a natural first step,
% since DMRG and, for time-evolution t-DMRG, tensor-techniques are
% computationally much easier  than in 2D topologies (grids) that
% require PEPS-type techniques. In particular, computing time evolutions
% on 2D lattices will be challenging.

In short, quantum control will be the
means to get the most performance out of an
imperfect system and combine challenging physics at the few-qubit level with
engineering at the multi-qubit level. This should aim for example at
enabling quantum simulations that are impossible on classical computers.
In addition,
the realization of the following long-term goals, using optimal control
techniques, seems challenging yet conceivable:
demonstrating the practical usefulness of engineered quantum states,
for example in quantum metrology;
implementing reliable strategies for the control of mesoscopic
systems;
exploring the dynamics of quantum many-body systems beyond
 equilibrium; and
understanding the microscopic origin of thermodynamic laws.
In other words, the long-term goal of quantum optimal control for quantum
technologies is to develop a software layer enhancing the performance
of quantum hardware for tasks in computing, simulation, communication,
metrology and sensing beyond what is achievable by classical means,
enabling the achievement of quantum supremacy~\cite{Preskill2012}.

\section{Prospects for applications and commercial exploitation}
\label{sec:prospects}

Quantum-control enabled technologies have potential for truly
revolutionary innovation.
More sophisticated quantum control techniques are making current technologies
more powerful and also
help to create novel technologies, e.g. in sensing with
super sensitive magnetic detectors, microscopic
temperature measurement devices,
molecular imaging etc.
Better control of quantum systems has the potential to significantly reduce
instrument costs, turning perhaps million dollar NMR spectrometers into small
and portable devices with many new fields of applications.

As visible above, quantum optimal control applications broadly fall into two
classes: Applications to novel {\em quantum technologies} and applications in
chemistry-related areas such as spin resonance.
The magnetic resonance industry as we know it today would not exist without
quantum control.
Novel optimal control strategies have already been implemented
in commercial NMR spectrometers and implementations of optimal control sequences
in MRI are being pursued. Clearly, with more and more improved sequences being
developed, this trend is expected to continue.
Other chemistry-related applications where coherent control plays out
powerfully are imaging, optical microscopy and various variants of
spectroscopy as well as
chemical analysis where shaped laser pulses improve the resolution and
enhance the specificity to a particular mo\-le\-cule. Areas of application
range from remote chemical detection all the way to cancer diagnosis.

On the other end of the spectrum, there is presently an emerging industrial
effort in Quantum Computing lead by IBM, Google, and Microsoft. The first two
companies have invested into the development of superconducting qubits and both of
them use optimal control techniques \cite{Kelly2014,Cross2015}. This is no
coincidence - optimal control can have impact in systems that have reached some
technical maturity in research laboratories, which is the point at which
industry gets interested. Further industrial perspectives will be linked to the
further development of a quantum technologies industry. Early convergence could
happen in quantum sensing, which takes sensing ideas similar to those appearing
in spin resonance and combines them with ideas from quantum technologies.

\section{Conclusions}
\label{sec:concl}

Quantum control is a key facilitator for spectroscopy and imaging as well as  AMO physics and
emerging quantum technologies for computation, simulation, metrology, sensing
and  communication.
For all these applications, it is crucial  to reach the required precision given
experimental limits on control amplitudes, power, timing, accuracy of
instruments as well as the ever-present interaction with the environment.
Optimal control theory provides a framework to identify which quantum tasks can
be accomplished with what precision in the presence of decoherence and
experimental imperfections and limitations.

Quantum control systems theory will require the
integration of control aspects at many different levels.
Future quantum technologies will rely on integrated
architectures of hybrid quantum systems \cite{Kurizki2015} with e.g. nuclear spins for long-term
storage,
quantum-nanomechanical devices for sensing
and photons for the communication of quantum states.
This will require also the integration of quantum mechanics in engineering
education and vice versa.
It will be necessary to establish strong links of quantum control experts to
quantum engineering and to the manufacturing of quantum devices.

Due to its interdisciplinary nature with applications in many diverse fields,
 future advances in the optimal control of quantum systems will require the
 combined
effort of people with expertise in a wide range of research
fields. Only the close link of basic research, development and applications will
open
scientifically and economically rewarding perspectives and will foster the
innovation potential of
emerging quantum technologies in an optimal way. The Virtual Facility for
Quantum Control (VF-QC)
%,
%which was recently established
under
the umbrella of  the EU Coordinated Action for Quantum Technologies in Europe
(QUTE-EUROPE) marks an important step in this direction.
The primary goal of the VF-QC is to provide a common structure for the growing
quantum control community in Europe, for the promotion of quantum control and
to
provide expertise to other scientific communities,  to policy makers and the
general public.
Establishing common terminology, common standards and common visions are crucial
prerequisites
to maximize the beneficial impact of optimal quantum control methods on current
and future technology, economics and society.

\section*{Acknowledgements}
This strategic report has grown out of the Pan-European Coordination
and support action on Optimal Control of Quantum Systems (QUAINT),
funded by the European Commission under grant number 297861.
We would like to thank all people who have participated in the community consultation for their comments and suggestions.
In particular we acknowledge contributions from Andreas Buchleitner, Tobias Brixner, Daniel Burgarth, Enrico Prati, David Tannor, and Zden\v{e}k To\v{s}ner. 
%\sgcomment{section editors, please add additional names if appropriate}.
We are indebted to Per J. Liebermann for assisting the writing process as
well as Frank Langbein, Peter K. Schuhmacher and Andrea Dumont for managing the
community consultation.

%\bibliography{../bibfiles/survey}
\bibliography{survey}

\begin{thebibliography}{100}

\bibitem{quantumcontrol}
The quaint website.
\newblock www.quantumcontrol.eu, 2015.

\bibitem{Abragam1961}
A.~Abragam.
\newblock {\em The Principles of Nuclear Magnetism}.
\newblock Oxford University Press, 1961.

\bibitem{Aeschlimann2007}
Martin Aeschlimann, Michael Bauer, Daniela Bayer, Tobias Brixner, F.~Javier
  Garc\'{i}a~de Abajo, Walter Pfeiffer, Martin Rohmer, Christian Spindler, and
  Felix Steeb.
\newblock Adaptive subwavelength control of nano-optical fields.
\newblock {\em Nature}, 446:301--304, 2007.

\bibitem{Agrachev2004}
A.~A. Agrachev and Y.~L. Sachkov.
\newblock {\em Control theory from the geometric viewpoint}.
\newblock Springer-Verlag, Berlin, vol. 87 of encyclopaedia of mathematical
  sciences, control theory and optimization, ii. edition, 2004.

\bibitem{Albertini2003}
F.~Albertini and D'{A}lessandro D.
\newblock Notions of controllability for bilinear multilevel quantum systems.
\newblock {\em IEEE Trans. Automat. Control}, 48:1399--1403, 2003.

\bibitem{Albertini2015}
F.~Albertini and D.~D'{A}lessandro.
\newblock Minimum time optimal synthesis for two level quantum systems.
\newblock {\em J. Math. Phys.}, 56:012106, 2015.

\bibitem{Altafini2012}
C.~Altafini and F.~Ticozzi.
\newblock Modeling and control of quantum systems: An introduction.
\newblock {\em IEEE Trans. Automat. Control}, 57:1898--1917, 2012.

\bibitem{Altafini2002}
Claudio Altafini.
\newblock Controllability of quantum mechanical systems by root space
  decomposition of su({N}).
\newblock {\em J. Math. Phys.}, 43(5):2051--2062, 2002.

\bibitem{Altafini2003}
Claudio Altafini.
\newblock Controllability properties for finite dimensional quantum markovian
  master equations.
\newblock {\em J. Math. Phys.}, 44(6):2357--2372, 2003.

\bibitem{AmShallem2014}
Morag Am-Shallem and Ronnie Kosloff.
\newblock The scaling of weak field phase-only control in {M}arkovian dynamics.
\newblock {\em J. Chem. Phys.}, 141:044121, 2014.

\bibitem{Amaran2013}
Saieswari Amaran, Ronnie Kosloff, Micha\l{} Tomza, Robert Moszynski, Leonid
  Rybak, Liat Levin, Zohar Amitay, J.~Martin Berglund, Daniel~M. Reich, and
  Christiane~P. Koch.
\newblock Femtosecond two-photon photoassociation of hot magnesium atoms: A
  quantum dynamical study using thermal random phase wavefunctions.
\newblock {\em J. Chem. Phys.}, 139:164124, 2013.

\bibitem{Katz2012}
Ori~Katz andEran Small and Yaron Silberberg.
\newblock Looking around corners and through thin turbid layers in real time
  with scattered incoherent light.
\newblock {\em Nature Photon.}, 6:549--553, 2012.

\bibitem{Anderson2015}
B.~E. Anderson, H.~Sosa-Martinez, C.~A. Riofr\'{i}o, Ivan~H. Deutsch, and
  Poul~S. Jessen.
\newblock Accurate and robust unitary transformations of a high-dimensional
  quantum system.
\newblock {\em Phys. Rev. Lett.}, 114(24):240401, 2015.

\bibitem{Annunziato2010}
M.~Annunziato and A.~Borzi.
\newblock Optimal control of probability density functions of stochastic
  processes.
\newblock {\em Math. Modelling Anal.}, 15(4):393--407, 2010.

\bibitem{Assemat2012}
E.~Ass{\'e}mat, L.~Attar, M.~J. Penouilh, M.~Picquet, A.~Tabard, Y.~Zhang,
  S.~J. Glaser, and D.~Sugny.
\newblock Optimal control of the inversion of two spins in nuclear magnetic
  resonance.
\newblock {\em Chem. Phys.}, 405:71--75, 2012.

\bibitem{Assemat2010}
E.~Ass\'emat, M.~Lapert, Y.~Zhang, M.~Braun, S.~J. Glaser, and D.~Sugny.
\newblock Simultaneous time-optimal control of the inversion of two
  spin-$\frac{1}{2}$ particles.
\newblock {\em Phys. Rev. A}, 82(1):013415, 2010.

\bibitem{Avisar2011}
David Avisar and David~J. Tannor.
\newblock Complete reconstruction of the wave function of a reacting molecule
  by four-wave mixing spectroscopy.
\newblock {\em Phys. Rev. Lett.}, 106(17):170405, 2011.

\bibitem{Avisar2012}
David Avisar and David~J. Tannor.
\newblock Multi-dimensional wavepacket and potential reconstruction by resonant
  coherent anti-stokes raman scattering: Application to {H2O} and {HOD}.
\newblock {\em J. Chem. Phys.}, 136(21):214107, 2012.

\bibitem{Awschalom2013}
D.~D. Awschalom, L.~C. Bassett, A.~S. Dzurak, E.~L. Hu, and J.~R. Petta.
\newblock Quantum spintronics: Engineering and manipulating atom-like spins in
  semiconductors.
\newblock {\em Science}, 339:1174--1179, 2013.

\bibitem{Ban2012}
Yue Ban, Xi~Chen, E.~Ya Sherman, and J.~G. Muga.
\newblock Fast and robust spin manipulation in a quantum dot by electric
  fields.
\newblock {\em Phys. Rev. Lett.}, 109(20):206602, 2012.

\bibitem{Barends2014}
R.~Barends, J.~Kelly, A.~Megrant, A.~Veitia, D.~Sank, E.~Jeffrey, T.~C. White,
  J.~Mutus, A.~G. Fowler, B.~Campbell, Y.~Chen, Z.~Chen, B.~Chiaro,
  A.~Dunsworth, C.~Neill, P.~O'Malley, P.~Roushan, A.~Vainsencher, J.~Wenner,
  A.~N. Korotkov, A.~N. Cleland, and J.~M. Martinis.
\newblock Logic gates at the surface code threshold: Superconducting qubits
  poised for fault-tolerant quantum computing.
\newblock {\em Nature}, 508:500--503, 2014.

\bibitem{Bartana1997}
A.~Bartana, R.~Kosloff, and D.~J. Tannor.
\newblock Laser cooling of internal degrees of freedom. {II}.
\newblock {\em J. Chem. Phys.}, 106:1435--1448, 1997.

\bibitem{Bartana2001}
Allon Bartana, Ronnie Kosloff, and David~J Tannor.
\newblock Laser cooling of molecules by dynamically trapped states.
\newblock {\em Chem. Phys.}, 267(1--3):195--207, 2001.

\bibitem{Bayer2014}
T.~Bayer, M.~Wollenhaupt, H.~Braun, and T.~Baumert.
\newblock Ultrafast and efficient control of coherent electron dynamics via
  {SPODS}.
\newblock {\em Adv. Chem. Phys.}, page~1, 2014.

\bibitem{Bayer2009}
T.~Bayer, M.~Wollenhaupt, C.~Sarpe-Tudoran, and T.~Baumert.
\newblock Robust photon locking.
\newblock {\em Phys. Rev. Lett.}, 102(2):023004, 2009.

\bibitem{Bayer2013}
Tim Bayer, Hendrike Braun, Cristian Sarpe, Robert Siemering, Philipp von~den
  Hoff, Regina de~Vivie-Riedle, Thomas Baumert, and Matthias Wollenhaupt.
\newblock Charge oscillation controlled molecular excitation.
\newblock {\em Phys. Rev. Lett.}, 110(12):123003, 2013.

\bibitem{Beauchard2005}
K.~Beauchard.
\newblock Local controllability of a 1-{D} {S}chr\"{o}dinger equation.
\newblock {\em J. Math. Pures Appl.}, 84(7):851--956, 2005.

\bibitem{Beauchard2010}
Karine Beauchard, Jean-Michel Coron, and Pierre Rouchon.
\newblock Controllability issues for continuous-spectrum systems and ensemble
  controllability of {B}loch equations.
\newblock {\em Commun. Math. Phys.}, 296(2):525--557, 2010.

\bibitem{Bennett2002}
C.~H. Bennett, I.~Cirac, M.~S. Leifer, D.~W. Leung, N.~Linden, S.~Popescu, and
  G.~Vidal.
\newblock Optimal simulation of two-qubit {H}amiltonians using general local
  operations.
\newblock {\em Phys. Rev. A}, 66(1):012305, 2002.

\bibitem{Bergholm2012}
V.~Bergholm and T.~Schulte-Herbr\"{u}ggen.
\newblock How to transfer between arbitrary n-qubit quantum states by coherent
  control and simplest switchable noise on a single qubit.
\newblock e-print: http://arXiv.org/pdf/1206.494, 2012.

\bibitem{Bergmann1998}
K.~Bergmann, H.~Theuer, and B.~W. Shore.
\newblock Coherent population transfer among quantum states of atoms and
  molecules.
\newblock {\em Rev. Mod. Phys.}, 70(3):1003--1025, 1998.

\bibitem{Bernstein2004}
M.~A. Bernstein, K.~F. King, and Zhou.
\newblock {\em Handbook of MRI Pulse Sequences}.
\newblock Elsevier, Burlington-San Diego-London, 2004.

\bibitem{Bohinski2014}
Timothy Bohinski, Katharine Moore~Tibbetts, Maryam Tarazkar, Dmitri~A. Romanov,
  Spiridoula Matsika, and Robert~J. Levis.
\newblock Strong field adiabatic ionization prepares a launch state for
  coherent control.
\newblock {\em J. Phys. Chem. Lett.}, 5(24):4305--4309, 2014.

\bibitem{Bonnard2003}
B.~Bonnard and M.~Chyba.
\newblock {\em Singular trajectories and their role in control theory}.
\newblock Springer-Verlag, Berlin, mathematics and applications, vol. 40
  edition, 2003.

\bibitem{Bonnard2009}
B.~Bonnard, M.~Chyba, and D.~Sugny.
\newblock Time-minimal control of dissipative two-level quantum systems: The
  generic case.
\newblock {\em IEEE Trans. Automat. Control}, 54(11):2598--2610, 2009.

\bibitem{Bonnard2014}
B.~Bonnard, O.~Cots, J.-B. Pomet, and N.~Shcherbakova.
\newblock Riemannian metrics on 2{D}-manifolds related to the {E}uler-{P}oinsot
  rigid body motion.
\newblock {\em ESAIM Control Optim. Calc. Var.}, 20(3):864--893, 2014.

\bibitem{Bonnard2013}
B.~Bonnard, O.~Cots, and N.~Shcherbakova.
\newblock The {S}erret-{A}ndoyer {R}iemannian metric and {E}uler-{P}oinsot
  rigid body motion.
\newblock {\em Math. Control Relat. Fields}, 3(3):287--302, 2013.

\bibitem{Boozer2012}
A.~D. Boozer.
\newblock Time-optimal synthesis of {SU}(2) transformations for a
  spin-$\frac{1}{2}$ system.
\newblock {\em Phys. Rev. A}, 85(1):013409, 2012.

\bibitem{Borneman2010}
T.~W. Borneman, M.~D. H{\"u}rlimann, and D.~G. Cory.
\newblock Application of optimal control to {CPMG} refocusing pulse design.
\newblock {\em J. Magn. Reson.}, 207(2):220--233, 2010.

\bibitem{Borzi2012}
A.~Borzi.
\newblock Quantum optimal control using the adjoint method.
\newblock {\em Nanoscale Syst. Math. Model. Theory Appl.}, 1:93--111, 2012.

\bibitem{Boscain2005}
U.~Boscain, T.~Chambrion, and G.~Charlot.
\newblock Nonisotropic 3-level quantum systems: Complete solutions for minimum
  time and minimum energy.
\newblock {\em Discrete Contin. Dyn. Syst. Ser. B}, 5(4):957--990, 2005.

\bibitem{Boscain2002}
U.~Boscain, G.~Charlot, J.-P. Gauthier, S.~Gu\'erin, and H.-R. Jauslin.
\newblock Optimal control in laser-induced population transfer for two- and
  three-level quantum systems.
\newblock {\em J. Math. Phys.}, 43:2107--2132, 2002.

\bibitem{Boscain2006}
U.~Boscain and P.~Mason.
\newblock Time minimal trajectories for a spin 1/2 particle in a magnetic
  field.
\newblock {\em J. Math. Phys.}, 47(6):062101, 2006.

\bibitem{Boscain2004}
U.~Boscain and B.~Piccoli.
\newblock {\em Optimal Synthesis for Control Systems on 2-{D} Manifolds}.
\newblock Springer, smai, vol. 43 edition, 2004.

\bibitem{Bose2003}
S.~Bose.
\newblock Quantum communication through an unmodulated spin chain.
\newblock {\em Phys. Rev. Lett.}, 91(20):207901, 2003.

\bibitem{Bouten2007}
Luc Bouten, Ramon. van {H}andel, and Matthew~R. James.
\newblock An introduction to quantum filtering.
\newblock {\em SIAM J. Control Optim.}, 46(6):2199, 2007.

\bibitem{Bouten2009}
Luc Bouten, Ramon van {H}andel, and Matthew~R. James.
\newblock A discrete invitation to quantum filtering and feedback control.
\newblock {\em SIAM Rev.}, 51(2):239--316, 2009.

\bibitem{Bowers1986}
C.~R. Bowers and D.~P. Weitekamp.
\newblock Transformation of symmetrization order to nuclear-spin magnetization
  by chemical reaction and nuclear magnetic resonance.
\newblock {\em Phys. Rev. Lett.}, 57(21):2645--2648, 1986.

\bibitem{Braakman2013}
F.~R. Braakman, P.~Barthelmy, C.~Reichl, W.~Wegscheider, and L.~M.~K.
  Vandersypen.
\newblock Long-distance coherent coupling in a quantum dot array.
\newblock {\em Nat. Nanotechnol.}, 8:432--437, 2013.

\bibitem{Braun2010}
M.~Braun and S.~J. Glaser.
\newblock Cooperative pulses.
\newblock {\em J. Magn. Reson}, 207(1):114--123, 2010.

\bibitem{Braun2014}
M.~Braun and S.~J. Glaser.
\newblock Concurrently optimized cooperative pulses in robust quantum control:
  Application to broadband {R}amsey-type pulse sequence elements.
\newblock {\em New J. Phys.}, 16:115002, 2014.

\bibitem{bretschneider2012}
Christian Bretschneider, Alexander Karabanov, Niels~Chr. Nielsen, and Walter
  K{\"o}ckenberger.
\newblock Conversion of parahydrogen induced longitudinal two-spin order to
  evenly distributed single spin polarisation by optimal control pulse
  sequences.
\newblock {\em J. Chem. Phys.}, 136(9):094201, 2012.

\bibitem{Breuer2009}
Heinz-Peter Breuer, Elsi-Mari Laine, and Jyrki Piilo.
\newblock Measure for the degree of non-markovian behavior of quantum processes
  in open systems.
\newblock {\em Phys. Rev. Lett.}, 103:210401, Nov 2009.

\bibitem{Brif2010}
Constantin Brif, Raj Chakrabarti, and Herschel Rabitz.
\newblock Control of quantum phenomena: Past, present and future.
\newblock {\em New J. Phys.}, 12(7):075008, 2010.

\bibitem{Brixner2003}
Tobias Brixner and Gustav Gerber.
\newblock Quantum control of gas-phase and liquid-phase femtochemistry.
\newblock {\em Chem. Phys. Chem.}, 4(5):418--438, 2003.

\bibitem{Brockett1972}
R.~W. Brockett.
\newblock System theory on group manifolds and coset spaces.
\newblock {\em SIAM J. Control}, 10(2):265--284, 1972.

\bibitem{Brockett1973}
R.~W. Brockett.
\newblock Lie theory and control systems defined on spheres.
\newblock {\em SIAM J. Appl. Math.}, 25(2):213--225, 1973.

\bibitem{Brumer2003}
P.~Brumer and M.~Shapiro.
\newblock {\em Principles and Applications of the Quantum Control of Molecular
  Processes}.
\newblock Wiley Interscience, 2003.

\bibitem{Brumer2007}
Paul Brumer, Kunihito Hoki, and Michael Spanner.
\newblock An analysis of two liquid-state adaptive feedback experiments.
\newblock {\em Isr. J. Chem.}, 47(1):111--114, 2007.

\bibitem{Brumer1989}
Paul Brumer and Moshe Shapiro.
\newblock One photon mode selective control of reactions by rapid or shaped
  laser pulses: An emperor without clothes?
\newblock {\em Chem. Phys.}, 139(1):221--228, 1989.

\bibitem{Bryson1975}
A.~Bryson and Y.~C. Ho.
\newblock {\em Applied Optimal Control}.
\newblock Hemisphere, New-York, 1975.

\bibitem{Buchler2008}
H.~P. B\"{u}chler, S.~Diehl, A.~Kantian, A.~Micheli, and P.~Zoller.
\newblock Preparation of entangled states by quantum {M}arkov processes.
\newblock {\em Phys. Rev. A}, 78(4):042307, 2008.

\bibitem{Burgarth2014}
D.~K. Burgarth, P.~Facchi, V.~Giovanetti, H.~Nakazato, S.~Pascazio, and
  K.~Yuasa.
\newblock Exponential rise of dynamical complexity in quantum computing through
  projections.
\newblock {\em Nat. Commun.}, 5:5173, 2014.

\bibitem{Burgarth2010a}
Daniel Burgarth, Koji Maruyama, Michael Murphy, Simone Montangero, Tommaso
  Calarco, Franco Nori, and Martin~B. Plenio.
\newblock Scalable quantum computation via local control of only two qubits.
\newblock {\em Phys. Rev. A}, 81(4):040303, 2010.

\bibitem{Buzek2006}
V.~Buzek, M.~Hillery, M.~Ziman, and M.~Rosko.
\newblock Programmable quantum processors.
\newblock {\em Quant. Inf. Proc.}, 5:313--420, 2006.

\bibitem{Cai2012}
J.-M. Cai, B.~Naydenov, R.~Pfeiffer, L.~P. Mc{G}uinness, K.~D. Jahnke,
  F.~Jelezko, M.~B. Plenio, and A.~Retzker.
\newblock Robust dynamical decoupling with concatenated continuous driving.
\newblock {\em New J. Phys.}, 14(11):113023, 2012.

\bibitem{Calarco2004a}
T.~Calarco, U.~Dorner, P.~S. Julienne, C.~J. Williams, and P.~Zoller.
\newblock Quantum computations with atoms in optical lattices: Marker qubits
  and molecular interactions.
\newblock {\em Phys. Rev. A}, 70(1):012306, 2004.

\bibitem{Caneva2009}
T.~Caneva, M.~Murphy, T.~Calarco, R.~Fazio, S.~Montangero, V.~Giovannetti, and
  G.~E. Santoro.
\newblock Optimal control at the quantum speed limit.
\newblock {\em Phys. Rev. Lett.}, 103(24):240501, 2009.

\bibitem{Caneva2011}
Tommaso Caneva, Tommaso Calarco, and Simone Montangero.
\newblock Chopped random-basis quantum optimization.
\newblock {\em Phys. Rev. A}, 84(2):022326, 2011.

\bibitem{Caneva2012}
Tommaso Caneva, Tommaso Calarco, and Simone Montangero.
\newblock Entanglement-storage units.
\newblock {\em New J. Phys.}, 14(9):093041, 2012.

\bibitem{Carlini2006}
A.~Carlini, A.~Hosoya, T.~Koike, and Y.~Okudaira.
\newblock Time-optimal quantum evolution.
\newblock {\em Phys. Rev. Lett.}, 96(6):060503, 2006.

\bibitem{Carlini2011}
A.~Carlini, A.~Hosoya, T.~Koike, and Y.~Okudaira.
\newblock Time-optimal {CNOT} between indirectly coupled qubits in a linear
  {I}sing chain.
\newblock {\em J. Phys. A}, 44(14):145302, 2011.

\bibitem{Carlini2012}
A.~Carlini and T.~Koike.
\newblock Time-optimal transfer of coherence.
\newblock {\em Phys. Rev. A}, 86(5):054302, 2012.

\bibitem{Carlini2013}
A.~Carlini and T.~Koike.
\newblock Time-optimal unitary operations in {I}sing chains: unequal couplings
  and fixed fidelity.
\newblock {\em J. Phys. A}, 46(4):045307, 2013.

\bibitem{Chakrabarti2007}
R.~Chakrabarti and H.~Rabitz.
\newblock Quantum control landscapes.
\newblock {\em Int. Rev. Phys. Chem.}, 26(4):671--735, 2007.

\bibitem{Chambrion2009}
Thomas Chambrion, Paolo Mason, Mario Sigalotti, and Ugo Boscain.
\newblock Controllability of the discrete-spectrum schr\"{o}dinger equation
  driven by an external field.
\newblock {\em Ann. I. H. Poincare-An.}, 26(1):329--349, 2009.

\bibitem{Chen2010}
Xi~Chen, I.~Lizuain, A.~Ruschhaupt, D.~Gu\'ery-{O}delin, and J.~G. Muga.
\newblock Shortcut to adiabatic passage in two- and three-level atoms.
\newblock {\em Phys. Rev. Lett.}, 105(12):123003, 2010.

\bibitem{Chow2010a}
J.~M. Chow, L.~DiCarlo, J.~M. Gambetta, F.~Motzoi, L.~Frunzio, S.~M. Girvin,
  and R.~J. Schoelkopf.
\newblock Optimized driving of superconducting artificial atoms for improved
  single-qubit gates.
\newblock {\em Phys. Rev. A}, 82(4):040305, 2010.

\bibitem{Ciaramella2015}
G.~Ciaramella, A.~Borz{\`\i}, G.~Dirr, and D.~Wachsmuth.
\newblock Newton methods for the optimal control of closed quantum spin
  systems.
\newblock {\em SIAM J. Sci. Comput.}, 37:A319--A346, 2015.

\bibitem{Clarke2008}
J.~Clarke and F.~K. Wilhelm.
\newblock Superconducting quantum bits.
\newblock {\em Nature}, 453:1031--1042, 2008.

\bibitem{Combes2006}
Joshua Combes and Kurt Jacobs.
\newblock Rapid state reduction of quantum systems using feedback control.
\newblock {\em Phys. Rev. Lett.}, 96(1):010504, 2006.

\bibitem{Combes2008}
Joshua Combes, Howard~M. Wiseman, and Kurt Jacobs.
\newblock Rapid measurement of quantum systems using feedback control.
\newblock {\em Phys. Rev. Lett.}, 100(16):160503, 2008.

\bibitem{Cong2014}
S.~Cong.
\newblock {\em Control of Quantum Systems: Theory and Methods}.
\newblock Wiley \& Sons, Singapore, 2014.

\bibitem{Conolly1986}
S.~Conolly, D.~Nishimura, and A.~Macovski.
\newblock Optimal control methods in {NMR} spectroscopy.
\newblock {\em IEEE Trans. Med. Imaging}, MI-5:106--115, 1986.

\bibitem{Coron2007}
J.~M. Coron.
\newblock {\em Control and Nonlinearity}.
\newblock Am.~Math.~Soc., Providence RI, 2007.

\bibitem{Courvoisier2008}
Francois Courvoisier, Luigi Bonacina, Veronique Boutou, Laurent Guyon,
  Christophe Bonnet, Benoit Thuillier, Jerome Extermann, Matthias Roth,
  Herschel Rabitz, and Jean-Pierre Wolf.
\newblock Identification of biological microparticles using ultrafast depletion
  spectroscopy.
\newblock {\em Faraday Discuss.}, 137(0):37--49, 2008.

\bibitem{Cozzini2006}
M.~Cozzini, T.~Calarco, A.~Recati, and P.~Zoller.
\newblock Fast {R}ydberg gates without dipole blockade via quantum control.
\newblock {\em Opt. Commun.}, 264(2):375--384, 2006.

\bibitem{Cross2015}
Andrew~W. Cross and Jay~M. Gambetta.
\newblock Optimized pulse shapes for a resonator-induced phase gate.
\newblock {\em Phys. Rev. A}, 91(3):032325, 2015.

\bibitem{Cruz2004}
Johanna M~Dela Cruz, Igor Pastirk, Matthew Comstock, Vadim~V Lozovoy, and
  Marcos Dantus.
\newblock Use of coherent control methods through scattering biological tissue
  to achieve functional imaging.
\newblock {\em P. Natl. Acad. Sci. USA}, 101(49):16996--17001, 2004.

\bibitem{Daems2013}
D.~Daems, A.~Ruschhaupt, D.~Sugny, and S.~Gu\'{e}rin.
\newblock Robust quantum control by a single-shot shaped pulse.
\newblock {\em Phys. Rev. Lett.}, 111:050404, 2013.

\bibitem{DAlessandro2008}
D.~D'{A}lessandro.
\newblock {\em Introduction to Quantum Control and Dynamics}.
\newblock Chapman and Hall, Boca Raton, applied mathematics and nonlinear
  science series edition, 2008.

\bibitem{DAlessandro2001}
D.~D'{A}lessandro and M.~Dahled.
\newblock Optimal control of two-level quantum systems.
\newblock {\em IEEE Trans. Automat. Control}, 46(6):866--876, 2001.

\bibitem{Daniel2003}
Chantal Daniel, J\"{u}rgen Full, Leticia Gonz\'{a}lez, Cosmin Lupulescu,
  J\"{o}rn Manz, Andrea Merli, \v{S}tefan Vajda, and Ludger W\"{o}ste.
\newblock Deciphering the reaction dynamics underlying optimal control laser
  fields.
\newblock {\em Science}, 299:536--539, 2003.

\bibitem{Dantus2004}
M.~Dantus and V.~V. Lozovoy.
\newblock Experimental coherent laser control of physicochemical processes.
\newblock {\em Chem. Rev.}, 104(4):1813--1860, 2004.

\bibitem{Danzl2008}
Johann~G. Danzl, Elmar Haller, Mattias Gustavsson, Manfred~J. Mark, Russell
  Hart, Nadia Bouloufa, Olivier Dulieu, Helmut Ritsch, and Hanns-Christoph
  N\"{a}gerl.
\newblock Quantum gas of deeply bound ground state molecules.
\newblock {\em Science}, 321:1062--1066, 2008.

\bibitem{DeChiara2008}
G.~De~{C}hiara, T.~Calarco, M.~Anderlini, S.~Montangero, P.~J. Lee, B.~L.
  Brown, W.~D. Phillips, and J.~V. Porto.
\newblock Optimal control of atom transport for quantum gates in optical
  lattices.
\newblock {\em Phys. Rev. A}, 77(5):052333, 2008.

\bibitem{Fouquieres2013}
P.~de~{F}ouquieres and S.~G. Schirmer.
\newblock Quantum control landscapes: A closer look.
\newblock {\em Infinite dimensional analysis, quantum probability and related
  topic}, 16:1350021, 2013.

\bibitem{Fouquieres2011}
P~de~{F}ouquieres, S.~G. Schirmer, S.~J. Glaser, and Ilya Kuprov.
\newblock Second order gradient ascent pulse engineering.
\newblock {\em J. Magn. Reson.}, 212(2):412--417, oct 2011.

\bibitem{Dementyev2008}
Anatoly~E. Dementyev, David~G. Cory, and Chandrasekhar Ramanathan.
\newblock Rapid diffusion of dipolar order enhances dynamic nuclear
  polarization.
\newblock {\em Phys. Rev. B}, 77(2):024413, 2008.

\bibitem{Devoret2014}
M.~Devoret.
\newblock {\em Quantum Machines: Measurement and Control of Engineered Quantum
  Systems}.
\newblock {\'E}cole de Physique des Houches 2011. Oxford University Press,
  Oxford, 2014.

\bibitem{Dirr2008}
G.~Dirr and U.~Helmke.
\newblock Lie theory for quantum control.
\newblock {\em GAMM-Mitt.}, 31(1):59--93, 2008.

\bibitem{Dirr2009}
G.~Dirr, U.~Helmke, I.~Kurniawan, and T.~Schulte-Herbr\"{u}ggen.
\newblock Lie-semigroup structures for reachability and control of open quantum
  systems: {K}ossakowski-{L}indblad generators form {L}ie wedge to {M}arkovian
  channels.
\newblock {\em Rep. Math. Phys.}, 64(1--2):93--121, 2009.

\bibitem{Doerk2010}
Hauke Doerk, Zbigniew Idziaszek, and Tommaso Calarco.
\newblock Atom-ion quantum gate.
\newblock {\em Phys. Rev. A}, 81(1):012708, 2010.

\bibitem{Doherty2000}
Andrew~C. Doherty, Salman Habib, Kurt Jacobs, Hideo Mabuchi, and Sze~M. Tan.
\newblock Quantum feedback control and classical control theory.
\newblock {\em Phys. Rev. A}, 62(1):012105, 2000.

\bibitem{Dolde2014}
F.~Dolde, V.~Bergholm, Y.~Wang, I.~Jakobi, B.~Naydenov, S.~Pezzagna, J.~Meijer,
  F.~Jelezko, P.~Neumann, T.~Schulte-Herbr{\"u}ggen, J.~Biamonte, and
  J.~Wrachtrup.
\newblock High-fidelity spin entanglement using optimal control.
\newblock {\em Nat. Commun.}, 5:3371, 2014.

\bibitem{Dong2010}
D.~Dong and I.~A. Petersen.
\newblock Quantum control theory and applications: A survey.
\newblock {\em IET Control Theory A.}, 4(12):2651--2671, 2010.

\bibitem{Doria2011a}
Patrick Doria, Tommaso Calarco, and Simone Montangero.
\newblock Optimal control technique for many-body quantum dynamics.
\newblock {\em Phys. Rev. Lett.}, 106:190501, 2011.

\bibitem{Dorner2005}
U~Dorner, T~Calarco, P~Zoller, and P~Grangier.
\newblock Quantum logic via optimal control in holographic dipole traps.
\newblock {\em J. Opt. B: Quantum Semiclass. Opt.}, 7(10):S341, 2005.

\bibitem{Dowling2003}
J.~P. Dowling and G.~Milburn.
\newblock Quantum technology: The second quantum revolution.
\newblock {\em Phil. Trans. R. Soc. Lond. A}, 361:1655--1674, 2003.

\bibitem{Dudovich2002}
Nirit Dudovich, Dan Oron, and Yaron Silberberg.
\newblock Single-pulse coherently controlled nonlinear {R}aman spectroscopy and
  microscopy.
\newblock {\em Nature}, 418:512--514, 2002.

\bibitem{Edwards2014}
Luke~J. Edwards, D.~V. Savostyanov, Z.~T. Welderufael, Donghan Lee, and Ilya
  Kuprov.
\newblock Quantum mechanical {NMR} simulation algorithm for protein-size spin
  systems.
\newblock {\em J. Magn. Reson.}, 243:107--113, 2014.

\bibitem{Egger2014a}
D.~J. Egger and F.~K. Wilhelm.
\newblock Adaptive hybrid optimal quantum control for imprecisely characterized
  systems.
\newblock {\em Phys. Rev. Lett.}, 112:240503, 2014.

\bibitem{Egger2014c}
D.~J. Egger and F.~K. Wilhelm.
\newblock Optimal control of a quantum measurement.
\newblock {\em Phys. Rev. A}, 90:052331, 2014.

\bibitem{Egger2014b}
D.~J. Egger and F.~K. Wilhelm.
\newblock Optimized controlled-{Z} gates for two superconducting qubits coupled
  through a resonator.
\newblock {\em Supercond. Sci. Technol.}, 27:014001, 2014.

\bibitem{Ehni2013a}
S.~Ehni.
\newblock {\em Optimal Control in High Resolution NMR Spectroscopy: Transfer
  Elements and their Application for Structure Elucidation}.
\newblock Dissertation, Karlsruhe, 2013.

\bibitem{Ehni2013}
Sebastian Ehni and Burkhard Luy.
\newblock {BEBE}tr and {BUBI}: {J}-compensated concurrent shaped pulses for
  {H}-1-{C}-13 experiments.
\newblock {\em J. Magn. Reson.}, 232:7--17, 2013.

\bibitem{Ehni2014}
Sebastian Ehni and Burkhard Luy.
\newblock Robust {INEPT} and refocused {INEPT} transfer with compensation of a
  wide range of couplings, offsets, and {B}-1-field inhomogeneities (cob3).
\newblock {\em J. Magn. Reson.}, 247:111--117, 2014.

\bibitem{Eitan2011}
Reuven Eitan, Michael Mundt, and David~J. Tannor.
\newblock Optimal control with accelerated convergence: Combining the {K}rotov
  and quasi-{N}ewton methods.
\newblock {\em Phys. Rev. A}, 83(5):053426, 2011.

\bibitem{Elliott2009}
D.~Elliott.
\newblock {\em Bilinear Control Systems: Matrices in Action}.
\newblock Springer, London, 2009.

\bibitem{Emary2014}
Clive Emary and John Gough.
\newblock Coherent feedback control in quantum transport.
\newblock {\em Phys. Rev. B}, 90(20):205436, 2014.

\bibitem{Engel2009a}
Volker Engel, Christoph Meier, and David~J. Tannor.
\newblock {\em Local Control Theory: Recent Applications to Energy and Particle
  Transfer Processes in Molecules}, volume 141, pages 29--101.
\newblock John Wiley \& Sons, Inc., feb 2009.

\bibitem{Englert2008}
L.~Englert, M.~Wollenhaupt, L.~Haag, C.~Sarpe-Tudoran, B.~Rethfeld, and
  T.~Baumert.
\newblock Material processing of dielectrics with temporally asymmetric shaped
  femtosecond laser pulses on the nanometer scale.
\newblock {\em Appl. Phys. A}, 92(4):749--753, 2008.

\bibitem{Enthart2008}
Andreas Enthart, J.~Christoph Freudenberger, Julien Furrer, Horst Kessler, and
  Burkhard Luy.
\newblock The {CLIP}/{CLAP}-{HSQC}: Pure absorptive spectra for the measurement
  of one-bond couplings.
\newblock {\em J. Magn. Reson.}, 192(2):314--322, 2008.

\bibitem{Ernst1992}
Richard~R Ernst.
\newblock Nuclear magnetic resonance fourier transform spectroscopy (nobel
  lecture).
\newblock {\em Angewandte Chemie International Edition in English},
  31(7):805--823, 1992.

\bibitem{Ernst1987}
Richard~R Ernst, Geoffrey Bodenhausen, and Alexander Wokaun.
\newblock {\em Principles of nuclear magnetic resonance in one and two
  dimensions}, volume~14.
\newblock Clarendon Press Oxford, 1987.

\bibitem{Ferraro2015}
E.~Ferraro, M.~De~{M}ichielis, M.~Fanciulli, and E.~Prati.
\newblock Coherent tunneling by adiabatic passage of an exchange-only spin
  qubit in a double quantum dot chain.
\newblock {\em Phys. Rev. B}, 91(7):075435, 2015.

\bibitem{Fisher2009}
R.~Fisher, H.~Yuan, A.~Sp\"{o}rl, and S.~J. Glaser.
\newblock Timeoptimal generation of cluster states.
\newblock {\em Phys. Rev. A}, 79:042304, 2009.

\bibitem{Floether2012}
Frederik~F. Floether, Pierre de~{F}ouquieres, and Sophie~G. Schirmer.
\newblock Robust quantum gates for open systems via optimal control: Markovian
  versus non-markovian dynamics.
\newblock {\em New J. Phys.}, 14:073023, jul 2012.

\bibitem{Fowler2014}
A.~G. Fowler and J.~M. Martinis.
\newblock Quantifying the effects of local many-qubit errors and non-local
  two-qubit errors on the surface code.
\newblock {\em Phys. Rev. A}, 89:032316, 2014.

\bibitem{Freeman1991}
R.~Freeman.
\newblock Selective excitation in high-resolution {NMR}.
\newblock {\em Chem. Rev.}, 91:1397--1312, 1991.

\bibitem{Freeman1998}
R.~Freeman.
\newblock Shaped radiofrequency pulses in high resolution {NMR}.
\newblock {\em Prog. Nucl. Magn. Reson. Spectrosc.}, 32:59--106, 1998.

\bibitem{Freeman1987}
R.~Freeman and X.~Wu.
\newblock Design of magnetic resonance experiments by genetic evolution.
\newblock {\em J. Magn. Reson.}, 75:184--189, 1987.

\bibitem{Frostig2015}
Hadas Frostig, Tim Bayer, Nirit Dudovich, Yonina~C Eldar, and Yaron Silberberg.
\newblock Single-beam spectrally controlled two-dimensional raman spectroscopy.
\newblock {\em Nature Photon.}, 9:339--343, 2015.

\bibitem{Furst2014}
H.~A. F\"{u}rst, M.~H. Goerz, U.~G. Poschinger, M.~Murphy, S.~Montangero,
  T.~Calarco, F.~Schmidt-Kaler, K.~Singer, and C.~P. Koch.
\newblock Controlling the transport of an ion: Classical and quantum mechanical
  solutions.
\newblock {\em New J. Phys.}, 16:075007, 2014.
\newblock arXiv:1312.4156.

\bibitem{Garon2013}
A.~Garon, S.~J. Glaser, and D.~Sugny.
\newblock Time-optimal control of {SU}(2) quantum operations.
\newblock {\em Phys. Rev. A}, 88:043422, 2013.

\bibitem{Garon2015}
A.~Garon, R.~Zeier, and S.~J. Glaser.
\newblock Visualizing operators of coupled spins systems.
\newblock {\em Phys. Rev. A}, 91:042122/1--28, 2015.

\bibitem{Matson1994}
Matson G.B.
\newblock An integrated program for amplitude-modulated rf pulse generation and
  re-mapping with shaped gradients.
\newblock {\em Magn. Reson. Imaging}, 12:1205--1225, 1994.

\bibitem{Gershenzon2007}
Naum~I. Gershenzon, Kyryl Kobzar, Burkhard Luy, Steffen~J. Glaser, and
  Thomas~E. Skinner.
\newblock Optimal control design of excitation pulses that accommodate
  relaxation.
\newblock {\em J. Magn. Reson.}, 188(2):330--336, 2007.

\bibitem{Gershenzon2008}
Naum~I. Gershenzon, Thomas~E. Skinner, Bernhard Brutscher, Navin Khaneja, Manoj
  Nimbalkar, Burkhard Luy, and Steffen~J. Glaser.
\newblock Linear phase slope in pulse design: Application to coherence
  transfer.
\newblock {\em J. Magn. Reson.}, 192(2):235--243, 2008.

\bibitem{Glaser1989}
S.~J. Glaser and G.~Drobny.
\newblock The tailored {TOCSY} experiment: Chemical shift selective coherence
  transfer.
\newblock {\em Chem. Phys. Lett.}, 164:456--462, 1989.

\bibitem{Glaser1990}
S.~J. Glaser and G.~P. Drobny.
\newblock Assessment and optimization of pulse sequences for homonuclear
  isotropic mixing.
\newblock {\em Advances in Magnetic Resonance, W.S. Warren, Ed.}, 14:35--58,
  1990.

\bibitem{Goerz2015}
M.~H. Goerz, G.~Gualdi, D.~M. Reich, C.~P. Koch, F.~Motzoi, K.~B. Whaley,
  J.~Vala, M.~M. M\"{u}ller, S.~Montangero, and T.~Calarco.
\newblock Optimizing for an arbitrary perfect entangler. ii. application.
\newblock {\em Phys. Rev. A}, 91, 2015.
\newblock arXiv:1412.7350.

\bibitem{Goerz2011}
Michael~H. Goerz, Tommaso Calarco, and Christiane~P. Koch.
\newblock The quantum speed limit of optimal controlled phasegates for trapped
  neutral atoms.
\newblock {\em J. Phys. B: At., Mol. Opt. Phys.}, 44:154011, jul 2011.

\bibitem{Goerz2014}
Michael~H. Goerz, Eli~J. Halperin, Jon~M. Aytac, Christiane~P. Koch, and
  K.~Birgitta Whaley.
\newblock Robustness of high-fidelity rydberg gates with single-site
  addressability.
\newblock {\em Phys. Rev. A}, 90:032329, sep 2014.
\newblock arXiv:1401.1858.

\bibitem{Goerz2014a}
Michael~H. Goerz, Daniel~M. Reich, and Christiane~P. Koch.
\newblock Optimal control theory for a unitary operation under dissipative
  evolution.
\newblock {\em New J. Phys.}, 16:055012, may 2014.
\newblock arXiv:1310.2271.

\bibitem{Gollub2008}
Caroline Gollub, Markus Kowalewski, and Regina de~Vivie-Riedle.
\newblock Monotonic convergent optimal control theory with strict limitations
  on the spectrum of optimized laser fields.
\newblock {\em Phys. Rev. Lett.}, 101:073002, 2008.

\bibitem{Gordon1997}
R.~J. Gordon and S.~A. Rice.
\newblock Active control of the dynamics of atoms and molecules.
\newblock {\em Annu. Rev. Phys. Chem.}, 48:601, 1997.

\bibitem{Gorshkov2008}
Alexey~V. Gorshkov, Tommaso Calarco, Mikhail~D. Lukin, and Anders~S.
  S\o{}rensen.
\newblock Photon storage in $\ensuremath{\Lambda}$-type optically dense atomic
  media. iv. optimal control using gradient ascent.
\newblock {\em Phys. Rev. A}, 77:043806, 2008.

\bibitem{Gough2009}
J.~E. Gough and S.~Wildfeuer.
\newblock Enhancement of field squeezing using coherent feedback.
\newblock {\em Phys. Rev. A}, 80:042107, oct 2009.

\bibitem{Grace2010}
Matthew~D. Grace, Jason Dominy, Robert~L. Kosut, Constantin Brif, and Herschel
  Rabitz.
\newblock Environment-invariant measure of distance between evolutions of an
  open quantum system.
\newblock {\em New J. Phys.}, 12(1):015001, jan 2010.

\bibitem{Greenman2015}
Loren Greenman, Christiane~P. Koch, and K.~Birgitta Whaley.
\newblock Laser pulses for coherent xuv raman excitation.
\newblock {\em Phys. Rev. A}, 92:in press, 2015.

\bibitem{Greentree2004}
Andrew~D. Greentree, Jared~H. Cole, A.~R. Hamilton, and Lloyd C.~L. Hollenberg.
\newblock Coherent electronic transfer in quantum dot systems using adiabatic
  passage.
\newblock {\em Phys. Rev. B}, 70(23):235317, 2004.

\bibitem{Griesinger2012}
C.~Griesinger, M.~Bennati, H.~M. Vieth, C.~Luchinat, G.~Parigi, P.~H{\"o}fer,
  F.~Engelke, S.~J. Glaser, V.~Denysenkov, and T.~F. Prisner.
\newblock Dynamic nuclear polarization at high magnetic fields in liquids.
\newblock {\em Prog. NMR Spectrosc.}, 64:4--28, 2012.

\bibitem{Grissom2009}
W.~A. Grissom, D.~Xu, A.~B. Kerr, and J.~A. Fessler.
\newblock Fast large-tip-angle multidimensional and parallel {RF} pulse design
  in {MRI}.
\newblock {\em IEEE Trans. Med. Imag.}, 28:1548--1559, 2009.

\bibitem{Grond2009}
Julian Grond, J\"org Schmiedmayer, and Ulrich Hohenester.
\newblock Optimizing number squeezing when splitting a mesoscopic condensate.
\newblock {\em Phys. Rev. A}, 79:021603, 2009.

\bibitem{Gross1992}
P.~Gross, D.~Neuhauser, and H.~Rabitz.
\newblock Optimal control of curve-crossing systems.
\newblock {\em J. Chem. Phys.}, 96(4):2834, 1992.

\bibitem{Gross1997}
Peter Gross and Marcos Dantus.
\newblock Femtosecond photoassociation: Coherence and implications for control
  in bimolecular reactions.
\newblock {\em J. Chem. Phys.}, 106:8013, feb 1997.

\bibitem{Guerin2003}
S.~Gu\'{e}rin and H.~R. Jauslin.
\newblock {\em Control of Quantum Dynamics by Laser Pulses: Adiabatic {F}loquet
  Theory}, volume 125, pages 147--267.
\newblock John Wiley \& Sons, Inc., 2003.

\bibitem{Haberle2013}
T.~H\"aberle, D.~Schmid-Lorch, K.~Karrai, F.~Reinhard, and J.~Wrachtrup.
\newblock High-dynamic-range imaging of nanoscale magnetic fields using optimal
  control of a single qubit.
\newblock {\em Phys. Rev. Lett.}, 111:170801, 2013.

\bibitem{Haeberlen1976}
U.~Haeberlen.
\newblock {\em The Principles of Nuclear Magnetism}.
\newblock Academic Press, New York, 1976.

\bibitem{Hahn1950}
E.~L. Hahn.
\newblock Spin echoes.
\newblock {\em Phys. Rev.}, 80:580--594, 1950.

\bibitem{Hamm2011}
Peter Hamm and Martin Zanni.
\newblock {\em Concepts and methods of 2D infrared spectroscopy}.
\newblock Cambridge University Press, 2011.

\bibitem{Hammerer2002}
K.~Hammerer, G.~Vidal, and J.~I. Cirac.
\newblock Charaterization of nonlocal gates.
\newblock {\em Phys. Rev. A}, 66(6):062321, 2002.

\bibitem{Hansen2012}
Anders~K. Hansen, Magnus~A. S{\o}rensen, Peter~F. Staanum, and Michael Drewsen.
\newblock Single-ion recycling reactions.
\newblock {\em Angewandte Chemie International Edition}, 51(32):7960--7962,
  2012.

\bibitem{Herek2002}
J.~Herek, M.~Wohlleben, X.~Cogdell, X.~Zeidler, and M.~Motzkus.
\newblock Quantum control of energy flow in light harvesting.
\newblock {\em Nature}, 417:533, 2002.

\bibitem{Hillery2009}
M.~Hillery and V.~Buzek.
\newblock Quantum machines.
\newblock {\em Contemp. Phys.}, 50:575--586, 2009.

\bibitem{Hincks2014}
I.N. Hincks, C.~Granade, T.W. Borneman, and D.G. Cory.
\newblock Accounting for classical hardware in the control of quantum devices.
\newblock {\em arXiv:1409.8178 [quant-ph]}, 2014.

\bibitem{Hodges2008}
J.~S. Hodges, J.~C. Yang, C.~Ramanathan, and D.~G. Cory.
\newblock Universal control of nuclear spins via anisotropic hyperfine
  interactions.
\newblock {\em Phys. Rev. A}, 78:010303, Jul 2008.

\bibitem{Hogben2011}
H.~J. Hogben, M.~Krzystyniak, G.~T.~P. Charnock, P.~J. Hore, and Ilya Kuprov.
\newblock Spinach -- a software library for simulation of spin dynamics in
  large spin systems.
\newblock {\em J. Magn. Reson.}, 208(2):179--194, feb 2011.

\bibitem{Hohenester2007}
Ulrich Hohenester, Per~Kristian Rekdal, Alfio Borz\`{i}, and J\"{o}rg
  Schmiedmayer.
\newblock Optimal quantum control of {B}ose-{E}instein condensates in magnetic
  microtraps.
\newblock {\em Phys. Rev. A}, 75:023602, 2007.

\bibitem{Hoque2011}
M.~Z. Hoque, M.~Lapert, E.~Hertz, F.~Billard, D.~Sugny, B.~Lavorel, and
  O.~Faucher.
\newblock Observation of laser-induced field-free permanent planar alignment of
  molecules.
\newblock {\em Phys. Rev. A}, 84:013409, 2011.

\bibitem{Hovav2014}
Y.~Hovav, A.~Feintuch, S.~Vega, and D.~Goldfarb.
\newblock Dynamic nuclear polarization using frequency modulation at 3.34 {T}.
\newblock {\em J. Magn. Reson.}, 238:94--105, 2014.

\bibitem{Iida2012}
S.~Iida, M.~Yukawa, H.~Yonezawa, N.~Yamamoto, and A.~Furusawa.
\newblock Experimental demonstration of coherent feedback control on optical
  field squeezing.
\newblock {\em IEEE Trans. Automat. Control}, 57(8):2045--2050, 2012.

\bibitem{Mao1986}
Mao J, Mareci TH, Scott KN, and Andrew ER.
\newblock Selective inversion radiofrequency pulses by optimal control.
\newblock {\em J. Magn. Reson.}, 70:310--318, 1986.

\bibitem{Rosenfeld1996}
Mao J, Mareci TH, Scott KN, and Andrew ER.
\newblock Design of adiabatic selective pulses using optimal control theory.
\newblock {\em Magn. Reson. Med.}, 36:401--409, 1996.

\bibitem{Jager2013}
Georg J\"ager and Ulrich Hohenester.
\newblock Optimal quantum control of {B}ose-{E}instein condensates in magnetic
  microtraps: Consideration of filter effects.
\newblock {\em Phys. Rev. A}, 88:035601, sep 2013.

\bibitem{Jager2014}
Georg J\"ager, Daniel~M. Reich, Michael~H. Goerz, Christiane~P. Koch, and
  Ulrich Hohenester.
\newblock Optimal quantum control of bose-einstein condensates in magnetic
  microtraps: Comparison of gradient-ascent-pulse-engineering and krotov
  optimization schemes.
\newblock {\em Phys. Rev. A}, 90:033628, 2014.
\newblock 1409.2976.

\bibitem{James2008}
M.~R. James, H.~I. Nurdin, and I.~R. Petersen.
\newblock $h^\infty$ control of linear quantum stochastic systems.
\newblock {\em IEEE Trans. Automat. Control}, 53(8):1787--1803, sep 2008.

\bibitem{Janich2012}
M.~A. Janich, M.~A. McLean, R.~Noeske, S.~J. Glaser, and R.~F. Schulte.
\newblock Slice-selective broadband refocusing pulses for the robust generation
  of crushed spin-echoes.
\newblock {\em J. Magn. Reson.}, 223:129--137, 2012.

\bibitem{Janich2011}
M.~A. Janich, R.~F. Schulte, M.~Schwaiger, and S.~J. Glaser.
\newblock Robust slice-selective broadband refocusing pulses.
\newblock {\em J. Magn. Reson.}, 213:126--135, 2011.

\bibitem{Jelezko2004}
F.~Jelezko, T.~Gaebel, I.~Popa, M.~Domhan, A.~Gruber, and J.~Wrachtrup.
\newblock Observation of coherent oscillation of a single nuclear spin and
  realization of a two-qubit conditional quantum gate.
\newblock {\em Phys. Rev. Lett.}, 93(13):130501, 2004.

\bibitem{Jelezko2006}
F.~Jelezko and J.~Wrachtrup.
\newblock Single defect centres in diamond: A review.
\newblock {\em Phys. Status Solidi A}, 203(13):3207--3225, 2006.

\bibitem{Jirari2009}
H.~Jirari, F.W.J. Hekking, and O.~Buisson.
\newblock Optimal control of superconducting n-level quantum systems.
\newblock {\em Europhys. Lett.}, 87:28004, 2009.

\bibitem{Johansson2013}
J.~R. Johansson, P.~D. Nation, and F.~Nori.
\newblock Qutip 2: A python framework for the dynamics of open quantum systems.
\newblock {\em Comp. Phys. Comm.}, 184:1234, 2013.

\bibitem{Jones2011}
J.~A. Jones.
\newblock Quantum computing with nmr.
\newblock {\em Prog. NMR Spectrosc.}, 59:91--120, 2011.

\bibitem{Joye2007}
Alain Joye.
\newblock General adiabatic evolution with a gap condition.
\newblock {\em Communications in Mathematical Physics}, 275(1):139--162, 2007.

\bibitem{Judson1992}
Richard~S. Judson and Herschel Rabitz.
\newblock Teaching lasers to control molecules.
\newblock {\em Phys. Rev. Lett.}, 68(10):1500--1503, 1992.

\bibitem{Jurdjevic1997}
V.~Jurdjevic.
\newblock {\em Geometric Control Theory}.
\newblock Cambridge University Press, Cambridge, Cambridge, 1997.

\bibitem{Jurdjevic1972}
V.~Jurdjevic and H.~Sussmann.
\newblock Control systems on lie groups.
\newblock {\em Journal of Differential Equations}, 12:313--329, sep 1972.

\bibitem{Kallush2006}
S.~Kallush and R.~Kosloff.
\newblock Quantum governor: Automatic quantum control and reduction of the
  influence of noise without measuring.
\newblock {\em Phys. Rev. A}, 73:032324, 2006.

\bibitem{Kalman1969}
R.~Kalman, P.~L. Falb, and M.~A. Arbib.
\newblock {\em Topics in Mathematical System Theory}.
\newblock McGraw-Hill, New York, 1969.

\bibitem{Kalman1960}
R.~E. Kalman.
\newblock A new approach to linear filtering and prediction problems.
\newblock {\em J. Basic Engineering}, 82:35--45, 1960.

\bibitem{Katsuki2006}
Hiroyuki Katsuki, Hisashi Chiba, Bertrand Girard, Christoph Meier, and Kenji
  Ohmori.
\newblock Visualizing picometric quantum ripples of ultrafast wave-packet
  interference.
\newblock {\em Science}, 311(5767):1589--1592, 2006.

\bibitem{Katz2010}
Gil Katz, Mark~A. Ratner, and Ronnie Kosloff.
\newblock Control by decoherence: weak field control of an excited state
  objective.
\newblock {\em New J. Phys.}, 12(1):015003, 2010.

\bibitem{Katz2011}
Ori Katz, Eran Small, Yaron Bromberg, and Yaron Silberberg.
\newblock Focusing and compression of ultrashort pulses through scattering
  media.
\newblock {\em Nature Photon.}, 5:372--377, 2011.

\bibitem{Kaufmann2013}
T.~Kaufmann, J.~M. Franck, T.~J. Keller, R.~P. Barnes, S.~J. Glaser, J.~M.
  Martinis, and S.~Han.
\newblock {DAC}-board based {X}-band {EPR} spectrometer with arbitrary waveform
  control.
\newblock {\em J. Magn. Reson.}, 235:95--108, 2013.

\bibitem{Kawakami2014}
E.~Kawakami, P.~Scarlino, D.~R. Ward, F.~R. Braakman, D.~E. Savage, M.~G.
  Lagally, and L.~M.~K. Vandersypen.
\newblock Electrical control of a long-lived spin qubit in a {S}i/{S}i{G}e
  quantum dot.
\newblock {\em Nat. Nanotechnol.}, 9:666--670, 2014.

\bibitem{Kehlet2004}
C.~Kehlet, A.~C. Sivertsen, M.~Bjerring, T.~O. Reiss, N.~Khaneja, S.~J. Glaser,
  and N.~C. Nielsen.
\newblock Improving solid-state nmr dipolar recoupling by optimal control.
\newblock {\em J. Am. Chem. Soc.}, 126:10202--10203, 2004.

\bibitem{Kelly2014}
J.~Kelly, R.~Barends, B.~Campbell, Y.~Chen, Z.~Chen, B.~Chiaro, A.~Dunsworth,
  A.~G. Fowler, I.-C. Hoi, E.~Jeffrey, A.~Megrant, J.~Mutus, C.~Neill, P.~J.~J.
  O'Malley, C.~Quintana, P.~Roushan, D.~Sank, A.~Vainsencher, J.~Wenner, T.~C.
  White, A.~N. Cleland, and John~M. Martinis.
\newblock Optimal quantum control using randomized benchmarking.
\newblock {\em Phys. Rev. Lett.}, 112:240504, jun 2014.

\bibitem{Keyl2014}
M.~Keyl, R.~Zeier, and T.~Schulte-Herbr{\"u}ggen.
\newblock Controlling several atoms in a cavity.
\newblock {\em New J. Phys.}, 16:065010, jun 2014.

\bibitem{Khaneja2001a}
N.~Khaneja and S.~J. Glaser.
\newblock Cartan decomposition of su(2n), constructive controllability of spin
  systems and universal quantum computing.
\newblock {\em Chem. Phys.}, 267:11, 2001.

\bibitem{Khaneja2002a}
N.~Khaneja and S.~J. Glaser.
\newblock Efficient transfer of coherence through {I}sing spin chains.
\newblock {\em Phys. Rev. A}, 66:060301(R), 2002.

\bibitem{Khaneja2007a}
N.~Khaneja, B.~Heitmann, A.~Sp{\"o}rl, H.~Yuan, T.~Schulte-Herbr{\"u}ggen, and
  S.~J. Glaser.
\newblock Shortest paths for efficient control of indirectly coupled qubits.
\newblock {\em Phys. Rev. A}, 75(1):012322, 2007.

\bibitem{Khaneja2005a}
N.~Khaneja, F.~Kramer, and S.~J. Glaser.
\newblock Optimal experiments for maximizing coherence transfer between coupled
  spins.
\newblock {\em J. Magn. Reson.}, 173(1):116--124, 2005.

\bibitem{Khaneja2003a}
N.~Khaneja, B.~Luy, and S.~J. Glaser.
\newblock Boundary of quantum evolution under decoherence.
\newblock {\em Proc. Natl. Acad. Sci. USA}, 100:13162--13166, 2003.

\bibitem{Khaneja2003}
N.~Khaneja, T.~Reiss, B.~Luy, and S.~J. Glaser.
\newblock Optimal control of spin dynamics in the presence of relaxation.
\newblock {\em J. Magn. Reson.}, 162:311--319, 2003.

\bibitem{Khaneja2007}
Navin Khaneja.
\newblock Switched control of electron nuclear spin systems.
\newblock {\em Phys. Rev. A}, 76:032326, sep 2007.

\bibitem{Khaneja2001}
Navin Khaneja, Roger Brockett, and Steffen~J. Glaser.
\newblock Time optimal control in spin systems.
\newblock {\em Phys. Rev. A}, 63:032308, 2001.

\bibitem{Khaneja2002}
Navin Khaneja, Steffen~J. Glaser, and Roger Brockett.
\newblock Sub-{R}iemannian geometry and time optimal control of three spin
  systems: Quantum gates and coherence transfer.
\newblock {\em Phys. Rev. A}, 65:032301, 2002.

\bibitem{Khaneja2005}
Navin Khaneja, Timo Reiss, Cindie Kehlet, Thomas Schulte-{H}erbr\"{u}ggen, and
  Steffen~J. Glaser.
\newblock Optimal control of coupled spin dynamics: design of {NMR} pulse
  sequences by gradient ascent algorithms.
\newblock {\em J. Magn. Reson.}, 172(2):296--305, 2005.

\bibitem{Kim2014}
D.~Kim, Z.~Shi, C.~B. Simmonds, D.~R. Ward, J.~R. Prance, T.~S. Koh, and M.~A.
  Eriksson.
\newblock Quantum control and process tomography of a semiconductor quantum dot
  hybrid qubit.
\newblock {\em Nature}, 511:70--74, 2014.

\bibitem{Kirchmair2013}
Gerhard Kirchmair, Brian Vlastakis, Zaki Leghtas, Simon~E. Nigg, Hanhee Paik,
  Eran Ginossar, Mazyar Mirrahimi, Luigi Frunzio, S.~M. Girvin, and R.~J.
  Schoelkopf.
\newblock Observation of quantum state collapse and revival due to the
  single-photon kerr effect.
\newblock {\em Nature}, 495(7440):205--209, mar 2013.

\bibitem{Kirk2004}
D.~E. Kirk.
\newblock {\em Optimal Control Theory: An introduction}.
\newblock Dover Publications, New York, 2004.

\bibitem{Kiselev2011}
Denis Kiselev, Luigi Bonacina, and Jean-Pierre Wolf.
\newblock Individual bioaerosol particle discrimination by multi-photon excited
  fluorescence.
\newblock {\em Opt. Express}, 19(24):24516--24521, 2011.

\bibitem{Kobzar2007}
K.~Kobzar.
\newblock {\em Optimal Control, Partial Alignment and More: The Design of Novel
  Tools for NMR Spectroscopy of Small Molecules}.
\newblock Dissertation, Munich, 2007.

\bibitem{Kobzar2012}
K.~Kobzar, S.~Ehni, T.~E. Skinner, S.~J. Glaser, and B.~Luy.
\newblock Exploring the limits of broadband 90$^\circ$ and 180$^\circ$
  universal rotation pulses.
\newblock {\em J. Magn. Reson.}, 225:142--160, 2012.

\bibitem{Kobzar2005}
K.~Kobzar, B.~Luy, N.~Khaneja, and S.~J. Glaser.
\newblock Pattern pulses: Design of arbitrary excitation profiles as a function
  of pulse amplitude and offset.
\newblock {\em J. Magn. Reson.}, 173:229--235, 2005.

\bibitem{Kobzar2008}
K.~Kobzar, T.~E. Skinner, N.~Khaneja, S.~J. Glaser, and B.~Luy.
\newblock Exploring the limits of excitation and inversion pulses ii. rf-power
  optimized pulses.
\newblock {\em J. Magn. Reson.}, 194:58--66, 2008.

\bibitem{Kobzar2004}
Kyryl Kobzar, Thomas~E. Skinner, Navin Khaneja, Steffen~J. Glaser, and Burkhard
  Luy.
\newblock Exploring the limits of broadband excitation and inversion pulses.
\newblock {\em J. Magn. Reson.}, 170(2):236--243, 2004.

\bibitem{Koch2004}
Christiane~P Koch, Jos{\'e}~P Palao, Ronnie Kosloff, and Fran{\c{c}}oise
  Masnou-Seeuws.
\newblock Stabilization of ultracold molecules using optimal control theory.
\newblock {\em Phys. Rev. A}, 70(1):013402, 2004.

\bibitem{Koch2012}
Christiane~P. Koch and Moshe Shapiro.
\newblock Coherent control of ultracold photoassociation.
\newblock {\em Chem. Rev.}, 212:4928, 2012.

\bibitem{Kocher2014}
S.~S. K\"ocher, T.~Heydenreich, and S.~J. Glaser.
\newblock Visualization and analysis of modulated pulses in magnetic resonance
  by joint time--frequency representations.
\newblock {\em J. Magn. Reson.}, 249(0):63--71, 2014.

\bibitem{Koroleva2013}
V.~D.~M. Koroleva, S.~Mandal, Y.-Q. Song, and M.~D. H{\"u}rlimann.
\newblock Broadband {CPMG} sequence with short composite refocusing pulses.
\newblock {\em J. Magn. Reson.}, 230:64--75, 2013.

\bibitem{Kosloff1989}
R.~Kosloff, S.~A. Rice, P.~Gaspard, S.~Tersigni, and D.~J. Tannor.
\newblock Wavepacket dancing: Achieving chemical selectivity by shaping light
  pulses.
\newblock {\em Chemical Physics}, 139(1):201--220, 1989.

\bibitem{Krauter2011}
Hanna Krauter, Christine~A. Muschik, Kasper Jensen, Wojciech Wasilewski,
  Jonas~M. Petersen, J.~Ignacio Cirac, and Eugene~S. Polzik.
\newblock Entanglement generated by dissipation and steady state entanglement
  of two macroscopic objects.
\newblock {\em Phys. Rev. Lett.}, 107:080503, 2011.

\bibitem{Krotov1996}
V.~F. Krotov.
\newblock {\em Global Methods in Optimal Control Theory}.
\newblock Dekker, New York, 1996.

\bibitem{Kuhn2013}
L.~T. Kuhn.
\newblock {\em Hyperpolarization Methods in {NMR} Spectroscopy}.
\newblock Springer, 2013.

\bibitem{Kuhn2007}
O.~K\"uhn and L.~W\"oste, editors.
\newblock {\em Analysis and control of ultrafast photoinduced reactions}.
\newblock Springer, Berlin, 2007.

\bibitem{Kuprov2013}
Ilya Kuprov.
\newblock Spin system trajectory analysis under optimal control pulses.
\newblock {\em J. Magn. Reson.}, 233(0):107--112, aug 2013.

\bibitem{Kurizki2015}
G.~Kurizki, P.~Bertet, Y.~Kubo, K.~M{\o}lmer, D.~Petrosyan, P.~Rabl, and
  J.~Schmiedmayer.
\newblock Quantum technologies with hybrid systems.
\newblock {\em Proc. Natl. Acad. Sci. USA}, 112:3866--3873, 2015.

\bibitem{Kurniawan2012}
I.~Kurniawan, G.~Dirr, and U.~Helmke.
\newblock Controllability aspects of quantum dynamics: A unified approach for
  closed and open systems.
\newblock {\em IEEE Trans. Autom. Contr. (IEEE-TAC)}, 57:1984--1996, 2012.

\bibitem{Kurnit1964}
N.~A. Kurnit, I.~D. Abella, and S.~R. Hartmann.
\newblock Observation of a photon echo.
\newblock {\em Phys. Rev. Lett.}, 13:567--568, 1964.

\bibitem{Lapert2013a}
M.~Lapert, E.~Ass{\'e}mat, S.~J. Glaser, and D.~Sugny.
\newblock Understanding the global structure of two-level quantum systems with
  relaxation: Vector fields organized through the magic plane and the
  steady-state ellipsoid.
\newblock {\em Phys. Rev. A}, 88:033407, 2013.

\bibitem{Lapert2014}
M.~Lapert, E.~Ass\'emat, S.~J. Glaser, and D.~Sugny.
\newblock Optimal control of the signal-to-noise ratio per unit time for a
  spin-1/2 particle.
\newblock {\em Phys. Rev. A}, 90:023411, 2014.

\bibitem{Lapert2012b}
M.~Lapert, J.~Salomon, and D.~Sugny.
\newblock Time-optimal monotonically convergent algorithm with an application
  to the control of spin systems.
\newblock {\em Phys. Rev. A}, 85:033406, 2012.

\bibitem{Lapert2012a}
M.~Lapert and D.~Sugny.
\newblock Field-free molecular orientation by terahertz laser pulses at high
  temperature.
\newblock {\em Phys. Rev. A}, 85:063418, 2012.

\bibitem{Lapert2009}
M.~Lapert, R.~Tehini, G.~Turinici, and D.~Sugny.
\newblock Monotonically convergent optimal control theory of quantum systems
  with spectral constraints on the control field.
\newblock {\em Phys. Rev. A}, 79:063411, 2009.

\bibitem{Lapert2010}
M.~Lapert, Y.~Zhang, M.~Braun, S.~J. Glaser, and D.~Sugny.
\newblock Singular extremals for the time-optimal control of dissipative spin
  $\frac{1}{2}$ particles.
\newblock {\em Phys. Rev. Lett.}, 104:083001, feb 2010.

\bibitem{Lapert2011}
M.~Lapert, Y.~Zhang, S.~J. Glaser, and D.~Sugny.
\newblock Towards the time-optimal control of dissipative spin 1/2 particles in
  nuclear magnetic resonance.
\newblock {\em J. Phys. B: At., Mol. Opt. Phys.}, 44:154014, 2011.

\bibitem{Lapert2012}
M.~Lapert, Y.~Zhang, M.~Janich, S.~J. Glaser, and D.~Sugny.
\newblock Exploring the physical limits of saturation contrast in magnetic
  resonance imaging.
\newblock {\em Scientific Reports}, 2:589, 2012.

\bibitem{Lee1967}
E.~B. Lee and L.~Markus.
\newblock {\em Foundations of Optimal Control Theory}.
\newblock Wiley, New York, 1967.

\bibitem{Lee2008}
J.-S. Lee, R.R. Regatte, and A.~Jerschow.
\newblock Optimal nuclear magnetic resonance excitation schemes for the central
  transition of a spin 3/2 in the presence of residual quadrupolar coupling.
\newblock {\em J. Chem. Phys.}, 129:224510, 2008.

\bibitem{Levin2015}
Liat Levin, Wojciech Skomorowski, Ronnie Kosloff, and Zohar Koch, Christiane
  P.and~Amitay.
\newblock Coherent control of bond making.
\newblock {\em Phys. Rev. Lett.}, 114:in press, jun 2015.

\bibitem{Levis2001}
R.~J. Levis, G.~M. Menkir, and H.~Rabitz.
\newblock Selective bond dissociation and rearrangement with optimally
  tailored, strong-field laser pulses.
\newblock {\em Science}, 292:709, 2001.

\bibitem{Levitt1986}
M.~H. Levitt.
\newblock Composite pulses.
\newblock {\em Prog. Nucl. Magn. Reson. Spectrosc.}, 18:61--122, 1986.

\bibitem{Levitt1996}
M.~H. Levitt.
\newblock {\em Composite pulses}.
\newblock Wiley, 1996.

\bibitem{Levitt2001}
M.~H. Levitt.
\newblock {\em Spin Dynamics: Basics of Nuclear Magnetic Resonance}.
\newblock Wiley, 2001.

\bibitem{Li2008}
Haowen Li, D.~Ahmasi Harris, Bingwei Xu, Paul~J. Wrzesinski, Vadim~V. Lozovoy,
  and Marcos Dantus.
\newblock Coherent mode-selective raman excitation towards standoff detection.
\newblock {\em Opt. Express}, 16(8):5499--5504, 2008.

\bibitem{Li2009}
J.-S. Li and N.~Khaneja.
\newblock Ensemble control of bloch equations.
\newblock {\em IEEE Trans. Autom. Control}, 54(3):528--536, 2009.

\bibitem{Li2011}
J.-S. Li, J.~Ruths, T.-Y. Yu, H.~Arthanari, and G.~Wagner.
\newblock Optimal pulse design in quantum control: A unified computational
  method.
\newblock {\em Proc. Natl. Acad. Sci. USA}, 108:1879--1884, 2011.

\bibitem{Li2006}
Jr-Shin Li and Navin Khaneja.
\newblock Control of inhomogeneous quantum ensembles.
\newblock {\em Phys. Rev. A}, 73:030302, 2006.

\bibitem{Li2010}
Xuan Li, Cian Menzel-Jones, David Avisar, and Moshe Shapiro.
\newblock Solving the spectroscopic phase: imaging excited wave packets and
  extracting excited state potentials from fluorescence data.
\newblock {\em Phys. Chem. Chem. Phys.}, 12:15760--15765, 2010.

\bibitem{Lien2014}
Chien-Yu Lien, Christopher~M Seck, Yen-Wei Lin, Jason~HV Nguyen, David~A Tabor,
  and Brian~C Odom.
\newblock Broadband optical cooling of molecular rotors from room temperature
  to the ground state.
\newblock {\em Nat. Commun.}, 5, 2014.

\bibitem{Liu1990}
H.~Liu, S.~J. Glaser, and G.~Drobny.
\newblock Development and optimization of multipulse propagators: Applications
  to homonuclear spin decoupling in solids.
\newblock {\em J. Chem. Phys.}, 93:7543--7560, 1990.

\bibitem{Liu2011}
H.~Liu and G.~B. Matson.
\newblock Radiofrequency pulse designs for three-dimensional {MRI} providing
  uniform tipping in inhomogeneous $b_1$ fields.
\newblock {\em Magn. Reson. Med.}, 66:1254--1266, 2011.

\bibitem{Lloyd2014}
S.~Lloyd and S.~Montangero.
\newblock Information theoretical analysis of quantum optimal control.
\newblock {\em Phys. Rev. Lett.}, 113:010502, 2014.

\bibitem{Lloyd2001}
S.~Lloyd and L.~Viola.
\newblock Engineering quantum dynamics.
\newblock {\em Phys. Rev. A}, 65:010101, 2001.

\bibitem{Lloyd2000}
Seth Lloyd.
\newblock Coherent quantum feedback.
\newblock {\em Phys. Rev. A}, 62:022108, jul 2000.

\bibitem{Loening2012}
N.~M. Loening, B.-J. van Rossum, and H.~Oschkinat.
\newblock Broadband excitation pulses for high-field solid-state nuclear
  magnetic resonance spectroscopy.
\newblock {\em Magn. Reson. Chem.}, 50:284--288, 2012.

\bibitem{Lucero2010a}
Erik Lucero, Julian Kelly, Radoslaw~C. Bialczak, Mike Lenander, Matteo
  Mariantoni, Matthew Neeley, A.~D. O'Connell, Daniel Sank, H.~Wang, Martin
  Weides, James Wenner, Tsuyoshi Yamamoto, A.~N. Cleland, and John~M. Martinis.
\newblock Reduced phase error through optimized control of a superconducting
  qubit.
\newblock {\em Phys. Rev. A}, 82:042339, 2010.

\bibitem{Lapert2015}
E.~Ass\'emat M.~Lapert, S.~J. Glaser, and D.~Sugny.
\newblock Optimal control of the signal-to-noise ratio per unit time of a spin
  1/2 particle: The crusher gradient and the radiation damping cases.
\newblock {\em J. Chem. Phys.}, 142:044202/1--9, 2015.

\bibitem{Mabuchi2008}
Hideo Mabuchi.
\newblock Coherent-feedback quantum control with a dynamic compensator.
\newblock {\em Phys. Rev. A}, 78:032323, 2008.

\bibitem{MacGregor2011}
A.~W. MacGregor, L.A. O'Dell, and R.W. Schurko.
\newblock New methods for the acquisition of ultra-wideline solid-state nmr
  spectra of spin-1/2 nuclides.
\newblock {\em J. Magn. Reson.}, 208:103--113, 2011.

\bibitem{Machnes2011}
S.~Machnes, U.~Sander, S.~J. Glaser, P.~de~Fouqui\`eres, A.~Gruslys,
  S.~Schirmer, and T.~Schulte-Herbr\"uggen.
\newblock Comparing, optimizing, and benchmarking quantum-control algorithms in
  a unifying programming framework.
\newblock {\em Phys. Rev. A}, 84:022305, 2011.
\newblock software avaibable at qlib.info.

\bibitem{Machnes2015}
S.~Machnes, D.~J. Tannor, F.~K. Wilhelm, and E.~Ass{\'e}mat.
\newblock Gradient optimization of analytic controls: the route to high
  accuracy quantum optimal control.
\newblock arXiv:1507.04261, 2015.

\bibitem{Maday2003}
Yvon Maday and Gabriel Turinici.
\newblock New formulations of monotonically convergent quantum control
  algorithms.
\newblock {\em J. Chem. Phys.}, 118:8191--8196, 2003.

\bibitem{Mandal2014}
S.~Mandal, T.~W. Borneman, V.~D.~M. Koroleva, and M.~D. H{\"u}rlimann.
\newblock Direct optimization of signal-to-noise ratio of cpmg-like sequences
  in inhomogeneous fields.
\newblock {\em J. Magn. Reson.}, 247:54--66, 2014.

\bibitem{Mansfield2004}
Peter Mansfield.
\newblock Snapshot magnetic resonance imaging (nobel lecture).
\newblock {\em Angewandte Chemie International Edition}, 43(41):5456--5464,
  2004.

\bibitem{Martin2015}
L.~Martin, F.~Motzoi, H.~Li, M.~Sarovar, and B.~Whaley.
\newblock Deterministic generation of remote entanglement with active quantum
  feedback.
\newblock {http://arxiv.org/abs/1506.07888}, 2015.

\bibitem{Marvet1995}
Una Marvet and Marcos Dantus.
\newblock Femtosecond photoassociation spectroscopy: coherent bond formation.
\newblock {\em Chem. Phys. Lett.}, 245(4-5):393--399, 1995.

\bibitem{Marx2010}
R.~Marx, A.~F. Fahmy, L.~Kauffman, S.~Lomonaco, A.~Sp{\"o}rl, N.~Pomplun,
  T.~Schulte-Herbr{\"u}ggen, J.~Myers, and S.J. Glaser.
\newblock Nuclear magnetic resonance quantum calculations of the jones
  polynomial.
\newblock {\em Phys. Rev. A}, 81:032319, 2010.

\bibitem{Massire2013}
A.~Massire, M.~A. Cloos, A.~Vignaud, D.~Le Bihan, A.~Amadon, and N.~Boulant.
\newblock Design of non-selective refocusing pulses with phase-free rotation
  axis by gradient ascent pulse engineering algorithm in parallel transmission
  at 7 t.
\newblock {\em J. Magn. Reson.}, 230:76--83, 2013.

\bibitem{Matthews2011}
Thomas~E. Matthews, Ivan~R. Piletic, M.~Angelica Selim, Mary~Jane Simpson, and
  Warren~S. Warren.
\newblock Pump-probe imaging differentiates melanoma from melanocytic nevi.
\newblock {\em Science Translational Medicine}, 3(71):71ra15, 2011.

\bibitem{Maune2012}
B.~M. Maune, M.~G. Borselli, B.~Huang, T.~D. Ladd, P.~W. Deelman, K.~S.
  Holabird, and A.~T. Hunter.
\newblock Coherent singlet-triplet oscillations in a silicon-based double
  quantum dot.
\newblock {\em Nature}, 481:344--347, 2012.

\bibitem{Maximov2008}
I.~Maximov, Z.~Tosner, and N.~C. Nielsen.
\newblock Optimal control design of {NMR} and dynamic nuclear polarization
  experiments using monotonically convergent algorithms.
\newblock {\em J. Chem. Phys.}, 128:184505, 2008.

\bibitem{Meister2014}
S.~Meister, J.~T. Stockburger, R.~Schmidt, and J.~Ankerhold.
\newblock Optimal control theory with arbitrary superpositions of waveforms.
\newblock {\em J. Phys. A}, 47:495002, 2014.

\bibitem{Merli2009}
Andrea Merli, Frauke Eimer, Fabian Weise, Albrecht Lindinger, Wenzel Salzmann,
  Terry Mullins, Simone G\"{o}tz, Roland Wester, Matthias Weidem\"{u}ller,
  Ruzin A\ifmmode \breve{g}\else \u{g}\fi{}ano\ifmmode~\breve{g}\else
  \u{g}\fi{}lu, and Christiane~P. Koch.
\newblock Photoassociation and coherent transient dynamics in the interaction
  of ultracold rubidium atoms with shaped femtosecond pulses. ii. theory.
\newblock {\em Phys. Rev. A}, 80:063417, 2009.

\bibitem{Meshulach1998}
Doron Meshulach and Yaron Silberberg.
\newblock Coherent quantum control of two-photon transitions by a femtosecond
  laser pulse.
\newblock {\em Nature}, 396:239--242, 1998.

\bibitem{Mirrahimi2004}
M.~Mirrahimi and P.~Rouchon.
\newblock Controllability of quantum harmonic oscillators.
\newblock {\em IEEE Trans. Automat. Control}, 49(5):745--747, may 2004.

\bibitem{Monmayrant2010}
Antoine Monmayrant, S\'{e}bastien Weber, and B\'{e}atrice chatel.
\newblock A newcomer's guide to ultrashort pulse shaping and characterization.
\newblock {\em J. Phys. B: At., Mol. Opt. Phys.}, 43(10):103001, may 2010.

\bibitem{Montangero2007}
Simone Montangero, Tommaso Calarco, and Rosario Fazio.
\newblock Robust optimal quantum gates for josephson charge qubits.
\newblock {\em Phys. Rev. Lett.}, 99:170501, 2007.

\bibitem{Mottonen2006}
M.~M\"ott\"onen, R.~de~Sousa, J.~Zhang, and K.~B. Whaley.
\newblock High-fidelity one-qubit operations under random telegraph noise.
\newblock {\em Phys. Rev. A}, 73:022332, 2006.

\bibitem{Motzoi2011}
F.~Motzoi, J.~M. Gambetta, S.~T. Merkel, and F.~K. Wilhelm.
\newblock Optimal control methods for rapidly time-varying {H}amiltonians.
\newblock {\em Phys. Rev. A}, 84:022307, 2011.

\bibitem{Motzoi2009}
F.~Motzoi, J.~M. Gambetta, P.~Rebentrost, and F.~K. Wilhelm.
\newblock Simple pulses for elimination of leakage in weakly nonlinear qubits.
\newblock {\em Phys. Rev. Lett.}, 103:110509, 2009.

\bibitem{Muhonen2015}
J.~T. Muhonen, A.~Laucht, S.~Simmons, J.~P. Dehollain, R.~Kalra, F.~E. Hudson,
  S.~Freer, K.~M. Itoh, D.~N. Jamieson, J.~C. Mc{C}allum, A.~S. Dzurak, and
  A.~Morello.
\newblock Quantifying the quantum gate fidelity of single-atom spin qubits in
  silicon by randomized benchmarking.
\newblock {\em J. Phys: Condens. Matter}, 27(15):154205, 2015.

\bibitem{Mukherjee2013}
V.~Mukherjee, A.~Carlini, A.~Mari, T.~Caneva, Montangerom S., T.~Carlarco,
  R.~Fazio, and V.~Giovannetti.
\newblock Speeding up and slowing down the relaxation of a qubit by optimal
  control.
\newblock {\em Phys. Rev. A}, 88:062326, 2013.

\bibitem{Muller2014}
C.~M{\"u}ller, X.~Kong, J.-M. Cai, K.~Melentijevi{\'c}, A.~Stacey, M.~Markham,
  D.~Twitchen, J.~Isoya, S.~Pezzagna, J.~Meijer, J.~F. Du, M.~B. Plenio,
  B.~Naydenov, L.~P. McGuinness, and F.~Jelezko.
\newblock Nuclear magnetic resonance spectroscopy with single spin sensitivity.
\newblock {\em Nat. Commun.}, 5:4703, 2014.

\bibitem{Muller2011}
M.~M. M\"uller, D.~M. Reich, M.~Murphy, H.~Yuan, J.~Vala, K.~B. Whaley,
  T.~Calarco, and C.~P. Koch.
\newblock Optimizing entangling quantum gates for physical systems.
\newblock {\em Phys. Rev. A}, 84:042315, 2011.

\bibitem{Murdoch1987}
J.~B. Murdoch, A.~H. Lent, and M.~R. Kritzer.
\newblock Computer-optimized narrowband pulses for multislice imaging.
\newblock {\em J. Magn. Reson.}, 74:226--263, 1987.

\bibitem{Murphy2010}
Michael Murphy, Simone Montangero, Vittorio Giovannetti, and Tommaso Calarco.
\newblock Communication at the quantum speed limit along a spin chain.
\newblock {\em Phys. Rev. A}, 82:022318, 2010.

\bibitem{Khaneja2004}
Navin N.~Khaneja, Jr-Shin J.-S.~Li, Cindie Kehlet, Burkhard Luy, and Steffen~J.
  Glaser.
\newblock Broadband relaxation-optimized polarization transfer in magnetic
  resonance.
\newblock {\em Proc. Natl. Acad. Sci. USA}, 101(41):14742--14747, 2004.

\bibitem{Neves2009}
J.~L. Neves, B.~Heitmann, N.~Khaneja, and S.~J. Glaser.
\newblock Heteronuclear decoupling by optimal tracking.
\newblock {\em J. Magn. Reson.}, 201:7--17, 2009.

\bibitem{Ni2008}
K.~K. Ni, S.~Ospelkaus, M.~H.~G. de~Miranda, A.~Pe'er, B.~Neyenhuis, J.~J.
  Zirbel, S.~Kotochigova, P.~S. Julienne, D.~S. Jin, and J.~Ye.
\newblock A high phase-space-density gas of polar molecules.
\newblock {\em Science}, 322(5899):231--235, 2008.

\bibitem{Ni2013}
Qing~Zhe Ni, Daviso E, Can TV, Markhasin E, Jawla SK, Swager TM, Temkin RJ,
  Herzfeld J, and Griffin RG.
\newblock High frequency dynamic nuclear polarization.
\newblock {\em Accounts in Chemical Research}, 46:1933--1941, 2013.

\bibitem{Nimbalkar2013}
M.~Nimbalkar, B.~Luy, T.~E. Skinner, J.~L. Neves, N.~I. Gershenzon, K.~Kobzar,
  W.~Bermel, and S.~J. Glaser.
\newblock The fantastic four: A plug 'n' play set of optimal control pulses for
  enhancing nmr spectroscopy.
\newblock {\em J. Magn. Reson.}, 228:16--31, 2013.

\bibitem{Nimbalkar2012}
M.~Nimbalkar, R.~Zeier, J.~L. Neves, H.~Elavarasi, S. B.~Yuan, N.~Khaneja,
  K.~Dorai, and S.~J. Glaser.
\newblock Multiple-spin coherence transfer in linear ising spin chains and
  beyond: Numerically optimized pulses and experiments.
\newblock {\em Phys. Rev. A}, 85:012325, 2012.

\bibitem{Nuernberger2007}
Patrick Nuernberger, Gerhard Vogt, Tobias Brixner, and Gustav Gerber.
\newblock Femtosecond quantum control of molecular dynamics in the condensed
  phase.
\newblock {\em Phys. Chem. Chem. Phys.}, 9:2470--2497, 2007.

\bibitem{Nuernberger2010}
Patrick Nuernberger, Daniel Wolpert, Horst Weiss, and Gustav Gerber.
\newblock Femtosecond quantum control of molecular bond formation.
\newblock {\em Proc. Natl. Acad. Sci. U.S.A}, 107(23):10366--10370, 2010.

\bibitem{Ogilvie2006}
Jennifer~P. Ogilvie, Delphine D\'{e}barre, Xavier Solinas, Jean-Louis Martin,
  Emmanuel Beaurepaire, and Manuel Joffre.
\newblock Use of coherent control for selective two-photon fluorescence
  microscopy in live organisms.
\newblock {\em Opt. Express}, 14(2):759--766, 2006.

\bibitem{OMeara2012}
C.~O'{M}eara, G.~Dirr, and T.~Schulte-Herbr{\"u}ggen.
\newblock Illustrating the geometry of coherently controlled unital open
  quantum systems.
\newblock {\em IEEE Trans. Autom. Contr. (IEEE-TAC)}, 57:2050--2054, 2012.
\newblock (an extended version can be found as arXiv:1103.2703).

\bibitem{Oron2005}
Dan Oron, Eran Tal, and Yaron Silberberg.
\newblock Scanningless depth-resolved microscopy.
\newblock {\em Opt. Express}, 13(5):1468--1476, 2005.

\bibitem{Overhauser1953}
A.W. Overhauser.
\newblock Polarization of nuclei in metals.
\newblock {\em Phys. Rev.}, 92:411--415, 1953.

\bibitem{Palao2002}
Jos\'e~P. Palao and Ronnie Kosloff.
\newblock Quantum computing by an optimal control algorithm for unitary
  transformations.
\newblock {\em Phys. Rev. Lett.}, 89:188301, 2002.

\bibitem{Palao2003}
Jos\'e~P. Palao and Ronnie Kosloff.
\newblock Optimal control theory for unitary transformations.
\newblock {\em Phys. Rev. A}, 68:062308, 2003.

\bibitem{Palao2013}
Jos\'e~P. Palao, Daniel~M. Reich, and Christiane~P. Koch.
\newblock Steering the optimization pathway in the control landscape using
  constraints.
\newblock {\em Phys. Rev. A}, 88:053409, 2013.

\bibitem{parthasarathy1992}
K.~R. Parthasarathy.
\newblock {\em An introduction to quantum stochastic calculus}.
\newblock Birkhauser, Basel, 1992.

\bibitem{Passante2009}
G.~Passante, O.~Moussa, C.~Ryan, and R.~Laflamme.
\newblock Experimental approximation of the jones polynomial by one quantum
  bit.
\newblock {\em Phys. Rev. Lett.}, 103:250501, 2009.

\bibitem{Pastirk2003}
Igor Pastirk, Johanna Dela~Cruz, Katherine Walowicz, Vadim Lozovoy, and Marcos
  Dantus.
\newblock Selective two-photon microscopy with shaped femtosecond pulses.
\newblock {\em Opt. Express}, 11(14):1695--1701, 2003.

\bibitem{Pechen2014}
A.~Pechen and H.~Rabitz.
\newblock Incoherent control of open quantum systems.
\newblock {\em Journal of Mathematical Sciences}, 199(6):695--701, jun 2014.

\bibitem{Pechen2011}
A.~N. Pechen and D.~J. Tannor.
\newblock Are there traps in quantum control landscapes?
\newblock {\em Phys. Rev. Lett.}, 106:120402, 2011.

\bibitem{Pechen2012}
A.~N. Pechen and D.~J. Tannor.
\newblock Comment on "are there traps in quantum control landscapes?": Reply.
\newblock {\em Phys. Rev. Lett.}, 108:198902, 2012.

\bibitem{Peirce1987}
A.~Peirce, M.~Dahleh, and H.~Rabitz.
\newblock Optimal control of quantum mechanical systems: Existence, numerical
  approximations and applications.
\newblock {\em Phys. Rev. A}, 37:4950--4962, 1987.

\bibitem{Petta2005}
J.~R. Petta, A.~C. Johnson, J.~M. Taylor, E.~A. Laird, A.~Yacoby, M.~D. Lukin,
  C.~M. Marcus, and A.~C. Gossard.
\newblock Coherent manipulation of coupled electron spins in semiconductor
  quantum dots.
\newblock {\em Science}, 309(5744):2180--2184, 2005.

\bibitem{Pfeiffer2013}
Walter Pfeiffer, Martin Aeschlimann, and Tobias Brixner.
\newblock {\em Optical Antennas}, chapter Coherent control of nano-optical
  excitations, pages 135--156.
\newblock Cambridge University Press, 2013.
\newblock Cambridge Books Online.

\bibitem{Pillai2009}
Rajesh~S. Pillai, Caroline Boudoux, Guillaume Labroille, Nicolas Olivier,
  Israel Veilleux, Emmanuel Farge, Manuel Joffre, and Emmanuel Beaurepaire.
\newblock Multiplexed two-photon microscopy ofdynamic biological samples with
  shapedbroadband pulses.
\newblock {\em Opt. Express}, 17(15):12741--12752, 2009.

\bibitem{Pla2012}
J.~J. Pla, K.~Y. Tan, J.~P. Dehollain, W.~H. Lim, J.~J. Morton, D.~N. Jamieson,
  and A.~Morello.
\newblock A single-atom electron spin qubit in silicon.
\newblock {\em Nature}, 489:541--545, 2012.

\bibitem{Pla2013}
J.~J. Pla, K.~Y. Tan, J.~P. Dehollain, W.~H. Lim, J.~J.~L. Morton, F.~A.
  Zwanenburg, D.~N. Jamieson, A.~S. Dzurak, and A.~Morello.
\newblock High-fidelity readout and control of a nuclear spin qubit in silicon.
\newblock {\em Nature}, 496:334--338, 2013.

\bibitem{Platzer2010}
Felix Platzer, Florian Mintert, and Andreas Buchleitner.
\newblock Optimal dynamical control of many-body entanglement.
\newblock {\em Phys. Rev. Lett.}, 105(2):020501, 2010.

\bibitem{Pomplun2010}
N.~Pomplun and S.~J. Glaser.
\newblock Exploring the limits of electron-nuclear polarization transfer
  efficiency in three-spin systems.
\newblock {\em Phys. Chem. Chem. Phys.}, 12:5791--5798, 2010.

\bibitem{Pomplun2008}
N.~Pomplun, B.~Heitmann, N.~Khaneja, and S.~J. Glaser.
\newblock Optimization of electron--nuclear polarization transfer.
\newblock {\em Applied Magnetic Resonance}, 34(3--4):331--346, 2008.

\bibitem{Pontryagin1964}
L.~S. Pontryagin, V.~G. Bol'tanskii, R.~S. Gamkrelidze, and E.~F. Mischenko.
\newblock {\em The Mathematical Theory of Optimal Processes}.
\newblock Pergamon Press, New York, 1964.

\bibitem{Pontryagin1962a}
L.~S. Pontryagin, V.~G. Boltyanskii, R.~V. Gamkrelidze, and E.~F. Mishchenko.
\newblock {\em The Mathematical Theory of Optimal Processes}.
\newblock Wiley, New York, 1962.

\bibitem{Poulsen2010}
Uffe~V. Poulsen, Shlomo Sklarz, David Tannor, and Tommaso Calarco.
\newblock Correcting errors in a quantum gate with pushed ions via optimal
  control.
\newblock {\em Phys. Rev. A}, 82:012339, jul 2010.

\bibitem{Preskill2012}
J.~Preskill.
\newblock Quantum computing and the entanglement frontier.
\newblock {\em arXiv:1203.5813}, 2012.
\newblock Rapporteur talk at the 25th Solvay Conference on Physics ("The Theory
  of the Quantum World"), 19-22 October 2011.

\bibitem{Prokhorenko2006}
Valentyn~I Prokhorenko, Andrea~M Nagy, Stephen~A Waschuk, Leonid~S Brown,
  Robert~R Birge, and RJ~Dwayne Miller.
\newblock Coherent control of retinal isomerization in bacteriorhodopsin.
\newblock {\em Science}, 313(5791):1257--1261, 2006.

\bibitem{Rabitz2000}
H.~Rabitz, R.~de~Vivie-Riedle, M.~Motzkus, and K.~Kompa.
\newblock Whither the future of controlling quantum phenomena?
\newblock {\em Science}, 288:824, may 2000.

\bibitem{Rabitz2012}
H.~Rabitz, T.~S. Ho, R.~Long, R.~Wu, and C.~Brif.
\newblock Comment on: Are there traps in quantum control landscapes?
\newblock {\em Phys. Rev. Lett.}, 108:198901, 2012.

\bibitem{Seideman2005}
S.~Ramakrishna and T.~Seideman.
\newblock Intense laser alignment in dissipative media as a route to solvent
  dynamics.
\newblock {\em Phys. Rev. Lett.}, 95:113001, 2005.

\bibitem{Rancan2015}
G.~Rancan, T.~T. Nguyen, and S.~J. Glaser.
\newblock Gradient ascent pulse engineering for rapid exchange saturation
  transfer.
\newblock {\em J. Magn. Reson.}, 252:1--9, 2015.

\bibitem{Rebentrost2009}
P.~Rebentrost, I.~Serban, T.~Schulte-Herbr\"uggen, and F.~K. Wilhelm.
\newblock Optimal control of a qubit coupled to a non-{M}arkovian environment.
\newblock {\em Phys. Rev. Lett.}, 102:090401, 2009.

\bibitem{Rehbinder2014}
Jean Rehbinder, Lukas Br\"{u}ckner, Alexander Wipfler, Tiago Buckup, and Marcus
  Motzkus.
\newblock Multimodal nonlinear optical microscopy with shaped 10 fs pulses.
\newblock {\em Opt. Express}, 22(23):28790--28797, 2014.

\bibitem{Reich2012}
Daniel Reich, Mamadou Ndong, and Christiane~P. Koch.
\newblock Monotonically convergent optimization in quantum control using
  {K}rotov's method.
\newblock {\em J. Chem. Phys.}, 136:104103, 2012.

\bibitem{Reich2015}
Daniel~M. Reich, Nadav Katz, and Christiane~P. Koch.
\newblock Exploiting non-markovianity for quantum control.
\newblock {\em Sci. Rep.}, 5:12430, 2015.

\bibitem{Reich2013}
Daniel~M. Reich and Christiane~P. Koch.
\newblock Cooling molecular vibrations with shaped laser pulses: Optimal
  control theory exploiting the timescale separation between coherent
  excitation and spontaneous emission.
\newblock {\em New J. Phys.}, 15(12):125028, 2013.
\newblock arXiv:1308.0803.

\bibitem{Reich2014}
Daniel~M. Reich, Jos\'{e}~P. Palao, and Christiane~P. Koch.
\newblock Optimal control under spectral constraints: Enforcing multi-photon
  absorption pathways.
\newblock {\em J. Mod. Opt.}, 61(10):822, 2014.

\bibitem{Reiner2004}
J.~E. Reiner, W.~P. Smith, L.~A. Orozco, H.~M. Wiseman, and J.~Gambetta.
\newblock Quantum feedback in a weakly driven cavity qed system.
\newblock {\em Phys. Rev. A}, 70:023819, 2004.

\bibitem{Reinsperger2014}
Tony Reinsperger and Burkhard Luy.
\newblock Homonuclear bird-decoupled spectra for measuring one-bond couplings
  with highest resolution: Clip/clap-reset and constant-time-clip/clap-reset.
\newblock {\em J. Magn. Reson.}, 239:110--120, 2014.

\bibitem{Reiss2002}
T.~O. Reiss, N.~Khaneja, and S.~J. Glaser.
\newblock Time-optimal coherence-order-selective transfer of in-phase coherence
  in heteronuclear is spin systems.
\newblock {\em J. Magn. Reson.}, 154(2):192--195, 2002.

\bibitem{Reiss2003}
T.~O. Reiss, N.~Khaneja, and S.~J. Glaser.
\newblock Broadband geodesic pulses for three spin systems: time-optimal
  realization of effective trilinear coupling terms and indirect {SWAP} gates.
\newblock {\em J. Magn. Reson.}, 165(1):95--101, 2003.

\bibitem{Rice2000a}
S.~Rice and M.~Zhao.
\newblock {\em Optimal control of quantum dynamics}.
\newblock Wiley, New York, 2000.

\bibitem{Rice2000}
S.~A. Rice and M.~Zhao.
\newblock {\em Optical control of molecular dynamics}.
\newblock John Wiley \& Sons, 2000.

\bibitem{Riste2012}
D.~Riste, C.~C. Bulting, K.~W. Lehnert, and L.~DiCarlo.
\newblock Feedback control of a solid-state qubit using high-fidelity
  projective measurement.
\newblock {\em Phys. Rev. Lett.}, 109:240502, 2012.

\bibitem{Riviello2015}
G.~Riviello, K.~Moore~Tibbetts, C.~Brif, R.~Long, R.-B. Wu, T.-S. Ho, and
  H.~Rabitz.
\newblock Searching for quantum optimal controls under severe constraints.
\newblock {\em Phys. Rev. A}, 91:043401, 2015.

\bibitem{Rojan2014}
Katharina Rojan, Daniel~M. Reich, Igor Dotsenko, Jean-Michel Raimond,
  Christiane~P. Koch, and Giovanna Morigi.
\newblock Arbitrary-quantum-state preparation of a harmonic oscillator
  viaoptimal control.
\newblock {\em Phys. Rev. A}, 90:023824, 2014.
\newblock arXiv:1406.6572.

\bibitem{Rosi2013}
S.~Rosi, A.~Bernard, N.~Fabbri, L.~Fallani, C.~Fort, M.~Inguscio, T.~Calarco,
  and S.~Montangero.
\newblock Fast closed-loop optimal control of ultracold atoms in an optical
  lattice.
\newblock {\em Phys. Rev. A}, 88:021601, aug 2013.

\bibitem{Rugango2015}
Rene Rugango, James~E Goeders, Thomas~H Dixon, John~M Gray, NB~Khanyile, Gang
  Shu, Robert~J Clark, and Kenneth~R Brown.
\newblock Sympathetic cooling of molecular ion motion to the ground state.
\newblock {\em New J. Phys.}, 17(3):035009, 2015.

\bibitem{Ruschhaupt2012}
A.~Ruschhaupt, Xi~Chen, D.~Alonso, and J.~G. Muga.
\newblock Optimally robust shortcuts to population inversion in two-level
  quantum systems.
\newblock {\em New J. Phys.}, 14(9):093040, 2012.

\bibitem{Ryan2008}
C.~A. Ryan, C.~Negrevergne, M.~Laforest, E.~Knill, and R.~Laflamme.
\newblock Liquid-state nuclear magnetic resonance as a testbed for developing
  quantum control methods.
\newblock {\em Phys. Rev. A}, 78:012328, 2008.

\bibitem{Rybak2011}
Leonid Rybak, Saieswari Amaran, Liat Levin, Micha\l{} Tomza, Robert Moszynski,
  Ronnie Kosloff, Christiane~P. Koch, and Zohar Amitay.
\newblock Generating molecular rovibrational coherence by two-photon
  femtosecond photoassociation of thermally hot atoms.
\newblock {\em Phys. Rev. Lett.}, 107:273001, 2011.

\bibitem{Rybar2012}
T.~Ryb{\'a}r, S.N. Filippov, M.~Ziman, and V.~Buzek.
\newblock Simulation of indivisible qubit channels in collision models.
\newblock {\em J. Phys. B: At. Mol. Opt. Phys.}, 45:154006, 2012.

\bibitem{Smith1994}
Smith SA, Levante TO, Meier BH, and Ernst RR.
\newblock Computer simulations in magnetic resonance. an object-oriented
  programming approach.
\newblock {\em J. Magn. Reson.}, 106:75--105, 1994.

\bibitem{Salomon2005}
J.~Salomon, C.~Dion, and G.~Turinici.
\newblock Optimal molecular alignment and orientation through rotational ladder
  climbing.
\newblock {\em J. Chem. Phys.}, 123:144310, 2005.

\bibitem{Salzmann2008}
W.~Salzmann, T.~Mullins, J.~Eng, M.~Albert, R.~Wester, M.~Weidem\"{u}ller,
  A.~Merli, S.~M. Weber, F.~Sauer, M.~Plewicki, F.~Weise, L.~W\"{o}ste, and
  A.~Lindinger.
\newblock Coherent transients in the femtosecond photoassociation of ultracold
  molecules.
\newblock {\em Phys. Rev. Lett.}, 100:233003, jun 2008.

\bibitem{Sarandy2005}
M.~S. Sarandy and D.~A. Lidar.
\newblock Adiabatic quantum computation in open systems.
\newblock {\em Phys. Rev. Lett.}, 95:250503, 2005.

\bibitem{Savostyanov2014}
D.~V. Savostyanov, S.~V. Dolgov, J.~M. Werner, and Ilya Kuprov.
\newblock Exact {NMR} simulation of protein-size spin systems using tensor
  train formalism.
\newblock {\em Phys. Rev. B}, 90:085139, 2014.

\bibitem{Sayrin2011}
Cl\'{e}ment Sayrin, Igor Dotsenko, Xingxing Zhou, Bruno Peaudecerf, Th\'{e}o
  Rybarczyk, S\'{e}bastien Gleyzes, Pierre Rouchon, Mazyar Mirrahimi, Hadis
  Amini, Michel Brune, Jean-Michel Raimond, and Serge Haroche.
\newblock Real-time quantum feedback prepares and stabilizes photon number
  states.
\newblock {\em Nature}, 477:73--77, sep 2011.

\bibitem{Duckett2012}
Duckett S.B. and Mewis R.E.
\newblock Application of parahydrogen induced polarization techniques in nmr
  spectroscopy and imaging.
\newblock {\em Acc. Chem. Res.}, 45:1247--1257, 2012.

\bibitem{Schaefer2012}
Ido Schaefer and Ronnie Kosloff.
\newblock Optimal-control theory of harmonic generation.
\newblock {\em Phys. Rev. A}, 86:063417, 2012.

\bibitem{Scheuer2014}
Jochen Scheuer, Xi~Kong, Ressa~S. Said, Jeson Chen, Andrea Kurz, Luca
  Marseglia, Jiangfeng Du, Philip~R. Hemmer, Simone Montangero, Tommaso
  Calarco, Boris Naydenov, and Fedor Jelezko.
\newblock Precise qubit control beyond the rotating wave approximation.
\newblock {\em New J. Phys.}, 16(9):093022, 2014.

\bibitem{Schilling2012}
F.~Schilling and S.~J. Glaser.
\newblock Tailored real-time scaling of heteronuclear couplings.
\newblock {\em J. Magn. Reson.}, 223:207--218, 2012.

\bibitem{Schilling2014}
F.~Schilling, N.~I. Warner, T.~E. Gershenzon, M.~Sattler, and S.~J. Glaser.
\newblock Next-generation heteronuclear decoupling for high-field biomolecular
  {NMR} spectroscopy.
\newblock {\em Angew. Chem. Int. Ed.}, 53:4475--4479, 2014.

\bibitem{Schirmer2001}
S.~G. Schirmer, H.~Fu, and A.~I. Solomon.
\newblock Complete controllability of quantum systems.
\newblock {\em Phys. Rev. A}, 63:063410, may 2001.

\bibitem{Schirmer2009}
S.~G. Schirmer and P.~J. Pemberton-Ross.
\newblock Fast, high-fidelity information transmission through spin-chain
  quantum wires.
\newblock {\em Phys. Rev. A}, 80:030301, 2009.

\bibitem{Schirmer2010}
S.~G. Schirmer and X.~Wang.
\newblock Stabilizing open quantum systems by markovian reservoir engineering.
\newblock {\em Phys. Rev. A}, 81:062306, 2010.

\bibitem{Schmidt2011}
R.~Schmidt, A.~Negretti, J.~Ankerhold, T.~Calarco, and J.T. Stockburger.
\newblock Optimal control of open quantum systems: Cooperative effects of
  driving and dissipation.
\newblock {\em Phys. Rev. Lett.}, 107:130404, 2011.

\bibitem{Schroder2009}
Markus Schr\"oder and Alex Brown.
\newblock Generalized filtering of laser fields in optimal control theory:
  application to symmetry filtering of quantum gate operations.
\newblock {\em New J. Phys.}, 11:105031, 2009.

\bibitem{SchulteHerbruggen2012}
T.~Schulte-Herbr{\"u}ggen, R.~Marx, A.~Fahmy, L.~Kauffman, S.~Lomonaco,
  N.~Khaneja, , and S.J. Glaser.
\newblock Control aspects of quantum computing using pure and mixed states.
\newblock {\em Phil. Trans. R. Soc. A}, 370:4651, 2012.

\bibitem{SchulteHerbruggen2011}
T.~Schulte-{H}erbr\"{u}ggen, A.~Sp\"{o}rl, N.~Khaneja, and S.~J. Glaser.
\newblock Optimal control for generating quantum gates in open dissipative
  systems.
\newblock {\em J. Phys. B: At., Mol. Opt. Phys.}, 44:154013, aug 2011.

\bibitem{SchulteHerbruggen2005}
T.~Schulte-Herbr{\"u}ggen, A.~K. Sp{\"o}rl, N.~Khaneja, and S.~J. Glaser.
\newblock Optimal control-based efficient synthesis of building blocks of
  quantum algorithms: A perspective from network complexity towards time
  complexity.
\newblock {\em Phys. Rev. A}, 72:042331, 2005.

\bibitem{SchulteHerbruggen2008}
T.~Schulte-Herbr{\"u}ggen, A.~K. Sp{\"o}rl, K.~Waldherr, T.~Gradl, S.~J.
  Glaser, and T.~Huckle.
\newblock {\em in: High-Performance Computing in Science and Engineering,
  Garching 2007}, chapter 'Using the {HLRB} Cluster as Quantum {CISC}
  Compiler', pages 517--533.
\newblock Springer, Berlin, 2008.
\newblock (an extended version can be found under:
  http://arXiv.org/pdf/0712.3227).

\bibitem{SchulzeSuenninghausen2014}
David Schulze-{S}uenninghausen, Johanna Becker, and Burkhard Luy.
\newblock Rapid heteronuclear single quantum correlation nmr spectra at natural
  abundance.
\newblock {\em J. Am. Chem. Soc.}, 136(4):1242--1245, 2014.

\bibitem{Schutjens2013}
R.~Schutjens, F.~Abu Dagga, D.~J. Egger, and F.~K. Wilhelm.
\newblock Single-qubit gates in frequency-crowded transmon systems.
\newblock {\em Phys. Rev. A}, 88:052330, 2013.

\bibitem{Schweiger2001}
A.~Schweiger and G.~Jeschke.
\newblock {\em Principles of pulse electron paramagnetic resonance}.
\newblock Oxford University Press, 2001.

\bibitem{Seideman2006}
T~Seideman and E.~Hamilton.
\newblock Nonadiabatic alignment by intense pulses: Concepts, theory and
  directions.
\newblock {\em Adv. At. Mol. Opt. Phys.}, 52:289, 2006.

\bibitem{Serban2005}
I.~Serban, J.~Werschnik, and E.~K.~U. Gross.
\newblock Optimal control of time-dependent targets.
\newblock {\em Phys. Rev. A}, 71:053810, may 2005.

\bibitem{Serrano2015}
Arnaldo~L Serrano, Ayanjeet Ghosh, Joshua~S Ostrander, and Martin~T Zanni.
\newblock Wide-field ftir microscopy using mid-ir pulse shaping.
\newblock {\em Opt. Express}, 23(14):17815--17827, 2015.

\bibitem{Shapiro2003}
M.~Shapiro and P.~Brumer.
\newblock {\em Principles of quantum control of mulecular processes}.
\newblock Wiley, New York, 2003.

\bibitem{Singer2010}
Kilian Singer, Ulrich Poschinger, Michael Murphy, Peter Ivanov, Frank Ziesel,
  Tommaso Calarco, and Ferdinand Schmidt-Kaler.
\newblock \textit{Colloquium} : Trapped ions as quantum bits: Essential
  numerical tools.
\newblock {\em Rev. Mod. Phys.}, 82:2609--2632, 2010.

\bibitem{Skinner2011}
T.~E. Skinner, M.~Braun, K.~Woelk, N.~I. Gershenzon, and S.~J. Glaser.
\newblock Design and application of robust rf pulses for toroid cavity nmr
  spectroscopy.
\newblock {\em J. Magn. Reson.}, 209:282--290, 2011.

\bibitem{Skinner2010}
T.~E. Skinner and N.~I. Gershenzon.
\newblock Optimal control design of pulse shapes as analytic functions.
\newblock {\em J. Magn. Reson.}, 204:248, 2010.

\bibitem{Skinner2012}
T.~E. Skinner, N.~I. Gershenzon, M.~Nimbalkar, W.~Bermel, B.~Luy, and S.~J.
  Glaser.
\newblock New strategies for designing robust universal rotation pulses:
  Application to broadband refocusing at low power.
\newblock {\em J. Magn. Reson.}, 216:78--87, 2012.

\bibitem{Skinner2012a}
T.~E. Skinner, N.~I. Gershenzon, M.~Nimbalkar, and S.~J. Glaser.
\newblock Optimal control design of band-selective excitation pulses that
  accommodate relaxation and rf inhomogeneity.
\newblock {\em J. Magn. Reson.}, 217:53--60, 2012.

\bibitem{Skinner2004}
T.~E. Skinner, T.~O. Reiss, B.~Luy, N.~Khaneja, and S.~J. Glaser.
\newblock Reducing the duration of broadband excitation pulses using optimal
  control with limited {RF} amplitude.
\newblock {\em J. Magn. Reson.}, 167:68--74, 2004.

\bibitem{Skinner2006}
TE~Skinner, K~Kobzar, B~Luy, MR~Bendall, W~Bermel, N.~Khaneja, and S.~J.
  Glaser.
\newblock Optimal control design of constant amplitude phase-modulated pulses:
  Application to calibration-free broadband excitation.
\newblock {\em J. Magn. Reson.}, 179(2):241--249, apr 2006.

\bibitem{Skinner2005}
TE~Skinner, TO~Reiss, B~Luy, N~Khaneja, and SJ~Glaser.
\newblock Tailoring the optimal control cost function to a desired output:
  application to minimizing phase errors in short broadband excitation pulses.
\newblock {\em J. Magn. Reson.}, 172(1):17--23, 2005.

\bibitem{Skinner2003}
Thomas~E. Skinner, Timo~O. Reiss, Burkhard Luy, Navin Khaneja, and Steffen~J.
  Glaser.
\newblock Application of optimal control theory to the design of broadband
  excitation pulses for high-resolution {NMR}.
\newblock {\em J. Magn. Reson.}, 163(1):8--15, 2003.

\bibitem{Sklarz2002}
Shlomo~E. Sklarz and David~J. Tannor.
\newblock Loading a {B}ose-{E}instein condensate onto an optical lattice: An
  application of optimal control theory to the nonlinear {S}chr\"{o}dinger
  equation.
\newblock {\em Phys. Rev. A}, 66(5):053619, 2002.

\bibitem{Sofikitis2009}
Dimitris Sofikitis, S{\'e}bastien Weber, Andr{\'e}a Fioretti, Ridha Horchani,
  Maria Allegrini, B{\'e}atrice Chatel, Daniel Comparat, and Pierre Pillet.
\newblock Molecular vibrational cooling by optical pumping with shaped
  femtosecond pulses.
\newblock {\em New J. Phys.}, 11(5):055037, 2009.

\bibitem{Somloi1993}
J.~Soml\'{o}i, V.~A. Kazakovski, and D.~J. Tannor.
\newblock Controlled dissociation of {I}$_2$ via optical transitions between
  the {X} and {B} electronic states.
\newblock {\em Chem. Phys.}, 172:85--98, 1993.

\bibitem{Sontag1998}
E.~Sontag.
\newblock {\em Mathematical Control Theory}.
\newblock Springer, New York, 1998.

\bibitem{Spindler2012}
Philipp~E. Spindler, Yun Zhang, Burkhard Endeward, Naum Gershernzon, Thomas~E.
  Skinner, Steffen~J. Glaser, and Thomas~F. Prisner.
\newblock Shaped optimal control pulses for increased excitation bandwidth in
  {EPR}.
\newblock {\em J. Magn. Reson.}, 218(0):49--58, may 2012.

\bibitem{Sporl2007}
A.~Sp\"orl, T.~Schulte-Herbr\"uggen, S.~J. Glaser, V.~Bergholm, M.~J. Storcz,
  J.~Ferber, and F.~K. Wilhelm.
\newblock Optimal control of coupled josephson qubits.
\newblock {\em Phys. Rev. A}, 75:012302, 2007.

\bibitem{Seideman2003}
H.~Stapelfeldt and T.~Seideman.
\newblock Aligning molecules with strong laser pulses.
\newblock {\em Rev. Mod. Phys.}, 75:543, 2003.

\bibitem{Stapelfeldt2003}
Henrik Stapelfeldt and Tamar Seideman.
\newblock \textit{Colloquium} : Aligning molecules with strong laser pulses.
\newblock {\em Rev. Mod. Phys.}, 75:543--557, apr 2003.

\bibitem{Stefanatos2004}
D.~Stefanatos, N.~Khaneja, and S.~J. Glaser.
\newblock Optimal control of coupled spins in presence of longitudinal and
  transverse relaxation.
\newblock {\em Phys. Rev. A}, 69:022319, 2004.

\bibitem{Stefanatos2005}
D.~Stefanatos, N.~Khaneja, and S.~J. Glaser.
\newblock Relaxation optimized transfer of spin order in ising chains.
\newblock {\em Phys. Rev. A}, 72:062320, 2005.

\bibitem{Steinbacher2015}
Andreas Steinbacher, Patrick Nuernberger, and Tobias Brixner.
\newblock Optical discrimination of racemic from achiral solutions.
\newblock {\em Phys. Chem. Chem. Phys.}, 17:6340--6346, 2015.

\bibitem{Sugawara2003}
M.~Sugawara.
\newblock General formulation of locally designed coherent control theory for
  quantum system.
\newblock {\em J. Chem. Phys.}, 118(15):6784, 2003.

\bibitem{Sugny2004}
D.~Sugny, A.~Keller, O.~Atabek, D.~Daems, C.~M. Dion, S.~Gu\'{e}rin, and H.~R.
  Jauslin.
\newblock Reaching optimally oriented molecular states by laser kicks.
\newblock {\em Phys. Rev. A}, 69:033402, mar 2004.

\bibitem{Sugny2007}
D.~Sugny, C.~Kontz, and H.~R. Jauslin.
\newblock Time-optimal control of a two-level dissipative quantum system.
\newblock {\em Phys. Rev. A}, 76:023419, 2007.

\bibitem{Sussman2006}
Benjamin~J. Sussman, Dave Townsend, Misha~Yu. Ivanov, and Albert Stolow.
\newblock Dynamic stark control of photochemical processes.
\newblock {\em Science}, 314(5797):278--281, 2006.

\bibitem{Sussmann1972}
H.~Sussmann and V.~Jurdjevic.
\newblock Controllability of nonlinear systems.
\newblock {\em Journal of Differential Equations}, 12:95--116, 1972.

\bibitem{Tannor1999}
D.~J. Tannor and A.~Bartana.
\newblock On the interplay of control fields and spontaneous emission in laser
  cooling.
\newblock {\em J. Phys. Chem. A}, 103:10359--10363, 1999.

\bibitem{Tannor2007}
David~J. Tannor.
\newblock {\em Introduction to Quantum Mechanics: A Time-Dependent
  Perspective}.
\newblock University Science Books, Sausalito, Sausalito, 2007.

\bibitem{Tannor1986}
David~J. Tannor, Ronnie Kosloff, and Stuart~A. Rice.
\newblock Coherent pulse sequence induced control of selectivity of reactions:
  Exact quantum mechanical calculations.
\newblock {\em J. Chem. Phys.}, 85(10):5805--5820, 1986.

\bibitem{Tannor1985}
David~J. Tannor and Stuart~A. Rice.
\newblock Control of selectivity of chemical reaction via control of wave
  packet evolution.
\newblock {\em J. Chem. Phys.}, 83(10):5013--5018, aug 1985.

\bibitem{Tannor1992}
D.J. Tannor, V~Kazakov, and V.~Orlov.
\newblock Control of photochemical branching: Novel procedures for finding
  optimal pulses and global upper bounds.
\newblock In J.~Broeckhove and L.~Lathouwers, editors, {\em Time-dependent
  quantum molecular dynamics}, pages 347--360. Plenum, 1992.

\bibitem{Tesch2002}
Carmen~M. Tesch and Regina de~Vivie-Riedle.
\newblock Quantum computation with vibrationally excited molecules.
\newblock {\em Phys. Rev. Lett.}, 89:157901, sep 2002.

\bibitem{Teufel2003}
S.~Teufel.
\newblock {\em Adiabatic perturbation theory in quantum dynamics}.
\newblock Springer-Verlag, Berlin, 2003.

\bibitem{Moore2015}
K.~Moore Tibbetts and H.~Rabitz.
\newblock Constrained control landscape for population transfer in a two-level
  system.
\newblock {\em Phys. Chem. Chem. Phys.}, 17:3164, 2015.

\bibitem{Tibbetts2015}
Katharine~Moore Tibbetts, Maryam Tarazkar, Timothy Bohinski, Dmitri~A Romanov,
  Spiridoula Matsika, and Robert~J Levis.
\newblock Controlling the dissociation dynamics of acetophenone radical cation
  through excitation of ground and excited state wavepackets.
\newblock {\em J. Phys. B: At., Mol. Opt. Phys.}, 48(16):164002, 2015.

\bibitem{Timoney2008}
N.~Timoney, V.~Elman, S.~Glaser, C.~Weiss, M.~Johanning, W.~Neuhauser, and Chr.
  Wunderlich.
\newblock Error-resistant single-qubit gates with trapped ions.
\newblock {\em Phys. Rev. A}, 77:052334, may 2008.

\bibitem{Torrontegui2013}
Erik Torrontegui, Sara Ib\'{a}\~{n}ez, Sofia Mart\'{i}nez-{G}araot, Michele
  Modugno, Adolfo del Campo, David Gu\'{e}ry-{O}delin, Andreas Ruschhaupt,
  Xi~Chen, and Juan~Gonzalo Muga.
\newblock Shortcuts to adiabaticity.
\newblock In Ennio Arimondo, Paul~R. Berman, and Chun~C. Lin, editors, {\em
  Advances in Atomic, Molecular, and Optical Physics}, volume~62 of {\em
  Advances In Atomic, Molecular, and Optical Physics}, pages 117--169. Academic
  Press, 2013.

\bibitem{Tosner2006}
Z.~Tosner, S.~J. Glaser, N.~Khaneja, and N.~C. Nielsen.
\newblock Effective hamiltonians by optimal control: Solid-state nmr
  double-quantum planar and isotropic dipolar recoupling.
\newblock {\em J. Chem. Phys.}, 125:184502/1--10, 2006.

\bibitem{Tosner2009}
Zden\v{e}k To\v{s}ner, Thomas Vosegaard, Cindie Kehlet, Navin Khaneja,
  Steffen~J. Glaser, and Niels~Chr. Nielsen.
\newblock Optimal control in {NMR} spectroscopy: Numerical implementation in
  {SIMPSON}.
\newblock {\em J. Magn. Reson.}, 197(2):120--134, apr 2009.

\bibitem{TralleroHerrero2005}
C.~Trallero-{H}errero, D.~Cardoza, T.~C. Weinacht, and J.~L. Cohen.
\newblock Coherent control of strong field multiphoton absorption in the
  presence of dynamic stark shifts.
\newblock {\em Phys. Rev. A}, 71:013423, 2005.

\bibitem{TralleroHerrero2006}
Carlos Trallero-{H}errero, J.~L. Cohen, and Thomas Weinacht.
\newblock Strong-field atomic phase matching.
\newblock {\em Phys. Rev. Lett.}, 96:063603, feb 2006.

\bibitem{Treutlein2006}
Philipp Treutlein, Theodor~W. H\"ansch, Jakob Reichel, Antonio Negretti,
  Markus~A. Cirone, and Tommaso Calarco.
\newblock Microwave potentials and optimal control for robust quantum gates on
  an atom chip.
\newblock {\em Phys. Rev. A}, 74:022312, aug 2006.

\bibitem{Turinici2000}
Gabriel Turinici.
\newblock On the controllability of bilinear quantum systems.
\newblock In {\em Mathematical Models and Methods for Ab Initio Quantum
  Chemistry}, volume~74 of {\em Lecture Notes in Chemistry}, pages 75--92.
  Springer Berlin Heidelberg, 2000.

\bibitem{Turinici2004}
Gabriel Turinici and Herschel Rabitz.
\newblock Optimally controlling the internal dynamics of a randomly oriented
  ensemble of molecules.
\newblock {\em Phys. Rev. A}, 70:063412, 2004.

\bibitem{Tzvetkova2007}
Pavleta Tzvetkova, Svetlana Simova, and Burkhard Luy.
\newblock Pehsqc: A simple experiment for simultaneous and sign-sensitive
  measurement of ((1)j(ch)+d-ch) and ((2)j(hh)+d-hh) couplings.
\newblock {\em J. Magn. Reson.}, 186(2):193--200, jun 2007.

\bibitem{Underwood2003}
Jonathan~G. Underwood, Michael Spanner, Misha~Yu. Ivanov, Jeff Mottershead,
  Benjamin~J. Sussman, and Albert Stolow.
\newblock Switched wave packets: A route to nonperturbative quantum control.
\newblock {\em Phys. Rev. Lett.}, 90:223001, jun 2003.

\bibitem{Untidt1998}
T~Untidt, T~Schulte-{H}erbr\"uggen, B~Luy, SJ~Glaser, C~Griesinger,
  OW~Sorensen, and NC~Nielsen.
\newblock Design of nmr pulse experiments with optimum sensitivity:
  coherence-order-selective transfer in i2s and i3s spin systems.
\newblock {\em Mol. Phys.}, 95(5):787--796, 1998.

\bibitem{Damme2014}
L.~Van~{D}amme, R.~Zeier, S.~J. Glaser, and D.~Sugny.
\newblock Application of the {P}ontryagin maximum principle to the time-optimal
  control in a chain of three spins with unequal couplings.
\newblock {\em Phys. Rev. A}, 90:013409, 2014.

\bibitem{Walle2009}
P.~van~der {W}alle, M.~T.~W. Milder, L.~Kuipers, and J.~L. Herek.
\newblock Quantum control experiment reveals solvation-induced decoherence.
\newblock {\em Proceedings of the National Academy of Sciences},
  106(19):7714--7717, 2009.

\bibitem{Frank2014}
S.~van {F}rank, A.~Negretti, T.~Berrada, R.~B\"{u}cker, S.~Montangero, J.-F.
  Schaff, T.~Schumm, T.~Calarco, and J.~Schmiedmayer.
\newblock Interferometry with non-classical motional states of a
  {B}ose--{E}instein condensate.
\newblock {\em Nat. Commun.}, 5:4009, may 2014.

\bibitem{Vandersypen2005}
L.~M.~K. Vandersypen and I.~L. Chuang.
\newblock Nmr techniques for quantum control and computation.
\newblock {\em Rev. Mod. Phys.}, 76:1037, 2005.

\bibitem{Verstraete2009}
F.~Verstraete, M.~M. Wolf, and J.~I. Cirac.
\newblock Quantum computation and quantum state engineering driven by
  dissipation.
\newblock {\em Nat. Phys.}, 5:633--636, 2009.

\bibitem{Veshtort2006}
M.~Veshtort and R.~G. Griffin.
\newblock {SPINEVOLUTION}: A powerful tool for the simulation of solid and
  liquid state nmr experiments.
\newblock {\em J. Magn. Reson.}, 178:248--282, 2006.

\bibitem{Vesterinen2014}
V.~Vesterinen, O.-P. Saira, A.~Bruno, and L.~DiCarlo.
\newblock Mitigating information leakage in a crowded spectrum of weakly
  anharmonic qubits.
\newblock arXiv:1405.0450 [cond-mat.mes-hall], may 2014.

\bibitem{Viale2010}
A.~Viale and S.~Aime.
\newblock Current concepts on hyperpolarized molecules in {MRI}.
\newblock {\em Curr. Opin. Chem. Biol.}, 14(1):90--96, 2010.

\bibitem{Vidal2002}
G.~Vidal, K.~Hammerer, and J.~I. Cirac.
\newblock Interaction cost of nonlocal gates.
\newblock {\em Phys. Rev. Lett.}, 88:237902, 2002.

\bibitem{Vijay2012}
R.~Vijay, C.~Macklin, D.~H. Slichter, S.~J. Weber, K.~W. Murch, R.~Naik, A.~N.
  Korotkov, and I.~Siddiqi.
\newblock Stabilizing {R}abi oscillations in a superconducting qubit using
  quantum feedback.
\newblock {\em Nature}, 490:77--80, 2012.

\bibitem{Vinding2013}
M.~S. Vinding, I.~I. Maximov, Z.~To\v{s}ner, and N.~Chr. Nielsen.
\newblock Fast numerical design of spatial-selective rf pulses in mri using
  krotov and quasi-newton based optimal control methods.
\newblock {\em J. Chem. Phys.}, 137:054203, 2012.

\bibitem{Vitanov2001}
N.~V. Vitanov, T.~Halfmann, B.~W. Shore, and K.~Bergmann.
\newblock Laser-induced population transfer by adiabatic passage techniques.
\newblock {\em Annu. Rev. Phys. Chem.}, 52:763--809, 2001.

\bibitem{Vosegaard2005}
T.~Vosegaard, C.~Kehlet, N.~Khaneja, S.~J. Glaser, and N.~C. Nielsen.
\newblock Improved excitation schemes for multiple-quantum magic-angle spinning
  for quadrupolar nuclei designed using optimal control theory.
\newblock {\em J. Am. Chem. Soc.}, 127:13768--13769, 2005.

\bibitem{Waldherr2014}
G.~Waldherr, Y.~Wang, S.~Zaiser, M.~Jamali, T.~Schulte-Herbr{\"u}ggen
  ad~H.~Abe, T.~Ohshima, J.~Isoya, J.F. Du, P.~Neumann, and J.~Wrachtrup.
\newblock Quantum error correction in a solid-state hybrid spin register.
\newblock {\em Nature}, 506:204, 2014.

\bibitem{Walther2009}
A.~Walther, B.~Julsgaard, L.~Rippe, Yan Ying, and S.~Kr\"oll und R. Fisher~und
  S.~Glaser.
\newblock Extracting high fidelity quantum computer hardware from random
  systems.
\newblock {\em Phys. Scr. T}, 137:014009, 2009.

\bibitem{Wang2010b}
X.~Wang and S.~G. Schirmer.
\newblock Analysis of lyapunov method for control of quantum systems.
\newblock {\em IEEE Trans. Autom. Control}, 55(10):2259--2270, 2010.

\bibitem{Wang2010}
Xiaoting Wang, Abolfazl Bayat, Sougato Bose, and Sophie~G. Schirmer.
\newblock Global control methods for {G}reenberger-{H}orne-{Z}eilinger-state
  generation on a one-dimensional {I}sing chain.
\newblock {\em Phys. Rev. A}, 82:012330, 2010.

\bibitem{Wang2010a}
Xiaoting Wang, Abolfazl Bayat, S.~G. Schirmer, and Sougato Bose.
\newblock Robust entanglement in antiferromagnetic {H}eisenberg chains by
  single-spin optimal control.
\newblock {\em Phys. Rev. A}, 81:032312, mar 2010.

\bibitem{Warren1993}
W.~S. Warren, H.~Rabitz, and M.~Dahlen.
\newblock Coherent control of quantum dynamics: The dream is alive.
\newblock {\em Science}, 259(5101):1581--1589, 1993.

\bibitem{Watts2015}
P.~Watts, J.~Vala, M.~M. M\"uller, T.~Calarco, K.~B. Whaley, D.~M. Reich, M.~H.
  Goerz, and C.~P. Koch.
\newblock Optimizing for an arbitrary perfect entangler: I. functionals.
\newblock {\em Phys. Rev. A}, 91, 2015.

\bibitem{Weiner2000}
A.~M. Weiner.
\newblock Femtosecond pulse shaping using spatial light modulators.
\newblock {\em Rev. Sci. Instrum.}, 71(5):1929--1960, 2000.

\bibitem{Werschnik2007}
J.~Werschnik and E.~K.~U. Gross.
\newblock Quantum optimal control theory.
\newblock {\em J. Phys. B: At., Mol. Opt. Phys.}, 40(18):175--211, 2007.

\bibitem{Wiseman1993}
H.~M. Wiseman and G.~J. Milburn.
\newblock Quantum theory of optical feedback via homodyne detection.
\newblock {\em Phys. Rev. Lett.}, 70:548--551, 1993.

\bibitem{Wiseman2010}
H.~M. Wiseman and G.~J. Milburn.
\newblock {\em Quantum measurement and control}.
\newblock Cambridge University Press, Cambridge, 2010.

\bibitem{Wollenhaupt2005}
M.~Wollenhaupt, V.~Engel, and T.~Baumert.
\newblock Femtosecond laser photoelectron spectroscopy on atoms and small
  molecules: Prototype studies in quantum control.
\newblock {\em Annu. Rev. Phys. Chem.}, 56:25, 2005.

\bibitem{Wright2005}
Matthew~J Wright, SD~Gensemer, J~Vala, R~Kosloff, and PL~Gould.
\newblock Control of ultracold collisions with frequency-chirped light.
\newblock {\em Phys. Rev. Lett.}, 95(6):063001, 2005.

\bibitem{Wu2012}
R.-B. Wu, R.~Long, J.~Dominy, T.-S. Ho, and H.~Rabitz.
\newblock Singularities of quantum control landscapes.
\newblock {\em Phys. Rev. A}, 86:013405, 2012.

\bibitem{Wuthrich2003}
Kurt W\"{u}thrich.
\newblock {NMR} studies of structure and function of biological macromolecules
  (nobel lecture).
\newblock {\em Angewandte Chemie International Edition}, 42(29):3340--3363,
  2003.

\bibitem{King2008}
D.~Xu, K.~F. King, Y.~Zhu, G.C. McKinnon, and Z.~P. Liang.
\newblock Designing multichannel, multidimensional, arbitrary flip angle rf
  pulses using an optimal control approach.
\newblock {\em Magn. Reson. Med.}, 59:547--560, 2008.

\bibitem{Yamamoto2014}
Naoki Yamamoto.
\newblock Coherent versus measurement feedback: Linear systems theory for
  quantum information.
\newblock {\em Phys. Rev. X}, 4:041029, 2014.

\bibitem{Yan1993}
Y.~J. Yan, R.~E. Gillilan, R.~M. Whitnell, K.~R. Wilson, and S.~Mukamel.
\newblock Optical control of molecular dynamics: Liouville-space theory.
\newblock {\em J. Phys. Chem.}, 97:2320--2333, 1993.

\bibitem{Yi2012}
T.~Yi-Peng, N.~Xin-Fang, L.~Jun, C.~Hong-Wei, Z.~Xian-Yi, P.~Xin-Hua, and
  Du~Jiang-Feng.
\newblock Preparing pseudo-pure states in a quadrupolar spin system using
  optimal control.
\newblock {\em Chinese Phys. Lett.}, 29:127601, 2012.

\bibitem{Yuan2007}
H.~Yuan, S.~J. Glaser, and N.~Khaneja.
\newblock Geodesics for efficient creation and propagation of order along
  {I}sing spin chains.
\newblock {\em Phys. Rev. A}, 76:012316, 2007.

\bibitem{Yuan2005}
H.~Yuan and N.~Khaneja.
\newblock Time optimal control of coupled qubits under nonstationary
  interactions.
\newblock {\em Phys. Rev. A}, 72(4):040301(R), 2005.

\bibitem{Yuan2011}
H.~Yuan and N.~Khaneja.
\newblock Efficient synthesis of quantum gates on a three-spin system with
  triangle topology.
\newblock {\em Phys. Rev. A}, 84:062301, 2011.

\bibitem{Yuan2014}
H.~Yuan, D.~Wei, Y.~Zhang, S.~Glaser, and N.~Khaneja.
\newblock Efficient synthesis of quantum gates on indirectly coupled spins.
\newblock {\em Phys. Rev. A}, 89:042315, 2014.

\bibitem{Yuan2008}
H.~Yuan, R.~Zeier, and N.~Khaneja.
\newblock Elliptic functions and efficient control of {I}sing spin chains with
  unequal couplings.
\newblock {\em Phys. Rev. A}, 77:032340, 2008.

\bibitem{Yuan2012}
Haidong Yuan, Christiane~P. Koch, Peter Salamon, and David~J. Tannor.
\newblock Controllability on relaxation-free subspaces: On the relationship
  between adiabatic population transfer and optimal control.
\newblock {\em Phys. Rev. A}, 85:033417, March 2012.
\newblock arXiv:1004.4050.

\bibitem{Zeier2004}
R.~Zeier, M.~Grassl, and Th. Beth.
\newblock Gate simulation and lower bounds on the simulation time.
\newblock {\em Phys. Rev. A}, 70:032319, 2004.

\bibitem{Zeier2011}
R.~Zeier and T.~Schulte-Herbr\"{u}ggen.
\newblock Symmetry principles in quantum systems theory.
\newblock {\em J. Math. Phys.}, 52:113510, 2011.

\bibitem{Zhang2013}
F.~Zhang, G. anf~Schilling, S.~J. Glaser, and C.~Hilty.
\newblock Chemical shift correlations from hyperpolarized {NMR} using a single
  shot.
\newblock {\em Anal. Chem.}, 85:2875--2881, 2013.

\bibitem{Zhang2011}
Y.~Zhang, M.~Lapert, D.~Sugny, M.~Braun, and S.~J. Glaser.
\newblock Time-optimal control of spin 1/2 particles in the presence of
  radiation damping and relaxation.
\newblock {\em J. Chem. Phys.}, 134:054103, 2011.

\bibitem{Zhu1998}
W.~Zhu, J.~Botina, and H.~Rabitz.
\newblock Rapidly convergent iteration methods for quantum optimal control of
  population.
\newblock {\em J. Chem. Phys.}, 108(5):1953--1963, 1998.

\bibitem{Zimboras2015}
Z.~Zimbor{\'a}s, R.~Zeier, T.~Schulte-Herbr{\"u}ggen, and D.~Burgarth.
\newblock Symmetry decides simulability of effective interactions.
\newblock {http://arxiv.org/abs/1504.07734}, 2015.

\end{thebibliography}

\end{document}